\documentclass[aps,showpacs,nofootinbib,superscriptaddress]{revtex4}
\usepackage{epsf}
\usepackage{graphicx}
\usepackage{amsmath}
\usepackage{rotating}
\usepackage{young}

\usepackage{amsfonts}
\usepackage{amssymb}
\usepackage{bm}

\newcommand{\bfn}{{\bm {n}}}
\newcommand{\bfJ}{{\bm {J}}}

\newcommand{\pion}{{\pi}}
\newcommand{\kaon}{{K}}
\newcommand{\kaonb}{{\bar{K}}}
\newcommand{\kaonS}{{K^*}}
\newcommand{\kaonSb}{{\bar{K}^*}}
\newcommand{\D}{{D}}
\newcommand{\Db}{{\bar{D}}}
\newcommand{\Ds}{{D_s}}
\newcommand{\Dsb}{{\bar{D}_s}}
\newcommand{\DS}{{D^*}}
\newcommand{\DSb}{{\bar{D}^*}}
\newcommand{\DsS}{{D_s^*}}
\newcommand{\DsSb}{{\bar{D}_s^*}}
\newcommand{\etap}{{\eta^\prime}}
\newcommand{\etac}{{\eta_c}}
\newcommand{\Jpsi}{{\psi}}
\newcommand{\Nucleon}{{N}}
\newcommand{\Lambdac}{{\Lambda_c}}
\newcommand{\SigmaS}{{\Sigma^*}}
\newcommand{\Sigmac}{{\Sigma_c}}
\newcommand{\SigmacS}{{\Sigma_c^*}}
\newcommand{\Cascada}{{\Xi}}
\newcommand{\Cascadac}{{\Xi_c}}
\newcommand{\CascadaS}{{\Xi^*}}
\newcommand{\CascadacS}{{\Xi_c^*}}
\newcommand{\Cascadacc}{{\Xi_{cc}}}
\newcommand{\CascadaccS}{{\Xi_{cc}^*}}
\newcommand{\Cascadacp}{{\Xi_c^\prime}}

\newcommand{\Omegac}{{\Omega_c}}
\newcommand{\Omegacc}{{\Omega_{cc}}}
\newcommand{\OmegacS}{{\Omega_c^*}}
\newcommand{\OmegaccS}{{\Omega_{cc}^*}}
\newcommand{\Omegaccc}{{\Omega_{ccc}}}

\newcommand{\Qur}{{Q^\dagger_{u\uparrow}}}
\newcommand{\Quj}{{Q^\dagger_{u\downarrow}}}
\newcommand{\Qdr}{{Q^\dagger_{d\uparrow}}}
\newcommand{\Qdj}{{Q^\dagger_{d\downarrow}}}
\newcommand{\Qsr}{{Q^\dagger_{s\uparrow}}}
\newcommand{\Qsj}{{Q^\dagger_{s\downarrow}}}
\newcommand{\Qcr}{{Q^\dagger_{c\uparrow}}}
\newcommand{\Qcj}{{Q^\dagger_{c\downarrow}}}

\newcommand{\Qubr}{{Q^\dagger_{\bar{u}\uparrow}}}
\newcommand{\Qubj}{{Q^\dagger_{\bar{u}\downarrow}}}
\newcommand{\Qdbr}{{Q^\dagger_{\bar{d}\uparrow}}}
\newcommand{\Qdbj}{{Q^\dagger_{\bar{d}\downarrow}}}
\newcommand{\Qsbr}{{Q^\dagger_{\bar{s}\uparrow}}}
\newcommand{\Qsbj}{{Q^\dagger_{\bar{s}\downarrow}}}
\newcommand{\Qcbr}{{Q^\dagger_{\bar{c}\uparrow}}}
\newcommand{\Qcbj}{{Q^\dagger_{\bar{c}\downarrow}}}

\newcommand{\uno}{{\bf 1}}
\newcommand{\dos}{{\bf 2}}
\newcommand{\tres}{{\bf 3}}
\newcommand{\cuatro}{{\bf 4}}
\newcommand{\ocho}{{\bf 8}}
\newcommand{\diez}{{\bf 10}}
\newcommand{\quince}{{\bf 15}}
\newcommand{\dieciseis}{{\bf 16}}
\newcommand{\veinte}{{\bf 20}}
\newcommand{\veintisiete}{{\bf 27}}

\newcommand{\cincuentayseis}{{\bf 56}}
\newcommand{\sesentaytres}{{\bf 63}}
\newcommand{\cientoveinte}{{\bf 120}}
\newcommand{\cientosesentayocho}{{\bf 168}}
\newcommand{\setecientosveinte}{{\bf 720}}
\newcommand{\novecientoscuarentaycinco}{{\bf 945}}
\newcommand{\mildoscientostreintaydos}{{\bf 1232}}
\newcommand{\dosmilquinientosveinte}{{\bf 2520}}
\newcommand{\cuatromilsetecientoscicuentaydos}{{\bf 4752}}
\newcommand{\trecemilcientocuatro}{{\bf 13104}}

\def\slashchar#1{\setbox0=\hbox{$#1$}
   \dimen0=\wd0 \setbox1=\hbox{/} \dimen1=\wd1
   \ifdim\dimen0>\dimen1 \rlap{\hbox to \dimen0{\hfil/\hfil}} #1
   \else  \rlap{\hbox to \dimen1{\hfil$#1$\hfil}} / \fi}

\newcommand{\beq}{\begin{equation}}
\newcommand{\eeq}[1]{\label{#1} \end{equation}}

\begin{document}

\title{The $s$-wave charmed baryon resonances from a coupled-channel approach 
with heavy quark symmetry}

\author{C.~Garc{\'\i}a-Recio}
\affiliation{Departamento de
 F{\'\i}sica At\'omica, Molecular y Nuclear,\\ Universidad de Granada,
 E-18071 Granada, Spain}
\author{V.K.~Magas} \affiliation{Departament d'Esctructura i
  Constituents de la Mat\`eria and Institut de Ci\`encies del Cosmos,
  \\ Universitat de Barcelona, E-08028 Barcelona, Spain}
\author{T.~Mizutani} 
\affiliation{Department of Physics, \\ Virginia
  Polytechnic Institute and State University, Blacksburg, VA 24061, USA}
\author{J.~Nieves}
\affiliation{Instituto de F{\'\i}sica Corpuscular (centro mixto CSIC-UV)\\
Institutos de Investigaci\'on de Paterna, Aptdo. 22085, 46071, Valencia, Spain
} 
\author{A. Ramos}\affiliation{Departament d'Esctructura i
  Constituents de la Mat\`eria and Institut de Ci\`encies del Cosmos,
  \\ Universitat de Barcelona, E-08028 Barcelona, Spain}
\author{L.L.~Salcedo}
\affiliation{Departamento de
 F{\'\i}sica At\'omica, Molecular y Nuclear,\\ Universidad de Granada,
 E-18071 Granada, Spain}
\author{L.~Tolos}
\affiliation{Frankfurt Institute for Advanced Studies, 
Johann Wolfgang Goethe Universitat, 60054 Frankfurt am Main,
Germany}
\affiliation{Theory Group, KVI, University of Groningen, Zernikelaan 25, 9747 AA
Groningen, The Netherlands}


\begin{abstract}

We study charmed baryon resonances which are generated dynamically
within a unitary meson-baryon coupled channel model that treats the
heavy pseudoscalar and vector mesons on equal footing as required by
heavy-quark symmetry.  It is an extension of recent SU(4) models with
$t$-channel vector meson exchanges to an SU(8) spin-flavor scheme, but
differs considerably from the SU(4) approach in how the strong
breaking of the flavor symmetry is implemented.  Some of our
dynamically generated states can be readily assigned to recently
observed baryon resonances, while others do not have a straightforward
identification and require the compilation of more data as well as an
extension of the model to $d$-wave meson-baryon interactions and
$p$-wave coupling in the neglected $s$- and $u$-channel diagrams.  Of
several novelties, we find that the $\Lambda_c(2595)$, which emerged
as a $ND$ quasi-bound state within the SU(4) approaches, becomes
predominantly a $ND^*$ quasi-bound state in the present SU(8) scheme.

\end{abstract}

\maketitle

\section{Introduction}

In hadron physics establishing whether a resonance has the usual
$q\bar q$ or $qqq$ structure or better qualifies as being dynamically
generated through the multiple scattering of their meson and/or baryon
components, has kept an active topic of research. In recent years, the
introduction of unitarity constraints in coupled channels methods has
led to extensions of the chiral effective theories that can be applied
at much higher energies and in the vicinity of resonances
\cite{report}. In the particular case of baryons, several resonances
are firm candidates to be identified with states generated dynamically
from the interaction of mesons of the $0^-$ octet Goldstone bosons
with the $1/2^+$ octet ground state baryons
\cite{Kai95,Kaiser:1995cy,Kaiser:1996js,OR98,kr98,Nacher:1999vg,Meissner:1999vr,Oller:2000fj,Nieves:2001wt,Inoue:2001ip,Lutz:2001yb,Garcia-Recio:2002td,ORB02,Ramos:2002xh,Oller:2005ig,Borasoy:2005ie,Borasoy:2006sr,Oller:2006jw,Hyodo:2008xr}.
Then interesting predictions such as the two-pole nature of the
$\Lambda(1405)$
\cite{Oller:2000fj,Garcia-Recio:2002td,Jido:2002yz,Jido:2003cb,Garcia-Recio:2003ks,Hyodo:2002pk,Nam:2003ch}
have found experimental confirmation \cite{Thomas,Prakhov}, as
discussed in Ref.~\cite{Magas:2005vu}. Later a number of works has
also been devoted to the study of dynamically generated $J^P=3/2^-$
states
\cite{Kolomeitsev:2003kt,Sarkar:2004jh,Sarkar:2005ap,Roca:2006sz,Hyodo:2006uw,Doring:2006ub,Doring:2005bx,Doring:2006pt,Garcia-Recio:2005hy,QNP06,Toki:2007ab}.
Early works considered the interaction of pseudoscalar $0^-$ mesons
with the baryons of the $3/2^+$ decuplet. The incorporation of vector
mesons into the above-mentioned coupled-channel picture was pursued in
studies of axial $1^+$ meson resonances of
Refs.~\cite{Roca:2005nm,Lutz:2003fm}, but until very recently those
degrees of freedom have not been considered in the baryon-meson
sector. In Ref.~\cite{Garcia-Recio:2005hy} the $(8_3,
1_3)$\footnote{Here and below we use the notation $\mu_{2J+1}$, for
the SU(3) multiplet $\mu$ of spin $J$.} nonet of lowest-lying vector
meson ($\rho(770), K(890),\bar K(890), \omega(780)$ and $\phi(1020)$)
degrees of freedom was incorporated into the Weinberg-Tomozawa (WT)
meson-baryon chiral Lagrangian by using a scheme which relies on the
SU(6) spin--flavor symmetry. The corresponding Bethe-Salpeter (BS)
equation successfully reproduces the previous SU(3)--flavor WT results
for the lowest-lying $s$- and $d$- wave, negative parity baryon
resonances ($N(1535)$, $N(1650)$, $\Lambda(1390)$, $\Lambda(1405)$,
$\Lambda(1520)$, $\Lambda(1690)$, $\Lambda(1670)$, $\Xi(1620)$,
$\Xi(1690)$, $\cdots$). It also provides some information on the
dynamics of the heavier ones such as $\Lambda(1800)$ or
$\Lambda(2325)$ which have sizable $N{\bar K}^*$ and $\Lambda \omega$
couplings, respectively~\cite{QNP06,Toki:2007ab}.

Recently the attention has turned towards studying resonances with
charm degrees of freedom, motivated by the discovery of quite a few
new states, as reported by the CLEO, Belle and BaBar Collaborations
\cite{2880-Artuso:2000xy,Mizuk:2004yu,Chistov:2006zj,Aubert:2006je,2940-Aubert:2006sp,2880-Abe:2006rz}.
At first sight, the physics in the charm $C=1$ sector bears a strong
resemblance to the phenomenology seen in the $\bar K N$ dynamics upon
replacing an $s$ quark by a $c$ quark\footnote{ Besides kinematical
similarities (open and closed channels near the resonance mass ...),
the underlying dynamics might be different as a strange quark is
related to chiral symmetry and a charm quark to heavy-quark symmetry,
as we will discuss below.}. This is reinforced by an apparent
similarity between the two $I=0$ $s$-wave $\Lambda(1405)$ and
$\Lambda_c(2595)$ resonances. The mass of the former lies in between
the $\pi\Sigma$ and $\bar K N$ channel thresholds, to which it couples
strongly.  The $\Lambda_c(2595)$ lies below the $DN$ and just slightly
above the $\pi\Sigma_c$ thresholds. This similarity was exploited in
the first exploratory work \cite{Tolos:2004yg} where free space
amplitudes were constructed from a set of separable coupled-channel
interactions obtained from chirally motivated Lagrangians upon
replacing the $s$ quark by the $c$ quark.  The $I=0$ $\Lambda_c(2595)$
was generated as a $DN$ $s$-wave resonant state of binding energy
$\approx 200$ MeV with a width of $\approx 3$ MeV. While giving the
first indication that the $\Lambda_c(2595)$ could have a dynamical
origin, this work ignored the strangeness degree of freedom due to its
very construction. Therefore, the $\pi$ and $K$ (Goldstone) mesons
were not treated on an equal footing, and the role of some channels
that would appear from the corresponding SU(4) meson and baryon
multiplets, such as $D_s \Lambda$, $D_s \Sigma$, $K\Xi_c$ and
$K\Xi_c^\prime$, was ignored.  A different approach which respected
the proper symmetries, was attempted in
Ref.~\cite{Lutz:2003jw}. There, charmed baryon resonances were
generated dynamically from the scattering of Goldstone bosons off the
ground-state $J^P=\frac{1}{2}^+$ charmed baryons. The $C=1$, $S=I=0$
resonance found at 2650 MeV was identified with the $\Lambda_c(2595)$
in spite of the fact that, due to the strong coupling to the
$\pi\Sigma_c$ states, its width came out to be more than twenty times
larger than the experimental width of about 4 MeV. The trouble with
this model is that, due to its very construction, it does not account
for the coupling to the $DN$ channel which contributes strongly to
generating the narrow $\Lambda_c(2595)$ according to
Ref.~\cite{Tolos:2004yg}.  A substantial improvement came in a recent
work \cite{Hofmann:2005sw} in which the above mentioned shortcomings
have been overcome by exploiting the universal vector meson coupling
hypothesis to break the SU(4) symmetry in a convenient and
well-defined manner. This is done by a $t$-channel exchange of vector
mesons between pseudoscalar mesons and baryons in such a way that
chiral symmetry is preserved in the light meson sector, while the
interaction is still of the WT type. We note that in the subsequent
sections this approach is referred as TVME (t-channel vector meson
exchange model).  The model generates a narrow $C=1$, $I=0$ resonance
that is identified with the $\Lambda_c(2595)$, together with other
resonances in other strangeness-charm sectors with $J=1/2^-$.  An
extension of the model to $d$-wave $J=3/2^-$ resonances,
generated by the interactions of pseudoscalar mesons with baryons of
the $J=3/2^+$ SU(4) multiplet, was developed in
Ref.~\cite{Hofmann:2006qx}.  Finally, some modifications over the
model in Ref.~\cite{Hofmann:2005sw}, both in the kernel and in the
regularization scheme, were implemented in
Ref.~\cite{Mizutani:2006vq}, obtaining qualitatively similar results
to the earlier approaches. In all these works the zero-range limit in
the $t$-channel exchange of vector mesons is identified as the driving
force for the $s$-wave scattering of pseudoscalar mesons off the
baryon ground states, and is used to carry out a coupled-channel
analysis.

A serious limitation of this SU(4) TVME model is that, whereas the
pseudoscalar mesons $D$ and $D_s$ are included in the coupled-channel
dynamics, their vector partners $D^*$ and $D_s^*$ are completely
left out.  This is not justified from the point of view of Heavy Quark Symmetry (HQS), which is a proper
QCD  spin-flavor symmetry~\cite{IW89,Ne94,MW00} when the
quark masses become much larger than the typical confinement scale,
$\Lambda_{\rm QCD}$.

HQS predicts that all types of spin interactions vanish for infinitely
massive quarks: the dynamics is unchanged under arbitrary
transformations in the spin of the heavy quark ($Q$).  The
spin-dependent interactions are proportional to the chromomagnetic
moment of the heavy quark, hence are of the order of $1/m_Q$. The
total angular momentum $\vec{J}$ of the hadron is a conserved
quantity, and the spin of the heavy quark $\vec{S}_Q$ is also
conserved in the $m_Q\to \infty$ limit. Consequently, the spin of the
light degrees of freedom $\vec{S}_l=\vec{J}-\vec{S}_Q$ is conserved in
that limit. Thus, heavy hadrons come in doublets (unless $s_l=0$, with
$\vec{S}_l^2=s_l(s_l+1)$) containing states with total spin
$j_\pm=s_l\pm 1/2$ obtained by combining the spin of the light degrees
of freedom with the spin of the heavy quark $s_Q=1/2$. These doublets
are degenerated in the $m_Q\to \infty$ limit. This is the case for the
ground state mesons $D$ and $D^*$ or $D_s$ and $D^*_s$ which are
composed of a charm quark with $s_Q=1/2$ and light degrees of freedom
with $s_l=1/2$, forming a multiplet of hadrons with spin 0 and 1 and
negative parity.  The entire multiplet of degenerate states should be
treated in any HQS inspired formalism as a single field that
transforms linearly under the heavy quark symmetries~\cite{Ne94,MW00}.
For finite charm quark mass, the pseudoscalar and vector $D$ meson
masses differ in just about one pion mass\footnote{One can easily
deduce that $m_D-m_{D^*} = {\cal O}\left(
\frac{1}{m_D+m_{D^*}}\right)$}, even less for the strange charmed
mesons, thus it is reasonable to expect, for instance, that the
coupling $DN-D^*N$ might play an important role, and could well be
bigger than that played by some of the other channel couplings
considered in the TVME model, say, the $DN-\Lambda_c\eta$.

The mass splitting between pseudoscalar
 and vector mesons is comparatively larger in the light
sectors, for example,
between the $\pi$ and $\rho$ mesons or the $K$ and the $K^*$.  This appears to
justify the usual approximation of ignoring vector
degrees of freedom in the SU(3) coupled-channel studies in
$s$-wave meson ($8_1$)--meson ($8_1$) and meson ($8_1$)--baryon
($8_2$, $10_4$) scattering and resonances, within different
renormalization schemes.
Nevertheless, as already mentioned, vector mesons have been
incorporated in the WT meson-baryon chiral Lagrangian
\cite{Garcia-Recio:2005hy,QNP06,Toki:2007ab,GarciaRecio:2006wb} through a scheme that
treats the six states of a light quark ($u$, $d$ or $s$ with spin up,
$\uparrow$, or down, $\downarrow$) as equivalent, leading to an SU(6)
symmetric scheme.  Here we will extend this model to SU(8)
spin--flavor symmetry to account also for the charm quark degree of
freedom.  Therefore, compared to the TVME approach, our
coupled-channel analysis will consider the additional role of vector
mesons ($D^*$, $D^*_s$, $\rho$, $K^*$,...). Moreover, both models
differ on how they implement the strong breaking of the SU(4) flavor
symmetry observed in nature.

We should clarify here that the extension of the WT
interaction to vector mesons~\cite{Garcia-Recio:2005hy} and to flavor
SU(4) is a model. Chiral symmetry is a symmetry of the light quark
sector, i.e. at best for flavor SU(3), but not for SU(4). Chiral
symmetry breaking fixes model independently the strength of the lowest
order interaction between Goldstone bosons and other hadrons (here
baryons) - the WT interaction. However, chiral symmetry
does not fix the interaction between vector mesons and baryons. On the
other hand, HQS connects vector and pseudoscalar
mesons which contain charm quarks. It does not tell anything about
mesons made out of light quarks. Nevertheless, it is clearly appealing
to have a model for four flavors and for pseudoscalar and vector
mesons which reduces to the WT interaction in the
sector where Goldstone bosons are involved and which incorporates
heavy-quark symmetry in the sector where charm quarks are
involved. 
The model assumption in the present extension does not appear to be
easy to justify, but we want to try it based on the reasonable 
outcome of the  SU(6)  extension in the three-flavor sectors
on the one hand, and formal plausibility of how the SU(4) WT
interaction in the charmed PS meson- SU(3) baryon interaction did come out in the
vector meson exchange picture as discussed in the TVME approach, on the other hand.
We just want to  fuse the two approaches.
Note than in our
case, we improve on previous models since we incorporate HQS in the charm sector.

In this exploratory work, we have studied only non-strange  single
charmed resonances. Our approach should work better close to the
relevant thresholds, and in particular to describe the lowest lying
resonances in each of the examined $IJ$ (meson--baryon isospin and total
angular momentum) sectors. Actually, working near threshold is the only
justification to ignore the contribution of $d$-wave meson-baryon
interactions and of $p$-wave couplings, through the neglected $s$- and
$u$-channel diagrams, which otherwise might play an important
role. These contributions have been also traditionally ignored in
the previous successful SU(3) flavor studies of lowest--lying
charmless $s-$ and $d-$wave
resonances~\cite{Kai95,Kaiser:1995cy,Kaiser:1996js,OR98,Nacher:1999vg,Meissner:1999vr,Oller:2000fj,Nieves:2001wt,Inoue:2001ip,Lutz:2001yb,Garcia-Recio:2002td,ORB02,Ramos:2002xh,Oller:2005ig,Borasoy:2005ie,Borasoy:2006sr,Oller:2006jw,Kolomeitsev:2003kt,Sarkar:2004jh,Sarkar:2005ap,Roca:2006sz,Hyodo:2006uw,Doring:2006ub,Doring:2005bx,Doring:2006pt,Garcia-Recio:2005hy,QNP06,Toki:2007ab},
which we aim to extend to the charm sector.

The present work is organized as follows. In Sec.~\ref{sec:su8} we
give the details of our SU(8) extension of the WT meson-baryon
Lagrangian. The basis of states and the general form of the Lagrangian
are presented in Sec.~\ref{subsec:su8a}, while the symmetry breaking
effects are discussed in Sec.~\ref{subsec:su8-break}. The
unitarization method and renormalization procedure are described in
Sect.~\ref{sec:BS}. Our SU(8) results are discussed in
Sec.~\ref{sec:su8results}, while Sec.~\ref{sec:su4results} is devoted
to a comparison with results obtained in a reduced SU(4) model.  A
summary of our conclusions is presented in Sec.~\ref{sec:conclusions}.

\section{SU(8) Extension of the WT Meson-Baryon Lagrangian}

\label{sec:su8}

\subsection{SU(8) symmetry}

\label{subsec:su8a}

Since we assume that the pure SU(4) (flavor) transformations
commute with the pure SU(2) (spin) transformations, it follows
that an SU(8) multiplet can be decomposed into SU(4) multiplets
each with  definite total spin. With the inclusion of spin there are
64 quark--antiquark ($q\bar{q}$) states, and the irreducible reduction of the SU(8) group
representation (denoting the SU(4) multiplets of
dimensionality $\bfn$ and spin $J$ by ${\bfn_{\dos\bfJ+\uno}}$)
reads
\begin{equation}
\ocho\otimes \ocho^* = \sesentaytres \oplus\uno =
\underbrace{(\quince_\uno \oplus \quince_\tres \oplus
  \uno_\tres)}_{\sesentaytres } \oplus \uno_\uno \,.
\end{equation}
Assuming that the lowest bound state is an $s$-state, and since the
relative parity of a fermion--antifermion pair is odd, the SU(4)
\quince-plet of pseudoscalar ($D_s, D, K, \pi,\eta,\eta_c, {\bar K},
{\bar D}, {\bar D}_s$) and the \dieciseis-plet of vector ($D_s^*,
D^*,K^*, \rho,\omega, J/\Psi, {\bar K}^{*}, {\bar D}^*, {\bar D}_s^*,
\phi$) mesons are placed in the $\sesentaytres$ representation.  Note
that the $\sesentaytres$ allows nine light vector mesons but only
eight $0^-$ light mesons. A ninth $0^-$ meson must go into the $\uno$
of SU(8).  We use pure $c\bar c$ wave functions for the charmonium
states $\eta_c$ and $J/\Psi$, and the usual quark content,
$\eta=\frac{1}{\sqrt 6}\left ( u\bar u + d\bar d-2 s\bar s\right )$,
$\eta'=\frac{1}{\sqrt 3}\left ( u\bar u + d\bar d+ s\bar s\right )$,
$\omega=\frac{1}{\sqrt 2}\left ( u\bar u + d\bar d\right)$, $\phi=-s
\bar s$, for the physical isoscalar light mesons. Such a specification
induces some mixing between the isoscalar SU(4) mathematical states to
build the physical states. Detailed spin-flavor wave functions,
specifying our conventions are given in Appendix~\ref{sec:spwf}.
Mesons of spin $J>1$ can be understood as states of the $q{\bar q}$
system with orbital angular momentum $L>0$, or molecular-type $q q
{\bar q}{\bar q}$ meson-meson states.

In the case of baryons, with the inclusion of the spin, one finds
512 three quark states: 
\begin{eqnarray}
&& \ocho\otimes \ocho \otimes \ocho = \cientoveinte \oplus \cincuentayseis
\oplus \cientosesentayocho \oplus \cientosesentayocho  =
 \\
&& \underbrace{(\veinte_\dos \oplus
  \veinte^\prime_\cuatro)}_{\cientoveinte}  \oplus \underbrace{(\cuatro_\cuatro\oplus \veinte_\dos)}_{\cincuentayseis}
\oplus 2\times \underbrace{(\veinte^\prime_\dos \oplus
\veinte_\cuatro\oplus \veinte_\dos \oplus
\cuatro_\dos)}_{\cientosesentayocho}  \ ,
\nonumber
\end{eqnarray}
where $\begin{Young} & \cr \cr \end{Young}$ and $\begin{Young} & & \cr 
\end{Young}$  are the $\veinte$ and $\veinte^\prime$ SU(4) representations,
respectively. It is natural to assign the lowest--lying baryons to
the $\cientoveinte$ of SU(8).  This is appropriate  because in the light sector it can accommodate
 an octet of spin--$1/2$ baryons and a decuplet of
spin--$3/2$ baryons which are precisely the SU(3)--spin
combinations of the low--lying baryon states ($N,\Sigma,\Lambda,
\Xi$ and $\Delta$, $\Sigma^*$, $\Xi^*$, $\Omega$). The remaining
states in the $\veinte_\dos$ and $\veinte^\prime_\cuatro$ are
completed with the charmed baryons: $\Lambda_c$, $\Sigma_c$,
$\Xi_c$, $\Xi'_c$ ,$\Omega_c$, $\Xi_{cc}$, $\Omega_{cc}$ and
$\Sigma^*_c$, $\Xi^*_c$, $\Omega_c^*$, $\Omega^*_{cc}$,
$\Xi^*_{cc}$, $\Omega_{ccc}$, respectively. Quantum numbers of the
charmed baryons are summarized in Table~\ref{tab:summ}. The
$\cientoveinte$ of SU(8) is totally symmetric, which allows the
baryon to be made of three quarks in $s$-wave (the color
wavefunction being antisymmetric).

\begin{table}

\begin{tabular}{ccccccc}\hline
Baryon &~~~~$S$~~~~&~~~~$J^P$~~~~&~~~~ $I$~~~~&~~~~$S_{\rm light}^\pi$~~~~&
Quark content
& $M_{\rm exp.}$ \cite{pdg06}
\\
       &       &         &   &          &               & [MeV]
\\\hline
$\Lambda_c$& 0 &$\frac12^+$& 0 &$0^+$&$udc$& $2286.46 \pm 0.14$
\\
$\Sigma_c$ & 0 &$\frac12^+$& 1 &$1^+$&$llc$& $2453.6 \pm 0.7$
\\
$\Sigma^*_c$ & 0 &$\frac32^+$& 1 &$1^+$&$llc$& $2518.0 \pm 0.5$
\\
$\Xi_c$ & $-$1 &$\frac12^+$&$\frac12$&$0^+$&$lsc$& $2469.5 \pm 1.5$
\\
$\Xi'_c$ & $-$1 &$\frac12^+$&$\frac12$&$1^+$&$lsc$& $2577 \pm 3$
\\
$\Xi^*_c$ &$-$1&$\frac32^+$&$\frac12$&$1^+$&$lsc$& $2646.3 \pm 1.4$
\\
$\Omega_c$ &$-$2 &$\frac12^+$& 0 &$1^+$&$ssc$& $2697.5 \pm 2.6$
\\
$\Omega^*_c$ &$-$2 &$\frac32^+$& 0 &$1^+$&$ssc$&
\\\hline
$\Xi_{cc}$& 0 &$\frac12^+$& $\frac12$ &$\frac12^+$&$ccl$
\\
$\Xi^*_{cc}$ & 0 &$\frac32^+$&$\frac12$  &$\frac12^+$&$ccl$
\\
$\Omega_{cc}$ & $-1$ &$\frac12^+$& 0 &$\frac12^+$&$ccs$
\\
$\Omega^*_{cc}$ & $-1$ &$\frac32^+$&$0$&$\frac12^+$&$ccs$
\\\hline
$\Omega^*_{ccc}$ & $0$ &$\frac32^+$&$0$& --- &$ccc$

\end{tabular}
\caption{Summary of the quantum numbers and experimental masses of
the baryons containing heavy quarks. $I$, and
  $S_{\rm light}^\pi$ are the isospin, and the spin parity of the light
  degrees of freedom and $S$, $J^P$ are strangeness and the spin parity
  of the baryon ($l$ denotes a light quark of flavor $u$ or
  $d$). Isospin averaged experimental masses are taken
  from Ref.~\protect\cite{pdg06}, with errors counting for the
  mass differences between the members of the same isomultiplet and
  the uncertainties quoted in Ref.~\protect\cite{pdg06}. }
\label{tab:summ}
\end{table}

Here we will focus on the $s$-wave interaction between the
lowest--lying meson ($\sesentaytres$) and the lowest--lying baryon
($\cientoveinte$) SU(8) multiplets at low energies. 
Note that at higher energies, higher
partial waves would become important, so
a suitable treatment of spin-orbit
effects in the SU(8) scheme should be considered.  Now, assuming that the
$s$-wave effective meson--baryon Hamiltonian is SU(8) invariant, and
since the SU(8) decomposition of the product of the $\sesentaytres$
and $\cientoveinte$  representations yields
\begin{eqnarray}
\sesentaytres \otimes \cientoveinte = \cientoveinte \oplus
\cientosesentayocho \oplus \dosmilquinientosveinte \oplus
\cuatromilsetecientoscicuentaydos \,,
\label{eq:su8}
\end{eqnarray}
we conclude that there are only four independent Wigner-Eckart irreducible matrix
elements (WEIME's) each of which being a function of the meson--baryon Mandelstam
variable $s$.  The WEIME's might be constrained by demanding that the
SU(8) amplitudes for the scattering of the Goldstone $J^P=0^-$ mesons
of the pion octet off the $J^P=\frac12^+$ baryons of the nucleon
octet  reduce to those from SU(3) chiral symmetry.  At
leading order in the chiral expansion, these latter amplitudes are
obtained from the WT Lagrangian which, besides hadron masses, only
depends on the pion weak decay constant $f\simeq 90\,$MeV in the
chiral limit.

This procedure is a natural extension of the one derived  
in Ref.~\cite{Garcia-Recio:2005hy} for SU(6) to the spin--flavor SU(8)
symmetry group. If one follows  the arguments in this 
reference, it is clear that not all the SU(3) invariant interactions in
the $(\ocho_\uno)$meson--$(\ocho_\dos)$baryon sector can be extended
to an SU(8) invariant interaction. This is easily  understood because the
number of independent  SU(3) WEIME's to describe the
$(\ocho_\uno)$meson--$(\ocho_\dos)$baryon interaction is six,
as follows trivially from the SU(3) decomposition
\begin{equation}
\ocho \otimes \ocho = \uno\oplus \ocho_s \oplus
\ocho_a \oplus \diez \oplus \diez^* \oplus \veintisiete \ ,  
\label{eq:su3}
\end{equation}
whereas the SU(8) spin-flavor symmetry requires the knowledge of four WEIME's,
as seen from Eq.~(\ref{eq:su8}). 
However, Chiral
Symmetry (CS) at leading order
is much more predictive than SU(3) symmetry, and
it predicts the values of the various SU(3) WEIME's which
otherwise would be totally arbitrary functions of $s$. Indeed, the
WT Lagrangian is not just SU(3)
symmetric but also chiral ($\text{SU}_L(3)\otimes\text{SU}_R(3)$)
invariant. Symbolically,  up to an overall constant,  the WT interaction is
\begin{equation}
{\cal L_{\rm WT}}= {\rm Tr} ( [M^\dagger, M][B^\dagger, B])
\,.
\end{equation}
This structure, dictated by CS, is most suitable to analyze
in the $t$-channel. The mesons $M$ fall in the SU(3) representation
\ocho~  which is also the adjoint representation. The commutator
$[M^\dagger,M]$ indicates a $t$-channel coupling to the $\ocho_a$
(antisymmetric) representation, thus
\begin{equation}
{\cal L_{\rm WT}}= \left((M^\dagger\otimes M)_{\ocho_a}\otimes
(B^\dagger\otimes B)_{\ocho_a}\right)_{\uno}
\,.
\end{equation}
Note that from the group point of view there would be as many different
$t-$channel SU(3) singlet Lagrangians as WEIME's.

For  the SU(8) spin-flavor symmetry, the mesons $M$ fall in the
$\sesentaytres$ which is the adjoint representation and the baryons $B$ 
are found in
the $\cientoveinte$, which is fully symmetric, and the group reductions
\begin{eqnarray}
\sesentaytres\otimes\sesentaytres &=& \uno\oplus\sesentaytres_s
\oplus\sesentaytres_a\oplus\setecientosveinte\oplus\novecientoscuarentaycinco\oplus\novecientoscuarentaycinco^*\oplus\mildoscientostreintaydos
\nonumber \\
\cientoveinte\otimes\cientoveinte^* &=&
\uno\oplus\sesentaytres\oplus\mildoscientostreintaydos\oplus\trecemilcientocuatro 
\end{eqnarray}
lead\footnote{The singlet representation  $\uno$  only appears in the reduction
of the product of one representation by its complex-conjugate. The
couplings of Eq.~(\ref{eq:coupl}) arise because  $\uno$, $\sesentaytres$ and $\mildoscientostreintaydos$ are self-complex conjugate
representations.} to the  total of four different $t-$channel SU(8)
singlet  couplings (that can be used to construct $s$-wave
meson-baryon  interactions)
\begin{eqnarray}
\left((M^\dagger\otimes M)_{\uno}\otimes
(B^\dagger\otimes B)_{\uno}\right)_{\uno}, &&
\left((M^\dagger\otimes M)_{\sesentaytres_a}\otimes
(B^\dagger\otimes B)_{\sesentaytres}\right)_{\uno},
\nonumber \\
\left((M^\dagger\otimes M)_{\sesentaytres_s}\otimes
(B^\dagger\otimes B)_{\sesentaytres}\right)_{\uno}, &&
\left((M^\dagger\otimes M)_{\mildoscientostreintaydos}\otimes
(B^\dagger\otimes B)_{\mildoscientostreintaydos}\right)_{\uno},
\label{eq:coupl}
\end{eqnarray}
which match in  number to that of the independent WEIME's expected
from the group reduction of  Eq.~(\ref{eq:su8}). To ensure that the
SU(8) amplitudes will reduce to those deduced from the
SU(3) WT Lagrangian in the $(\ocho_\uno)$meson--$(\ocho_\dos)$baryon subspace,
we set  all the couplings in Eq.~(\ref{eq:coupl}) to be zero except for
\begin{equation}
{\cal L_{\rm WT}^{\rm SU(8)}}=
\left((M^\dagger\otimes M)_{\sesentaytres_a}\otimes
(B^\dagger\otimes B)_{\sesentaytres}\right)_{\uno}  \,
\label{eq:NOcoupl}
\end{equation}
which is the natural and unique SU(8) extension of the usual SU(3) WT
Lagrangian.\footnote{
Alternatively, the unique spin-flavor symmetric extension of the
    WT Lagrangian  is derived in
    \cite{GarciaRecio:2006wb} for arbitrary number of flavors and
    colors assuming only a non-relativistic reduction for the
    baryons.}  The reduction of this Lagrangian to the SU(6) sector 
reproduces the ${\cal L_{\rm WT}^{\rm SU(6)}}$ found in
Ref.~\cite{Garcia-Recio:2005hy}. The corresponding BS approximation,
which we will discuss below, successfully reproduces 
the previous SU(3)--flavor WT results for the lowest-lying $s$- and
$d$-wave negative parity baryon resonances obtained from scattering of
the mesons of the pion octet off baryons of the nucleon
octet and delta decuplet  ~\cite{QNP06}. To compute the matrix elements of the SU(6)
WT interaction ${\cal L_{\rm WT}^{\rm SU(6)}}$, the SU(6)--multiplet
coupling factors found in~\cite{Carter:1965} were used in
\cite{Garcia-Recio:2005hy, QNP06}. Here,  because of the lack of these
factors for the SU(8) group, we use quark model constructions of hadrons 
with field theoretical methods to express everything
in tensor representations as
described in  Appendix~\ref{sec:spwf}.
Thus, we get the tree level
amplitudes (we use the convention $V=-{\cal L}$):
\begin{equation}
V^{IJSC}_{ab}(\sqrt{s})=
D^{IJSC}_{ab} \frac{\sqrt{s}-M}{2\,f^2}
\left(\sqrt{\frac{E+M}{2M}}\right)^2 \ ,
\label{eq:vsu8}
\end{equation}
where the last factor is due to the spinor normalization convention:
$\bar u u =\bar v v =1$, as in Refs.~\cite{ORB02,Jido:2003cb}.  In the
above expression $IJSC$ are the meson--baryon isospin, total angular
momentum, strangeness and charm quantum numbers, $M~(E)$ the common
mass (CM energy) of the baryons placed in the $\cientoveinte$ SU(8)
representation, and $D^{IJSC}$ a matrix in the coupled channel
space. For instance in the $J=3/2, I=2, S=0, C=1$ sector we consider 5
channels, $\Sigma_c\rho,\Delta D, \Sigma_c^* \pi, \Delta D^*$, and
$\Sigma_c^* \rho$, without including channels with double charmed
baryons since they are much higher in energy. The $D$ matrices
relevant for this work are presented in
Tables~\ref{tab:i0j12s0c1}--\ref{tab:i2j52s0c1} of Appendix
\ref{app:tables}. Note that we truncate the coupled channel space by
not including either double charmed baryon entries, or $\eta_c$, or
$J/\Psi$ mesons, since in this work we are just interested in the
charm $C=1$ sector. Double charmed baryons or $\eta_c$, or $J/\Psi$
mesons will contribute to this sector only in conjunction to another
charmed hadron, and the threshold of the corresponding channel would
be much higher in energy than those of channels involving a single 
charmed hadron.

\subsection{SU(8)-flavor symmetry breaking effects}

\label{subsec:su8-break}

 The SU(8) spin-flavor symmetry is strongly broken in
nature. So we have taken into account mass breaking effects by
adopting the physical hadron masses both in the tree level
interaction of Eq.~(\ref{eq:vsu8}) and in the evaluation of the
kinematical thresholds of  different channels.

Next, we have considered spin-flavor symmetry breaking effects due
to the difference between the weak non-charmed and charmed, as well as
pseudoscalar and vector meson decay constants.  These symmetry
breaking effects are sizable because, for instance, the ratios
$f_D/f_\pi$ of $f_\rho/f_\pi$ deviate from one,  and actually they are of the
order of 1.7. The pseudoscalar meson ($P$) decay constants, $f_P$, are
defined by
\begin{equation}
\langle 0 | {\bar q}_1 \gamma^\mu \gamma_5 q_2 (0) | P(p)\rangle =
-{\rm i} \sqrt{2} f_P p^\mu \label{eq:fp}
\end{equation}
and vector meson ($V$) decay constants, $f_V$, by
\begin{equation}
\langle 0 | {\bar q}_1 \gamma^\mu  q_2 (0) | V(p,\epsilon)\rangle
= \sqrt{2} m_{V}f_{V}\epsilon^\mu \ ,\label{eq:fv}
\end{equation}
where ${\bar q}_1$, $q_2$ are the quark fields, $\epsilon_\mu$  is the polarization vector of the meson,  and
$m_{V}$ its mass.  With the above definitions, HQS predicts
$f_{D_{(s)}}=f_{D^*_{(s)}}$, up to $\Lambda_{\rm QCD}/m_c$
corrections~\cite{MW00}, which guarantees that the normalizations
of the coupling constants in Eqs.~(\ref{eq:fp}) and~(\ref{eq:fv})
are consistent. For light mesons there exist sizable corrections
to the HQS-type relation $f_P=f_{V}$.
\begin{table}
\begin{center}
\begin{tabular}{ccc}\hline
Decay Constant & Exp. Value [MeV] & Theor. Estimates [MeV] \\\hline
$f_{\pi}$ & $92.4 \pm 0.1 \pm 0.3$~\cite{pdg06} & $-$ \\
$f_{K}$ & $113.0 \pm 1.0 \pm 0.3$~\cite{pdg06} & $-$ \\
$f_{\eta}$ & $\simeq 1.2 f_\pi$~\cite{pdg06} & $-$ \\
$f_{\rho}$ & $\simeq 153$\footnote{From $\Gamma(\rho\to
  e^+e^-)=4\pi\alpha^2 f^2_\rho/3 m_\rho$ and
  $\Gamma(\tau\to \rho
  \nu_\tau)=G^2\cos^2\theta_C f^2_\rho(m^2_\tau-m^2_\rho)^2(2m^2_\rho +
  m^2_\tau)/8\pi m^3_\tau$.\\}   & $-$ \\
$f_{K^*}$ & $\simeq 153$\footnote{From $\Gamma(\tau\to K^*
  \nu_\tau)=G^2\sin^2\theta_C f^2_{K^*}(m^2_\tau-m^2_{K^*})^2(2m^2_{K^*} +
  m^2_\tau)/8\pi m^3_\tau$.\\}   & $-$ \\
$f_{\omega}$ & $\simeq 138$\footnote{From $\Gamma(\omega\to e^+e^-)=4\pi\alpha^2 f^2_\omega/27 m_\omega$.\\}
& $-$ \\
$f_{\phi}$ & $\simeq 163$\footnote{From $\Gamma(\phi\to e^+e^-)=8\pi\alpha^2 f^2_\phi/27 m_\phi$.\\}
& $-$ \\
$f_D$      & $157.4 \pm 11.8 \pm 2.2$~\cite{pdg06,cleofd} & $-$ \\
$f_{D_s}$      & $193.7 \pm 9.2 \pm 4.9$~\cite{CLEOfds} & $-$\\
               & $193.0 \pm 11.3 \pm 4.9$~\cite{CLEOfds2} & $-$\\
               & $200.1 \pm 12.0 \pm 9.9$~\cite{BABARfds} & $-$\\
$f_{D^*}$      & $-$& $165\pm 6 \pm 13$~\cite{UKQCD}\footnote{Quenched
  LQCD.} \\
              &$-$ & $223 \pm 18$~\cite{qr}\footnote{Relativistic
                constituent quark model.} \\
              & $-$& $164 \pm 15$~\cite{eli}\footnote{Non-relativistic
                constituent quark model} \\
$f_{D^*_s}$    & $-$& $180\pm 4 \pm 11$~\cite{UKQCD}\footnote{Quenched
  LQCD.} \\
              & $-$& $237 \pm 18$~\cite{qr}\footnote{Relativistic
                constituent quark model.} \\
              & $-$& $231 \pm 13$~\cite{eli}\footnote{Non-relativistic
                constituent quark model} \\\hline
Decay Constant &  & Value used in this work \\\hline
$f_{\eta'}$ & & $ f_\eta$ \\
$f_{D_s}$ & & $193.7$ MeV \\
$f_{D^*}=f_{D^*_s}$ & &  $f_D$ \\
\hline
\end{tabular}
\end{center}
\caption{Meson decay constants. We use $\Gamma(\rho\to
  e^+e^-) = 7.04 \pm 0.06$ KeV, $\Gamma(\omega\to
  e^+e^-) = 0.60 \pm 0.02$ KeV, $\Gamma(\phi\to
  e^+e^-) = 1.27 \pm 0.04$ KeV, $\Gamma(\tau\to \rho
  \nu_\tau)= (5.71 \pm 0.07) \times 10^{-10}$ MeV and $\Gamma(\tau\to
  K ^*  \nu_\tau)= (3.06 \pm 0.45) \times 10^{-11}$ MeV from 
\protect\cite{pdg06}. Besides, $G=1.1664 \times 10^{-11}$ MeV$^{-2}$
  and  $\alpha=1/137.036$ are the Fermi and fine-structure constants,
  and $\cos\theta_C=0.974$,  the cosine of the Cabibbo angle.}
\label{tab:fpfv}
\end{table}

Within our present approach, besides the use of physical hadron masses, we
further break the SU(8) symmetry of Eq.~(\ref{eq:vsu8}) by
the replacement: $f^2 \to f_af_b$, depending on the nature of the
 mesons involved.  In Table~\ref{tab:fpfv} we compiled the values
of the decay constants used throughout  this work.  Whenever possible, we
use the experimental values for the decay constants.

Taking into account all the SU(8) breaking effects stated above, our tree level
amplitudes now read
\begin{equation}
V^{IJSC}_{ab}(\sqrt{s})= D^{IJSC}_{ab}
\frac{2\sqrt{s}-M_a-M_b}{4\,f_a f_b} \sqrt{\frac{E_a+M_a}{2M_a}}
\sqrt{\frac{E_b+M_b}{2M_b}} \,, \label{eq:vsu8break}
\end{equation}
where $M_a$ ($M_b$) and $E_a$ ($E_b$) are, respectively, the mass and
the CM energy of the baryon in the $a$ ($b$) channel. Though the
way  these flavor breaking effects  have been 
introduced is intuitive, we should acknowledge here again that  
this constitutes a further assumption of our model, and that we can
not establish  a clear connection to any kind of first principles, as
it is the case for all previous models.

We finish this section by comparing our present scheme with the one in
TVME. The latter relies on SU(4) symmetry to construct the effective
interaction between pseudoscalar mesons in the \dieciseis-plet with
the $\veinte_\dos$ and $\veinte_\cuatro$ plet representations of
$J^P=1/2^+$ and $J^P=3/2^+$ baryons through a $t-$channel exchange of
the \dieciseis-plet of vector mesons.  Note that vector mesons do not
enter in the $s$-channel coupled space.  Then the universal vector
meson coupling hypothesis provides the global interaction strength
among the above SU(4)-multiplets.  However,  deviations up to 50\%
from the universal value might be expected, as acknowledged in
Ref.~\cite{Hofmann:2005sw} from an analysis of the $D^* \to D \pi$
decay rate. Then, aided by the KSFR relation, which is consistent with
chiral symmetry at very low energy and momentum transfer, the
resultant lowest order meson-baryon interaction in the SU(4) limit is
found to take the WT form of Eq.~(\ref{eq:vsu8}) in the zero range
limit ~\cite{Mizutani:2006vq}\footnote{A small $p-$wave contribution
of the TVME model was neglected to reduce it to the WT
form~\cite{Mizutani:2006vq}.}. SU(4) symmetry is broken in the TVME
model by using the physical hadron masses, without adopting different
values for meson decay constants. The use of the physical masses for
the pseudoscalar mesons and the baryons (external legs in the coupled
channel formalism of the TVME model) is analogous to the mass breaking
effects included in the present scheme upon replacing $(\sqrt{s}-M)$
by $(2\sqrt{s}-M_a-M_b)/2$ in Eq.~(\ref{eq:vsu8}). However, the use of
physical masses for the vector mesons in the $t$-channel in the TVME
model leads to an additional symmetry breaking effect which is
different from those induced by the use of different meson decay
constants in our scheme. For instance, let us consider the $\Sigma_c
\pi \to \Sigma_c\pi $, $ND \to ND$, and $\Sigma_c \pi \to ND$
transitions.  Within the TVME model the first two amplitudes are
driven by the $t$-channel exchange of light vector mesons such as
$\rho$, while the last one is by the exchange of a charmed $D^*$
vector meson.  Thus, besides different SU(4) coupling coefficients,
the last channel obtains an additional suppression factor of $\sim
m^2_\rho/m_{D^*}^2 = 1/6.8$ relative to the first two
~\cite{Mizutani:2006vq}.  Within our scheme, these three channels will
scale as $1/f_\pi^2$, $1/f_D^2$ and $1/(f_Df_\pi)$, respectively. Then
on top of the matrix elements ($D_{ab}^{IJSC}$) we obtain suppression
factors $f_\pi^2/f_D^2 \sim 1/ 2.9$ and $f_\pi/f_D \sim 1/ 1.7$ for
the second and third amplitudes, respectively, relative to the first
one. In summary, in the present work diagonal transitions involving
charmed mesons are suppressed by about a factor of 3 with respect to
the TVME SU(4) models \cite{Hofmann:2005sw,Mizutani:2006vq}, while the
charm-exchange amplitudes will not be suppressed by the mass factor
$\sim m_{D^*}^2/m^2_\rho = 6.8$ but only by a factor of about two due
to the use of physical values for the decay constants.  Thus, it is
clear that the difference in the pattern of SU(4) symmetry breaking in
the two schemes discussed above is even qualitative.


\section{BS Meson--Baryon Scattering Matrix}

\label{sec:BS}

We solve the coupled channel BS equation with the interaction
kernel determined by Eq.~(\ref{eq:vsu8break}) which incorporates
the SU(8) breaking effects discussed in
Subsect.~\ref{subsec:su8-break}.  For a given $IJSC$ sector, the
solution for the coupled channel $s$-wave scattering amplitude
$T^{IJSC}(\sqrt{s})$, in the on-shell
scheme~\cite{OR98,Meissner:1999vr,Oller:2000fj,Jido:2003cb,Oller:1998zr,EJmeson,EJmeson2}
 is,
\begin{eqnarray}
T^{IJSC}(\sqrt{s}) &=& \frac{1}{1-
V^{IJSC}(\sqrt{s})\,G^{IJSC}(\sqrt{s})}\,V^{IJSC}(\sqrt{s}).
 \label{eq:scat-eq}
\end{eqnarray}
Here $G^{IJSC}(\sqrt{s})$ is a diagonal matrix consisting of 
loop functions. The loop function for  channel $i$ reads,
\begin{eqnarray}
G^{IJSC}_{ii}(\sqrt{s}) &=& {\rm i} 2 M_i \int \frac{d^4 q}{(2
\pi)^4} \, \frac{1}{(P-q)^2 - M_i^2 + i \varepsilon} \, \frac{1}{q^2
- m^2_i + {\rm i} \varepsilon} ~ , \label{eq:gprop}
\end{eqnarray}
where $M_i$ and $m_i$ are the masses of the baryon and meson in the channel, 
respectively. This quantity is logarithmically divergent,  hence must be
regularized. For example,  the regularization may be done by one-subtraction  
 at the subtraction point
$\sqrt{s}=\mu^{IJSC}_i$ such that
\begin{equation}
G^{IJSC}_{ii}(\sqrt{s}=\mu^{IJSC}_i) = 0\,,
\label{eq:subs}
\end{equation}
with index $i$ running in the coupled channel space. 
The TVME model of Refs.~\cite{Hofmann:2005sw,Hofmann:2006qx} uses the
Renormalization Scheme (RS) of Eq.~(\ref{eq:subs}), and we adopt here
its choice for the value of the subtraction point $\mu^{IJSC}_i$. It
is taken to be independent of the spin $J$, then set identical within
a given sector $ISC$ to $\sqrt{m_{\text{th}}^2+M^2_{\text{th}}}$,
where $m_{\text{th}}+M_{\text{th}}$ is the mass of the lightest
hadronic channel. For the SU(6) sector, this RS recovers previous
results for the lowest-lying $1/2^-$ and $3/2^-$ baryon resonances
appearing in the scattering of the octet of Goldstone bosons off the
lowest baryon octet and decuplet
~\cite{OR98,Meissner:1999vr,Oller:2000fj,Nieves:2001wt,Garcia-Recio:2002td,Jido:2003cb,Garcia-Recio:2003ks,
Kolomeitsev:2003kt,Sarkar:2004jh}, and leads to new predictions for
higher energy resonances.  According to the authors of
Refs.~\cite{Hofmann:2005sw,Hofmann:2006qx}, such a choice guarantees
an approximate crossing symmetry although such a claim appears
somewhat dubious because crossing symmetry involves isospin mixtures,
thus choosing an alternative subtraction point might lead to yet
another reasonable result. Nevertheless, one should bear in mind an
apparent correlation between the renormalization procedure and/or
subtraction point, and the values for the meson decay constants used
in the potential $V^{IJSC}$. For instance, it should be useful to
mention the SU(6) results which are somewhat complementary to the
present approach in the choice of the subtraction points and meson
decay constants Ref.~\cite{QNP06}.  By taking a different value of the
subtraction point for each channel $i$ of a given $IJSC$ sector in the
form $\mu^{IJSC}_i=\sqrt{M^2_i+m^2_i}$, the model gives a good
agreement with a common Goldstone boson decay constant value of 90 MeV
(pion decay constant in the chiral limit), instead of using the
different values quoted in Table~\ref{tab:fpfv}.

Given these uncertainties, in the present work we allow for
slight changes in the choice of the subtraction point by introducing a
parameter $\alpha$, such that
\begin{equation}
\left (\mu^{ISC}\right)^2=\alpha
\left(m_{\text{th}}^2+M^2_{\text{th}}\right)\ ,
\label{eq:sp}
\end{equation}
which will be fitted to data. In principle, $\alpha$ could have a
different value in each $IJSC$ sector. Unfortunately we have
insufficient empirical information. Therefore, we will adjust the
value of $\alpha$ to reproduce the position of the well established
$\Lambda_c(2595)$ resonance with $IJSC=(0,1/2,0,1)$, then the same value
will be used in all  other sectors.  

The mass and widths of the dynamically generated resonances in
each $IJSC$ sector are determined from the positions of the poles, $z_R$,
in the second Riemann sheet of the corresponding scattering
amplitudes, namely $M_R={\rm Re}\, (z_R)$ and $\Gamma_R= - 2{\rm Im}\,(z_R)$.
The coupling constants of each resonance to the
various baryon-meson states are obtained from the residues 
by fitting the amplitudes to the expression
\begin{equation}
T^{IJSC}_{ij}(z)=\frac{g_i e^{i \phi_i} g_j e^{i \phi_j}}{(z-z_R)} \ , \label{eq:pole}
\end{equation}
for complex energy values $z$ close to the pole, 
where the complex couplings are written in terms of the absolute value, $g_k$, and phase, $\phi_k$. 
These dimensionless couplings determine the corresponding decaying
branching ratios to the open baryon-meson channels. Since the
dynamically generated states may couple differently to their
baryon-meson components, we will examine the $ij$-channel
independent quantity
\begin{equation}
\label{obs} \tilde{T}^{IJSC}(z) \equiv \max_j\ \sum_i\,
|T^{IJSC}_{ij}(z)|\ ,
\end{equation}
which allows us to identify all the resonances within a given sector
at once.

\section{SU(8) results}
\label{sec:su8results}

In this work we have focused on the non-strange ($S=0$),  singly charmed ($C=1$)
 baryon resonances. The experimental status in this
sector is summarized in Table \ref{tab:exp}.
In the following tables and figures, we collect results of resonances 
up to 3.5 GeV, which is basically the energy of the heaviest channels used in the
present approach. Note, however, that the position and width of dynamical
states generated $\sim 500$ MeV beyond the lowest threshold
for each $IJSC$ sector must be taken with caution. 
In reality, they will be strongly influenced by 
other ingredients not included in the present model, such as the coupling to
three-body meson-meson-baryon states or interactions of higher angular momentum.
This observation  applies especially to those states higher than 3 GeV 
that couple strongly to low lying baryon-meson channels.

\begin{table}
\begin{center}
\begin{tabular}{lcccc}\hline
Resonance & $I (J^P)$ & Status & Mass [MeV] & $\Gamma$ [MeV]
\\\hline
$\Lambda_c(2595)$ &  $0(1/2^-)$ & *** & $2595.4 \pm 0.6$ & $3.6+ 2.0- 1.3$ \\
$\Lambda_c(2625)$ &  $0(3/2^-)$ & *** & $2628.1 \pm 0.6$ & $<1.9$ \\
$\Lambda_c(2765)$ &  $?(?^?)$ & * & $2766.6 \pm 2.4$ & $50$ \\
or $\Sigma_c(2765)$ & & & & \\
$\Lambda_c(2880)$ & $0(5/2^+)$ & *** & $2881.9 \pm 0.5$ & $5.8 \pm 1.9$\\
$\Lambda_c(2940)$ & $0(?^?)$ & *** & $2939.8 \pm 1.6$ & $18 \pm 8$  \\
$\Sigma_c(2800)^{++}$ & $1(?^?)$ & *** & $2801+ 4- 6$ & $75+ 22-
17$ \\
$\Sigma_c(2800)^{+}$ & $1(?^?)$ & *** & $2792+ 14- 5$ & $62+ 60-
40$ \\
$\Sigma_c(2800)^{0}$ & $1(?^?)$ & *** & $2802+ 4- 7$ & $61+ 28-
18$ \\
\hline
\end{tabular}
\end{center}
\caption{Summary of experimental data for baryon resonances with
charm 1, zero strangeness and negative (or unknown) parity
as compiled in Ref.~\cite{pdg06}. The
$\Lambda_c$ and $\Sigma_c$ ground states are omitted.}
\label{tab:exp}
\end{table}

\subsection{$I=0$, $J=1/2$}

In this sector the following 16 channels are
involved:
\begin{center}
\begin{tabular}{@{\extracolsep{1mm}}cccccccccccccccccccc}
 $ \Sigmac \pion $ &  $ \Nucleon \D 
  $ &  $ \Lambdac \eta $ 
 &  $ \Nucleon \DS $ &  $ \Cascadac 
  \kaon $ &  $ \Lambdac \omega $ 
 &  $ \Cascadacp \kaon $ 
 &  $ \Lambda \Ds $\\
 2591.6  &  2806.15   &  2833.91   
 &  2947.27   &  2965.12   &  3069.03   
 &  3072.52   &  3084.18  \\
 $ \Lambda \DsS $ &  $ \Sigmac \rho 
  $ &  $ \Lambdac \etap $ 
 &  $ \SigmacS \rho $ &  $ \Lambdac 
  \phi $ &  $ \Cascadac \kaonS $ 
 &  $ \Cascadacp \kaonS $ 
 &  $ \CascadacS \kaonS $\\
 3227.98  &  3229.05   &  3244.24   
 &  3293.46   &  3305.92   &  3363.33   
 &  3470.73   &  3540.23  \\
\end{tabular}
\end{center}
where the second line gives the channel thresholds in MeV.

\begin{figure}[htb]
\includegraphics[width=0.65\textwidth]{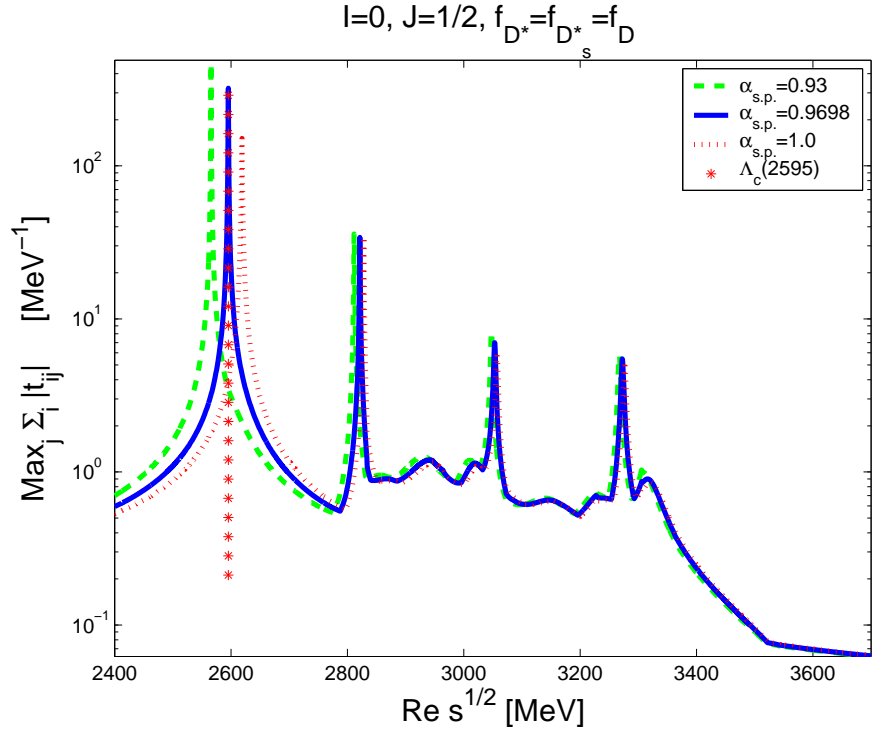}
 \caption{\label{fit_DR}
The function $\tilde{T} \equiv \max_j\ \sum_i\, |T_{ij}|$ in the
$I=0$, $J=1/2$ sector
 calculated along the  scattering line  within our broken SU(8)
 model, for
different values of the subtraction point. The stars denote the
nominal position of the  $\Lambda_c(2595)$ resonance.}
\end{figure}


\begin{figure}[htb]
\includegraphics[width=0.65\textwidth]{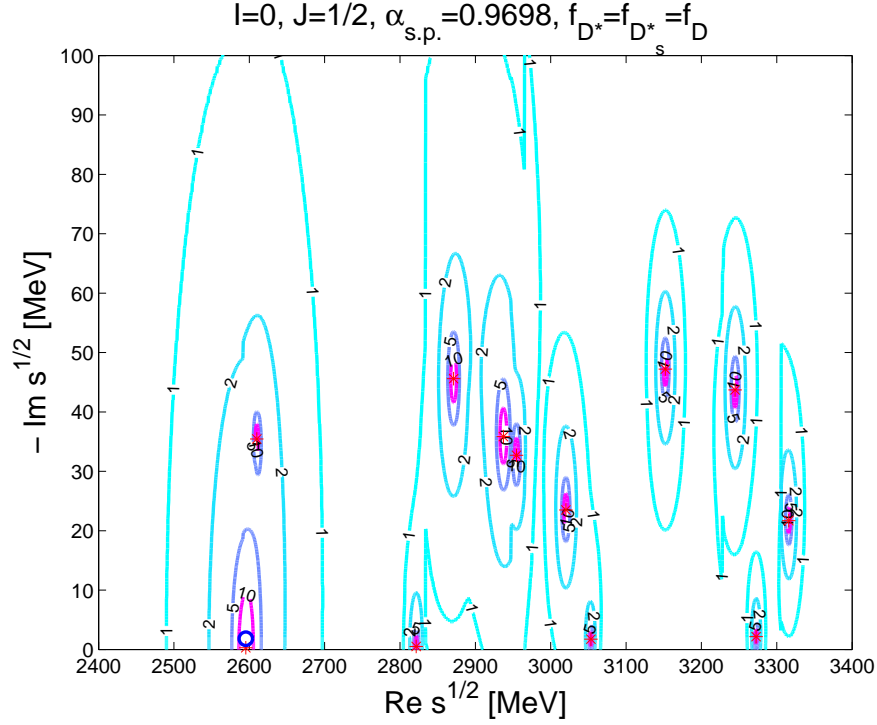}
 \caption{\label{fig:I0J12}
The function $\tilde{T} \equiv \max_j\ \sum_i\, |T_{ij}|$ in the
$I=0$, $J=1/2$ sector, calculated within the present broken SU(8) model
with $\alpha=0.9698$. Stars show the positions of the resonances obtained, while open cycles denote those of the experimentally
measured resonances, the $\Lambda_c(2595)$ in this figure.}
\end{figure}

\begin{table}[htb]
\caption{Mass [MeV], width [MeV] and couplings of resonances
with $S=0$ and $C=1$, generated in the $I=0$, $J=1/2$ sector using
our broken SU(8) model. See text for more details. The third
column quotes only  the values of the most important couplings, and those shown in bold  correspond to channels open for decay.}
\medskip
\begin{tabular}{|c|c|l|}
  \hline
\multicolumn{3}{|c|}{$I=0$, $J=1/2$} \\
 \hline
  $M_{R}$     &  $\Gamma_{R}$  & Couplings to main channels  \\
  \hline
 2595.4  & 0.58  & {\boldmath $g_{\Sigma_c \pi}=0.36$}, $g_{ND}=3.69$, $g_{ND^*}=5.70$,  $g_{\Lambda D_s}=1.42$,  $g_{\Lambda D_s^*}=2.94$ \\
  \hline
 2610.0   & 70.9  & {\boldmath  $g_{\Sigma_c \pi}=2.25$}, $g_{ND}=1.47$,  $g_{ND^*}=1.81$,  $g_{\Sigma_c \rho}=1.22$ \\
  \hline
 2821.5  & 1.0   & {\boldmath $g_{ND}=0.32$}, $g_{\Lambda_c \eta}=1.2$, $g_{\Xi_c K}=1.79$,  $g_{\Lambda D_s^*}=1.11$, $g_{\Sigma_c \rho}=1.23$,  $g_{\Sigma_c^* \rho}=1.15$ \\
   \hline
 2871.2  & 91.2   & {\boldmath $g_{ND}=2.0$}, $g_{\Lambda D_s}=1.15$, $g_{N D^*}=2.15$,  $g_{\Lambda D_s^*}=1.92$, $g_{\Lambda_c \omega}= 1.01$, $g_{\Sigma_c \rho}=2.56$, $g_{\Sigma_c^* \rho}=0.94$ \\
   \hline
 2937.2 & 71.7  &  {\boldmath $g_{\Lambda_c \eta}=1.34$},  $g_{\Lambda D_s}=1.4$,  $g_{N D^*}= 1.51$, $g_{\Lambda D_s^*}=3.41$, $g_{\Sigma_c^* \rho}= 2.23$\\
  \hline
 2954.7  & 65.4  &   {\boldmath $g_{\Sigma_c \pi}=1.02$}, $g_{{\Xi}'_c K}=1.2$, $g_{\Lambda D_s}=0.85$, $g_{\Lambda_c \omega}=2.46$, $g_{\Sigma_c \rho}=1.16$, $g_{\Sigma_c^* \rho}=0.91$ \\
  \hline
 3020.1  & 47.0  &   {\boldmath $g_{\Xi_c K}=1.13$},  $g_{\Lambda D_s}=1.07$, $g_{\Lambda D_s^*}=1.46$,  
  $g_{\Sigma_c \rho}=1.51$, $g_{\Sigma_c^* \rho}=2.49$,  $g_{\Xi_c^* K^*}=0.95$ \\
  \hline
 3053.3  & 3.49  &   {\boldmath $g_{\Lambda_c \omega}=0.11$},  $g_{\Lambda D_s}=1.43$, $g_{{\Xi}'_c K}= 1.49$, $g_{\Lambda_c \phi}=1.27$,   $g_{\Sigma_c^* \rho}=1.02$ \\
  \hline
 3152.1 & 94.4  &     {\boldmath $g_{\Lambda D_s}= 1.74$}, $g_{\Lambda D_s^*}= 2.02$, $g_{\Lambda_c \phi}=2.17$, $g_{{\Xi}'_c K^*}= 0.98$, $g_{\Xi_c K^*}=1.38$, $g_{\Xi_c^* K^*}=1.16$\\
  \hline
 3244.7  & 87.4  &   {\boldmath $g_{\Lambda D_s}=0.72$}, {\boldmath $g_{\Lambda D_s^*}=0.74$},  {\boldmath $g_{\Xi_c K}=0.68$}, $g_{{\Xi}'_c K^*}=2.32$, $g_{\Xi_c^* K^*}=2.63$  \\
  \hline
 3272.6  & 4.31  &   {\boldmath $g_{{\Xi}'_c K}=0.17$}, {\boldmath $g_{\Sigma_c \rho}=0.17$}, {\boldmath $g_{\Lambda D_s}=0.15$}, $g_{\Lambda_c \phi}=1.37$, $g_{\Xi_c K^*}=2.26$ \\
  \hline
 3315.9  &43.4  &   {\boldmath $g_{\Sigma_c \rho}=0.47$}, {\boldmath $g_{\Lambda D_s}=0.62$}, {\boldmath $g_{\Lambda_c \phi}=0.66$}, $g_{\Xi_c K^*}=1.2$, $g_{{\Xi}'_c K^*}=1.91$,  $g_{\Xi_c^* K^*}=2.38$  \\
  \hline
\end{tabular}
\label{table:I0J12}
\end{table}

\begin{table}[htb]
\caption{Positions of poles (in MeV) generated in
the $J=1/2$, $I=0$ sector from uncoupled calculations in the present
broken SU(8) model with $\alpha=0.9698$. An asterisk $^*$
denotes resonances. }
\medskip

\begin{tabular}{|c|c|c|c|c|c|c|c|c|c|}
\hline
Channel $(M_{\text{thres}})$   & $2600$    & $2700$ & $2800$ & $2900$ & $3000$ & $3100$ &
$3200$ & $3300$ & $3400$ \\
 \hline
 \hline
$\Sigma_c \pi$ $(2592)$            & $2621.6^*$ &       &        &        &        &        &        &         &     \\
 \hline
$ND$ $(2806)$                      &     & $2775.4$ &        &        &        &        &        &         &        \\
 \hline
$\Lambda_c \eta$ $(2834)$          &     &    &        &        &        &        &        &         &     \\
 \hline
$N D^*$ $(2947)$                   & $2687.3$  &    &         &        &        &    &        &        &     \\
 \hline
$\Xi_c K$ $(2965)$                 &     &    &        & $2947.7$ &        &        &        &         &     \\
 \hline
$\Lambda_c \omega$ $(3069)$        &     &    &        &        &        &        &        &         &     \\
 \hline
${\Xi}'_c K$ $(3073)$              &     &    &        &        & $3042.8$ &        &        &         &     \\
 \hline
$\Lambda D_s$ $(3084)$             &     &    &        &        &        &        &        &         &     \\
 \hline
$\Lambda D_s^*$ $(3228)$           &     &     &        &        &  $3057.3$&       &        &         &     \\
 \hline
 $\Sigma_c \rho$ $(3229)$          &     &    &         &$2962.6$  &       &        &        &         &     \\
 \hline
$\Lambda_c \eta'$ $(3244)$         &     &    &        &        &        &        &        &         &     \\
 \hline
$\Sigma_c^* \rho$ $(3293)$         &     &    &        & $2988.5$  &       &        &        &         &     \\
 \hline
$\Lambda_c \phi$ $(3306)$          &     &     &        &        &        &    &        &        &     \\
 \hline
$\Xi_c K^*$ $(3363)$               &     &     &        &        &        &    & $3293.3$     &        &     \\
 \hline
${\Xi}'_c K^*$ $(3471)$            &     &     &        &        &        &    &       & $3382.6$  &     \\
 \hline
$\Xi^*_c K^*$ $(3540)$             &     &    &        &        &        &        &        &         & $3440.2$  \\
\hline
\end{tabular}
\label{table:unc}
\end{table}

In Fig.~\ref{fit_DR} we show our results for $\tilde{T}$ as a
function of the C.M. energy $\sqrt{s}$, where we have used the
experimentally known meson decay constants and have chosen 
$f_{D_s}=193.7$ MeV. For  $\eta^\prime$,  $D^*$ and $D^*_s$, we have taken $f_{\eta^\prime}=f_\eta$ 
and $f_{D^*}=f_{D^*_s}=f_{D}$ (see Table~\ref{tab:fpfv}).
The dotted line shows that a narrow resonance is produced very close
to the position of the nominal $\Lambda_c(2595)$ which is denoted by the
starred line in the figure. By slightly changing  the value of the
subtraction point in the renormalization scheme, viz. setting the scaling
factor $\alpha$ to 0.9698 [see Eq.~(\ref{eq:sp})],
we can reproduce 
the position of the $\Lambda_c(2595)$. 

There are several other narrow resonances and bumps in
Fig.~\ref{fit_DR}, which can be better disentangled by
investigating the behavior of the function ${\tilde T} $ and $T$ in the
complex plane. We obtain the poles of $T$ in the second
Riemann sheet and determine the coupling constants to the various
baryon-meson channels through the residues of the corresponding
amplitudes, as indicated in Eq.~(\ref{eq:pole}).
The positions of the poles of $T$ are visualized in
Fig.~\ref{fig:I0J12} as peaks of ${\tilde T} $, while the values of the mass and width of
the corresponding resonances, together with the values of the couplings
to the baryon-meson components to which they couple most strongly, are
collected in Table~\ref{table:I0J12}.

The width of the $\Lambda_c(2595)$ resonance turns out to be small: $0.58$
MeV, smaller than the experimental width, $\Gamma=3.6^{+2.0}_{-1.3}$. 
 However,  we have not  included here the three-body decay
channel $\Lambda_c \pi\pi$ which  already represents  almost one
third of the decay events \cite{pdg06}.  
The narrowness of this resonance
is due to the fact that it lies only a few MeV above its unique two-body
decaying channel $\Sigma_c \pi$ to which it actually couples very weakly, as
indicated by the low value of the corresponding coupling constant in Table
\ref{table:I0J12}.

Here a remarkable feature  is
  that there is a second resonance very close to $\Lambda_c(2595)$. This is
  precisely the same pattern found in the charmless $I=0$, $S=-1$
  sector for the $\Lambda(1405)$ \cite{Jido:2003cb}.

It is illustrative  to compare the coupled channel results with
those from  an uncoupled calculation as shown in
Table.~\ref{table:unc}.
In the lower energy region we find poles at 2622 MeV, 2687 MeV and
2775 MeV, which are, respectively, bound $\Sigma_c\pi$, $ND^*$ and
$ND$ states. Note that the bound states in $ND$ and $ND^*$
channels are quite close. This emphasizes once more the importance
of including the vector mesons in the description.
The effect of the channel coupling may be easily traced by taking into
account the largest coefficients in Table~\ref{tab:i0j12s0c1},  
together with the channel threshold energies. We conclude that the
resonance at  2595.4  MeV appears to be basically the remnant of the $ND^*$
 bound state found at 2687 MeV in the uncoupled calculation, but with the
important additional binding effects from the $ND$ channel, to which
it couples very strongly, along with some moderate modifications induced
by  coupling to  $\Lambda D_s$ and $\Lambda D_s^*$ the latter 
lying more than 400 MeV away. As emphasized in the next
section, we observe here that the nature of this resonance is
substantially different from that found in
Refs.~\cite{Hofmann:2005sw,Mizutani:2006vq}, 
where it is identified
as a $ND$ bound state. A wider resonance at 2610 MeV originates
from the $\Sigma_c \pi$ resonance found at 2622 MeV in the
uncoupled calculation, mildly modified by its coupling to $ND$ as well as to
some channels
involving vector-mesons  such as $ND^*$ and  
$\Sigma_c\rho$. We do not find a clear remnant of the uncoupled
$ND$ state at 2775 MeV, apart from the already commented
influence on the properties of the $\Lambda_c(2595)$ resonance.

A narrow resonance at 2822 MeV originates mainly from the $\Xi_c
K$ bound state at 2947 MeV, which is substantially influenced by
the coupling to the $\Lambda_c \eta$ channel as well as to a few other
channels ($\Lambda D_s^*$, $\Sigma_c\rho$ and $\Sigma_c^*\rho$) involving vector
mesons. The two other narrow resonances found in this sector, at 3053 MeV and
3273 MeV, can be traced back, respectively, to the $\Xi^\prime_c K$ bound state at 
3043 MeV and the $\Xi_c K^*$ bound state at 3293 MeV of the uncoupled
calculation. The resonance at 3053 MeV shows also a strong coupling to a few 
other channels, especially to the neighboring $\Lambda D_s$ state. 

Most of the remaining wider resonances show also a stronger coupling to a
baryon-meson system for which the uncoupled calculation produces a bound state.
They are simply wider because they also couple significantly to meson-baryon
states lying below their mass. However, we identify two resonances that occur
genuinely as a result of the coupled channel formalism, since their main
baryon-meson component ($\Lambda_c \omega$ for the resonance at 2955 MeV and $\Lambda_c \phi$
for the resonance at 3152 MeV) does not have a bound state in the uncoupled
calculation.

In order to study possible variations of the result due to input parameters, we have also 
performed calculations by adopting the values for the unknown decay
constants of the charmed vector mesons 
from a quenched lattice QCD, 
 LQCD \cite{UKQCD}:
 $f_{D^*}=165$ MeV, $f_{D^*_s}=180$ MeV. Then the
$\Lambda_c(2595)$ is reproduced with a slightly smaller scaling
factor ($\alpha=0.9558$), but the qualitative features for the
positions, widths and couplings of all the other resonances
generated in this sector do not change drastically.

In comparing with data, we look  at the possibility of identifying some of our
narrow states with a resonance seen experimentally. Our predictions are limited
by the fact that 
we have implemented neither the coupling to three-body states 
nor $p$-, $d$- or higher-multipolarity interactions.
In the present  $I=0$, $J=1/2$ sector, we have
already identified our first narrow state with the $\Lambda_c(2595)$ 
and have established it as being a quasi-bound $ND^*$ system.  A resonance at
2880 MeV with a width of $\Gamma \sim 6$ MeV, decaying into $\Lambda_c \pi \pi$
states with some fraction of resonant decay through $\Sigma_c \pi$ states, was
reported in \cite{2880-Artuso:2000xy}. One might
identify it with our narrow resonance at
2822 MeV, especially because if this state
were moved to the experimental energy of 2880 MeV, it would appear above the
threshold of the $\Lambda_c\eta$ channel to which it couples significantly. 
However, the Belle collaboration determined recently the
spin of the $\Lambda_c(2880)$ to be $J=5/2$ from the 
$\Sigma_c(2455)\pi$ decay angular distribution, and the parity positive 
from agreement of the
$\Gamma(\Sigma_c(2520)\pi)/\Gamma(\Sigma_c(2455)\pi)$ branching ratio with a 
prediction of HQS \cite{Capstick:1986bm,Isgur:1991wq,Cheng:2006dk,Zhong:2007gp}. These
assignments do not match the spin-parity $1/2^-$ of our dynamically 
generated state. 

There is yet another
resonance in this sector, the $\Lambda_c(2940)$ of width $\Gamma \sim 18\pm 8$ 
MeV  which has been
seen in the $D p$ invariant mass distributions, but its spin and parity
have not been determined \cite{2940-Aubert:2006sp}. If it were a
$J=1/2^-$ state, one might be tempted to identify it with one of
our narrow states. However, none of the resonances obtained by our
model having a width smaller than 50 MeV, except for the lowest-lying one already identified with the
$\Lambda_c(2595)$, couples significantly to the $ND$ states from
which the invariant mass of the $\Lambda_c(2940)$ is
reconstructed. Our model does not give either any resonance in
this energy region that couples preferentially to $ND^*$ states, the
threshold of which is only a few MeV above the $\Lambda_c(2940)$
mass. Therefore, we do not expect the $\Lambda_c(2940)$ to be a
molecular $ND^*$ system, as claimed in the literature
\cite{He:2006is}. 
We finally note that the $p D^0$ histogram shown in 
Ref.~\cite{2940-Aubert:2006sp} is not incompatible with the existence
of a very narrow state just above the $ND$ threshold MeV, as the one
at 2822 MeV and width 1 MeV found in the present work, as well as in
the dynamical model of Ref.~\cite{Hofmann:2005sw}.

\subsection{$I=1$, $J=1/2$}
In the $I=1$, $J=1/2$ sector there are 22 channels involved:
\begin{center}
\begin{tabular}{@{\extracolsep{1mm}}cccccccccccccccccccc}
 $ \Lambdac \pion $ &  $ \Sigmac \pion 
  $ &  $ \Nucleon \D $ 
 &  $ \Nucleon \DS $ &  $ \Cascadac 
  \kaon $ &  $ \Sigmac \eta $ 
 &  $ \Lambdac \rho $ &  $ \Cascadacp 
  \kaon $ &  $ \Sigma \Ds $ 
 &  $ \Delta \DS $ &  $ \Sigmac \rho 
  $\\
 2424.5  &  2591.6   &  2806.15   
 &  2947.27   &  2965.12   &  3001.01   
 &  3061.95   &  3072.52   &  3161.65   
 &  3218.35   &  3229.05  \\
 $ \Sigmac \omega $ &  $ \SigmacS \rho 
  $ &  $ \SigmacS \omega $ 
 &  $ \Sigma \DsS $ &  $ \Cascadac 
  \kaonS $ &  $ \Sigmac \etap $ 
 &  $ \Cascadacp \kaonS $ 
 &  $ \Sigmac \phi $ &  $ \SigmaS \DsS 
  $ &  $ \SigmacS \phi $ 
 &  $ \CascadacS \kaonS $\\
 3236.13  &  3293.46   &  3300.54   
 &  3305.45   &  3363.33   &  3411.34   
 &  3470.73   &  3473.02   &  3496.87   
 &  3537.43   &  3540.23  \\
\end{tabular}
\end{center}


\begin{figure}[htb]
\includegraphics[width=0.65\textwidth]{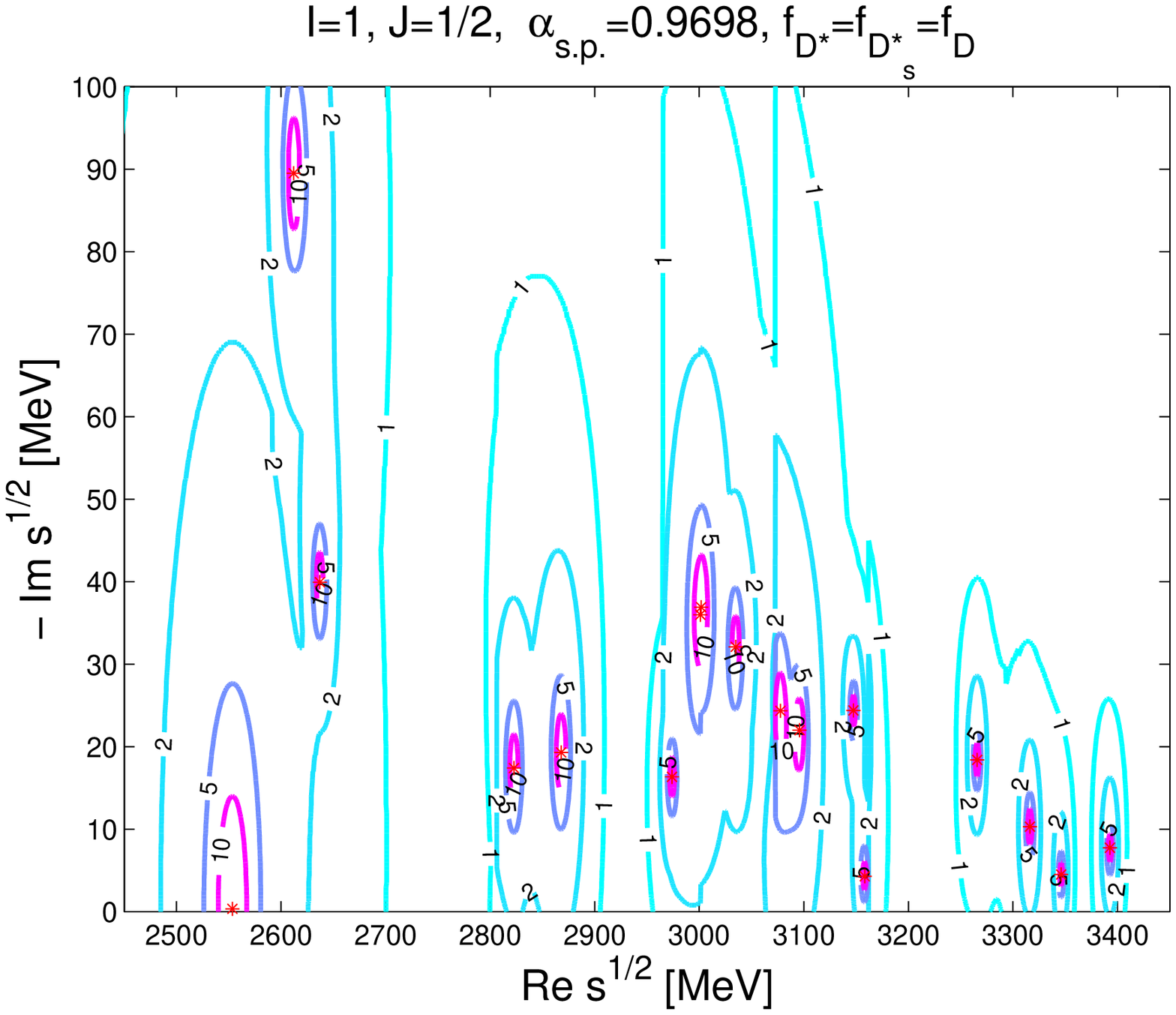}
 \caption{\label{fig:I1J12}
The same as Fig.~\protect\ref{fig:I0J12} for the $I=1$, $J=1/2$
sector. }
\end{figure}

\begin{table}[htb]
\caption{The same as
Table~\protect\ref{table:I0J12}, for the $I=1$, $J=1/2$ sector.}
\medskip
\begin{tabular}{|c|c|l|}
  \hline
\multicolumn{3}{|c|}{$I=1$, $J=1/2$} \\
  \hline
  $M_{R}$     &  $\Gamma_{R}$  & Couplings to main channels  \\
  \hline
2553.6  & 0.67  &  {\boldmath $g_{\Lambda_c \pi}=0.15$}, $g_{ND}=2.28$,  $g_{\Delta D^*}=6.74$, $g_{\Sigma^* D^*_s}=2.89$\\
 \hline
2612.2  & 179.0  &    {\boldmath $g_{\Lambda_c \pi}=1.95$}, $g_{ND^*}=3.78$, $g_{\Sigma_c \rho}=1.27$, $g_{\Sigma_c^* \rho}=1.4$\\
  \hline
2637.1  & 79.9  &   {\boldmath $g_{\Sigma_c \pi}=1.98$}, $g_{N D}=2.35$, $g_{N D^*}=1.69$, $g_{\Sigma^* D_s^*}=1.24$ \\
  \hline
 2822.8  &  34.8  &   {\boldmath $g_{N D}=1.55$}, $g_{\Sigma D_s}=1.01$,  $g_{\Xi_c K}= 1.04$, $g_{N D^*}=1.41$,  $g_{\Lambda_c \rho}=1.37$, $g_{\Delta D^*}=1.81$,$g_{\Sigma_c^* \rho}=2.27$, $g_{\Xi_c^* K^*}= 0.94$\\
 \hline
 2868.0  &  38.6  &   {\boldmath $g_{N D}=0.76$}, $g_{N D^*}=2.75$, 
 $g_{\Sigma D_s^*}=1.05$,   $g_{\Lambda_c \rho}=1.3$, $g_{\Sigma_c \rho}=1.54$, $g_{\Sigma_c \omega}=1.02$, $g_{\Delta D^*}= 1.78$\\
 \hline
 2974.0   &  37.2  &   
 {\boldmath $g_{N D}=0.72$}, {\boldmath $g_{\Sigma_c \pi}=0.63$}, 
 {\boldmath $g_{N D^*}=0.63$},  $g_{\Sigma_c \eta}=1.67$,   $g_{\Lambda_c \rho}=1.29$, 
 $g_{\Sigma_c \omega}=1.13$, $g_{\Delta D^*}= 1.21$\\
 \hline
3001.9  &  73.8  &   {\boldmath $g_{\Xi_c K}= 1.42$},  
$g_{\Sigma D_s}=1.68$,   
$g_{\Sigma D_s^*}=4.15$, $g_{\Sigma_c \rho}=1.42$, 
$g_{\Sigma_c^* \rho}= 1.26$\\
 \hline
3034.9  &  64.2  &   {\boldmath $g_{\Sigma_c \pi}=0.72$},  {\boldmath $g_{\Sigma_c \eta}=0.67$},   
{\boldmath $g_{N D^*}= 0.76$}, $g_{\Sigma D_s}=1.16$,   $g_{\Sigma_c \omega}=2.32$, $g_{\Sigma^* D_s^*}=2.44$,
$g_{\Sigma_c^* \rho}=1.16$, $g_{\Sigma_c^* \omega}=0.87$\\
 \hline
3077.7  &  48.7 &   {\boldmath $g_{N D^*}=0.85$}, $g_{\Sigma_c \omega}=2.40$, $g_{\Sigma^* D_s^*}=3.20$, $g_{\Sigma_c^* \omega}=1.43$\\
 \hline
3095.9  &  44.0  &    {\boldmath $g_{\Lambda_c \pi}=0.39$}, {\boldmath $g_{{\Xi}'_c K}= 0.88$}, {\boldmath $g_{N D^*}= 0.45$}, $g_{\Sigma D_s}=1.34$, $g_{\Sigma_c \rho}=1.09$,  $g_{\Sigma^* D_s^*}=2.80$, $g_{\Sigma_c^* \omega}=2.36$ \\
 \hline
3147.5  & 48.8 &    {\boldmath $g_{\Lambda_c \rho}=1.02$}, $g_{\Sigma_c \rho}=1.62$, $g_{\Sigma_c \omega}=0.94$,  $g_{\Sigma_c^* \rho}=1.67$, $g_{\Sigma_c^* \omega}=1.57$\\
 \hline
3158.2  & 8.6 &   {\boldmath $g_{\Sigma_c \eta}= 0.3$},  {\boldmath $g_{\Lambda_c \rho}= 0.27$},  $g_{\Sigma D_s}=1.54$,
$g_{\Sigma D_s^*}=1.51$,  $g_{\Xi_c K^*}= 1.53$\\
 \hline
3265.8  & 36.8  &  {\boldmath  $g_{\Sigma D_s}= 0.69$},  $g_{\Sigma D_s^*}=1.54$, $g_{\Sigma_c \phi}= 1.32$, $g_{{\Xi}'_c K^*}= 1.49$, $g_{\Sigma_c^* \phi}= 1.67$, $g_{\Xi_c^* K^*}= 1.34$\\
 \hline
3316.2  & 20.6  &  {\boldmath  $g_{{\Xi}'_c K}= 0.44$}, $g_{\Sigma_c \phi}= 1.66$,  $g_{{\Xi}'_c K^*}= 2.22$, $g_{\Sigma^* D_s^*}= 1.80$, $g_{\Xi_c^* K^*}= 1.01$\\
 \hline
3346.1  & 9.0  &   {\boldmath  $g_{\Sigma D_s}= 0.45$}, $g_{\Sigma_c \phi}= 1.81$, $g_{\Xi_c K^*}= 1.12$, $g_{\Sigma_c^* \phi}= 1.38$, $g_{\Xi_c^* K^*}= 1.14$\\
 \hline
3392.8  & 15.5  &   {\boldmath  $g_{\Sigma_c^* \rho}= 0.31$}, $g_{\Sigma_c^* \phi}= 2.17$, $g_{\Xi_c^* K^*}= 1.95$\\
  \hline
\end{tabular}
\label{table:I1J12}
\end{table}

Our results are presented in Fig.~\ref{fig:I1J12} and
Table~\ref{table:I1J12}. First, we find a very narrow resonance at
2554 MeV, which is
basically composed of the $\Delta D^*$ component (somewhat mixed with
 the $N D$ and $\Sigma^* D^*_s$ states). Hence, this resonance would be
 absent in models
 not including channels with a vector meson and a $3/2^+$ baryon.   
 Kinematically it can only decay  to $\Lambda_c \pi$, but the corresponding
coupling is very weak. 

At higher energies and up to 3.4 GeV, our model generates
many resonances in this sector, most of them having
rather appreciable widths. Some of these resonances find their analogue
within the states obtained by the pioneer
work of Lutz and Kolomeitsev \cite{Lutz:2003jw}, where only
the meson-baryon channels composed by an octet Goldstone boson plus
 a member of either the anti-triplet or 
 the sextet of open charm ground state baryons were considered. 
The resonances at 2612 MeV, 2637 MeV and 2974 MeV, would correspond
to the states in Ref.~\cite{Lutz:2003jw} found, respectively, at 2800 MeV 
(coupling mostly to $\Lambda_c \pi$),
2700 MeV (coupling mostly to $\Sigma_c\pi$) and 
2985 MeV (coupling mostly to  $\Sigma_c\eta$).
The states obtained here appear at a lower mass and show a different width
due to their coupling
to many of the new baryon-meson components considered in this work.

From the experimental side, an isotriplet
of excited charmed baryons, $\Sigma_c(2800)$ has been reported
\cite{Mizuk:2004yu}, decaying mainly to $\Lambda_c^+\pi^-$,
$\Lambda_c^+\pi^0$ and $\Lambda_c^+\pi^+$ pairs with a large
width of around $60$ MeV  (but  with a  $ > 50$\% error). This resonance
has been tentatively assumed to decay to $\Lambda_c \pi$  in
$d$-wave with its spin parity assumed as $J^P=3/2^-$ relying on quark models,  
but angular distributions have not yet been measured. 
Of the predicted states discussed above
having a moderate width,
 only the one at
 3096 MeV shows a sizable coupling to $\Lambda_c\pi$
states, but it lies 300 MeV away of the experimental mass. 
Then what about identifying the predicted 3035 MeV state 
as the $\Sigma_c(2800)$? 
Note first, that this 3035 MeV state decays significantly to $\Sigma_c \pi$
pairs.
Now if this state were to move to the
experimental position, it would become somewhat narrower, especially because 
two of the possible decay channels, $\Sigma_c\eta$ and $N D^*$, would be closed.
Its decay to 
$\Lambda_c \pi$ pairs, as observed experimentally, would require 
an implementation in the present model of an additional 
meson-baryon interaction in $d$-wave or/and $p$-wave coupling in the neglected $s$- and $u$-channel diagrams. 
However, according to the sizable coupling
to $\Sigma_c\pi$ states, the identification of our state as  
$\Sigma_c(2800)$ would be ruled 
out if the later were not observed in
experiments  looking at the $\Lambda_c\pi\pi$ systems. On inspecting 
the histograms
of Ref.~\cite{2880-Artuso:2000xy}, one finds signals around 
$M(\Lambda_c^+\pi^+\pi^-)-M(\Lambda_c^+)\sim 500$ MeV, 
corresponding to the mass of the $\Sigma_c(2800)$, although they are probably
too feeble.

It seems unlikely that any of the three states with energies lying between 
2800 and 3000 MeV could qualify for the observed isotriplet, basically because 
they couple quite significantly to $ND$ states. Hence, if the parameters of our
model would allow us to move one of these states below the $ND$ threshold, it
would become significantly narrower than the experimental width of $\sim 60$ MeV. 
However, on inspecting the histogram of $I=1$ $D^+ p$ pairs, shown in
Ref.~\cite{2940-Aubert:2006sp} to confirm the $\Lambda_c(2940)$ as being an
isosinglet, one observes a clear enhancement around 2860 MeV of width $\sim 10$
MeV, which we could identify with one of the states we see decaying into $ND$
pairs. The most likely candidate is the state at 2974 MeV because, if it
was to be moved to 2860 MeV, its width would be reduced from 37 MeV 
to a value of around 10 MeV due to
the closing of two out of three decaying channels.

Above  3 GeV we find a few states that couple
strongly to some vector meson--baryon channels.  We thus expect
them to be absent in the SU(4) models
\cite{Hofmann:2005sw,Mizutani:2006vq}. It may be of interest to note that  the 
relatively narrow resonances at 3158 and 3346 MeV could be seen from
the invariant mass of $\Sigma_c \eta$ and $\Sigma D_s$ states, respectively.

\subsection{$I=2$, $J=1/2$}
In the $ I=2$, $ J=1/2$ sector there are 4 channels involved:
\begin{center}
\begin{tabular}{@{\extracolsep{1mm}}cccccccccccccccccccc}
 $ \Sigmac \pion $ &  $ \Delta \DS 
  $ &  $ \Sigmac \rho $ 
 &  $ \SigmacS \rho $\\
 2591.6  &  3218.35   &  3229.05   
 &  3293.46  
\end{tabular}
\end{center}

\begin{figure}[htb]
\includegraphics[width=0.65\textwidth]{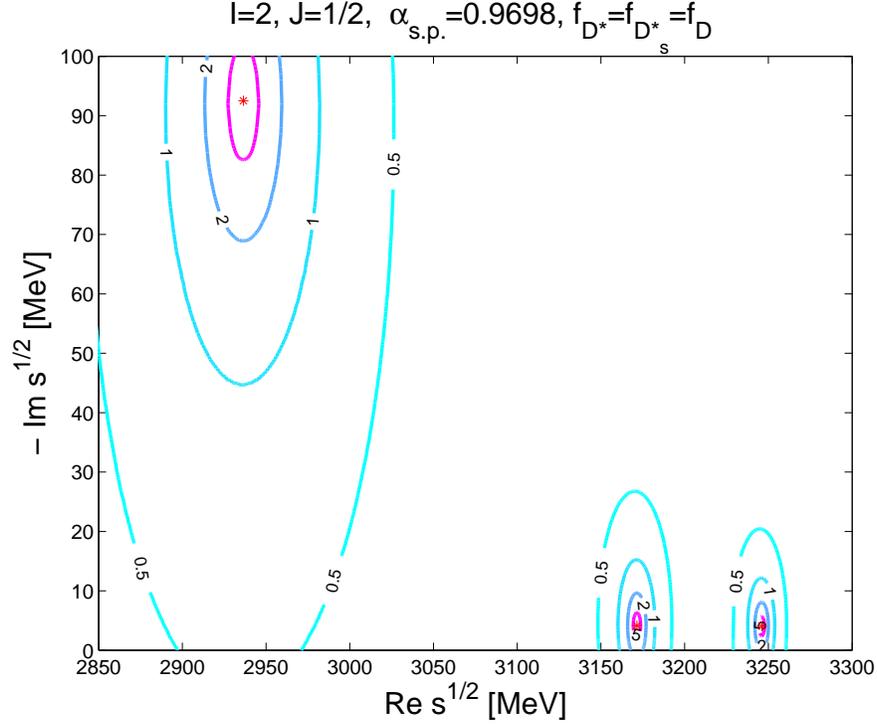}
 \caption{\label{fig:I2J12}
The same as Fig.~\protect\ref{fig:I0J12} for the $I=2$, $J=1/2$
sector. }
\end{figure}

\begin{table}[hbt]
\caption{The same as
Table~\protect\ref{table:I0J12}, but for the $I=2$, $J=1/2$ sector.}
\medskip
\begin{tabular}{|c|c|l|}
  \hline
\multicolumn{3}{|c|}{$I=2$, $J=1/2$} \\
  \hline
  $M_{R}$     &  $\Gamma_{R}$  & Couplings to all channels  \\
  \hline
  2936.5   & 185.0  &  {\boldmath $g_{\Sigma_c \pi}=1.62$},  $g_{\Delta D^*}=5.37$,  $g_{\Sigma_c \rho}=1.12 $, $g_{\Sigma_c^* \rho}=0.66$ \\
 \hline
   3171.5   &8.26  &  {\boldmath $g_{\Sigma_c \pi}=0.33$},  $g_{\Delta D^*}=1.61$,  $g_{\Sigma_c \rho}=2.26$, $g_{\Sigma_c^* \rho}=0.76$ \\
 \hline
    3246.1   &8.09  &     {\boldmath $g_{\Sigma_c \pi}=0.18$,  $g_{\Delta D^*}=0.33$,  $g_{\Sigma_c \rho}=0.47 $}, $g_{\Sigma_c^* \rho}=2.39$\\
  \hline
  \end{tabular}
 \label{table:I2J12}
 \end{table}

As seen from the results shown in Table \ref{table:I2J12}, our model generates
two narrow structures, at 3172 MeV and 3246 MeV which qualify, respectively, as
$\Sigma_c \rho$ and $\Sigma_c^*\rho$ quasi-bound states.

\subsection{ $I=0$, $J=3/2$}
Of various  $J=3/2$ resonance candidates,
we first consider  those in the $I=0$ sector, with 11 coupled
channels:
\begin{center}
\begin{tabular}{@{\extracolsep{1mm}}cccccccccccccccccccc}
 $ \SigmacS \pion $ &  $ \Nucleon \DS 
  $ &  $ \Lambdac \omega $ 
 &  $ \CascadacS \kaon $ 
 &  $ \Lambda \DsS $ &  $ \Sigmac \rho 
  $ &  $ \SigmacS \rho $ 
 &  $ \Lambdac \phi $ &  $ \Cascadac 
  \kaonS $ &  $ \Cascadacp \kaonS $ 
 &  $ \CascadacS \kaonS $\\
 2656.01  &  2947.27   &  3069.03   
 &  3142.02   &  3227.98   &  3229.05   
 &  3293.46   &  3305.92   &  3363.33   
 &  3470.73   &  3540.23  
\end{tabular}
\end{center}

\begin{figure}[htb]
\includegraphics[width=0.65\textwidth]{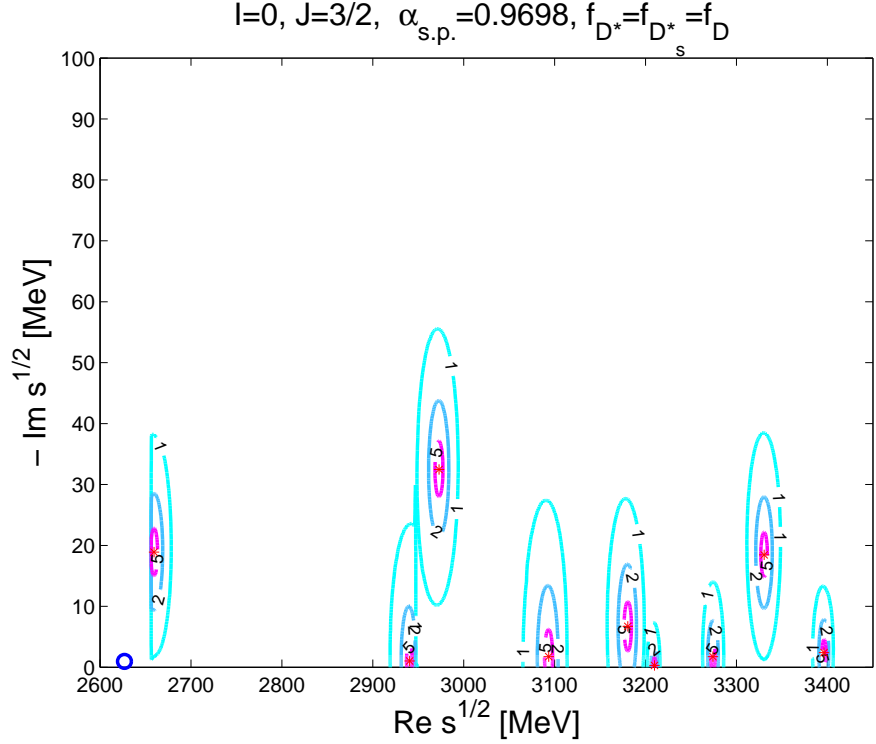}
 \caption{\label{fig:I0J32}
The same as Fig.~\protect\ref{fig:I0J12} for the $I=0$, $J=3/2$
sector. Open cycles denote the positions of the experimentally
measured resonances, $\Lambda_c(2625)$  in this case.}
\end{figure}

\begin{table}[htb]
\caption{The same as
Table~\protect\ref{table:I0J12},  for the $I=0$, $J=3/2$ sector.}
\medskip
\begin{tabular}{|c|c|l|}
  \hline
\multicolumn{3}{|c|}{$I=0$, $J=3/2$} \\
  \hline
  $M_{R}$     &  $\Gamma_{R}$  & Couplings to main channels  \\
  \hline
2659.5  & 37.8  &  {\boldmath $g_{\Sigma_c^* \pi}=2.23$}, $g_{ND^*}=2.11$,  $g_{\Sigma_c^* \rho}=1.34$\\
 \hline
2940.5 & 2.06  & {\boldmath $g_{\Sigma_c^* \pi}=0.21$},  $g_{ND^*}=2.21$,  $g_{\Sigma_c^* \rho}=1.40$\\
 \hline
2972.8 & 64.9  & {\boldmath $g_{\Sigma_c^* \pi}=1.05$}, $g_{\Lambda_c \omega}=2.42$, $g_{\Xi_c^* K}=0.95$, $g_{\Sigma_c^* \rho}=1.28$\\
 \hline
3093.5 &  3.48  & {\boldmath $g_{\Lambda_c \omega}=0.29$},  {\boldmath $g_{ND^*}=0.28$}, 
 $g_{\Lambda D_s^*}=2.89$,  $g_{\Xi_c^* K}=1.88$\\
  \hline
3180.5  &  13.4  & {\boldmath $g_{\Xi_c^* K}=0.69$}, $g_{\Lambda D_s^*}=2.49$, $g_{\Lambda_c \phi}=1.83$, 
$g_{\Xi_c K^*}=1.00$,  $g_{\Xi_c^* K^*}=0.79$\\
  \hline
3209.8  &0.6 & {\boldmath $g_{ND^*}=0.08$}, {\boldmath $g_{\Xi_c^* K}=0.10$},  $g_{\Sigma_c \rho}=1.81$,  $g_{\Sigma_c^* \rho}=1.09$\\
 \hline
3274.1  & 3.48 &  {\boldmath $g_{\Lambda D_s^*}=0.19$}, {\boldmath $g_{\Xi_c^* K}=0.2$}, $g_{\Lambda_c \phi}=1.30$, $g_{\Xi_c K^*}=2.31$\\
 \hline
3330.4  & 36.9  &   {\boldmath $g_{\Lambda D_s^*}=0.50$}, {\boldmath $g_{\Lambda_c \phi}=0.68$}, {\boldmath $g_{\Sigma_c^* \rho}=0.57$}, $g_{\Xi_c K^*}=0.85$, $g_{{\Xi}'_c K^*}=1.85$, $g_{\Xi_c^* K^*}=2.34$\\
 \hline
3396.3  & 4.8  & {\boldmath $g_{\Lambda D_s^*}=0.17$}, {\boldmath $g_{\Lambda_c \phi}=0.22$}, {\boldmath $g_{\Sigma_c^* \rho}=0.15$}, $g_{{\Xi}'_c K^*}=2.17$, $g_{\Xi_c^* K^*}=1.86$\\
 \hline
\end{tabular}
\label{table:I0J32}
\end{table}

The results displayed in Fig.~\ref{fig:I0J32} and
Table~\ref{table:I0J32} show that, in this sector, our model generates
nine resonances, four of which have a width smaller than 5 MeV.

Experimentally, there is a three star $\Lambda_c(2625)$ resonance
with $J^P=3/2^-$, which has a very narrow width, $\Gamma<1.9$ MeV,
and decays mostly to $\Lambda_c \pi \pi$ \cite{pdg06}. It has been identified as the counterpart of the $\Lambda(1520)$ which, in dynamical
models, shows a strong coupling to $\Sigma^*\pi$
\cite{Kolomeitsev:2003kt,Sarkar:2004jh,Sarkar:2005ap,QNP06,Toki:2007ab}. In
the present  $S=0$, $C=1$ sector, we
indeed find a $I=0$, $J=3/2$ $d$-wave $\Lambda_c$ resonance at 2660
MeV with its width $\Gamma=38$ MeV, which couples very strongly to
the $\Sigma^*_c\pi$ channel. We may therefore identify this
resonance as a strong candidate for the $d$-wave
$\Lambda_c(2625)$. A small change in  the subtraction point $\mu^{IJSC}$ could
easily move the resonance down by 40 MeV to the nominal position.
By so doing it should also be possible to reduce the width considerably, hence
closer to the experimental value because
the resonance position would get below the threshold of the channel
$\Sigma_c^*\pi$ to which
 this resonance  couples strongly. 
We note  that a similar
resonance was found at 2660 MeV in the SU(4) model of
Ref.~\cite{Hofmann:2006qx}.  The  novel feature in the
recent  approach is  that the resonance in question here has non-negligible
baryon-vector meson  components such as $N D^*$ and
$\Sigma^*_c\rho$. Other narrow resonances observed in this  channel 
are a state at 3094 MeV coupling strongly to
$\Xi^*_c K$, as in Ref.~\cite{Hofmann:2006qx}, but also to 
the vector-meson channel $\Lambda D_s^*$ and three narrow
resonances which couple mostly to channels with vector meson
components. Out of the resonances
obtained in this sector around 3 GeV and beyond, the one at 2941 MeV could be
a candidate
for the $\Lambda_c(2940)$ of width $\sim 18 $ MeV and unknown
$J^P$, observed recently  from the invariant mass of $D^0 p$ pairs
\cite{2940-Aubert:2006sp}. Its decay into $D^0 p$ pairs would imply the
additional implementation of $p$-wave interactions in our model. The strong
coupling of this resonance to $N D^*$ hints also for a substantial coupling into
$p$-wave $ND$ states.

\subsection{$I=1$, $J=3/2$}
\label{sec:i1j32}

In the $I=1$, $J=3/2$ sector one finds 20 channels:
\begin{center}
\begin{tabular}{@{\extracolsep{1mm}}cccccccccccccccccccc}
 $ \SigmacS \pion $ &  $ \Nucleon \DS 
  $ &  $ \Lambdac \rho $ 
 &  $ \SigmacS \eta $ &  $ \Delta \D 
  $ &  $ \CascadacS \kaon $ 
 &  $ \Delta \DS $ &  $ \Sigmac \rho 
  $ &  $ \Sigmac \omega $ 
 &  $ \SigmacS \rho $\\
 2656.01  &  2947.27   &  3061.95
 &  3065.42   &  3077.23   &  3142.02   
 &  3218.35   &  3229.05   &  3236.13   
 &  3293.46  \\
 $ \SigmacS \omega $ &  $ \Sigma \DsS 
  $ &  $ \SigmaS \Ds $ 
 &  $ \Cascadac \kaonS $ &  $ \Cascadacp 
  \kaonS $ &  $ \Sigmac \phi $ 
 &  $ \SigmacS \etap $ &  $ \SigmaS 
  \DsS $ &  $ \SigmacS \phi $ 
 &  $ \CascadacS \kaonS $\\
 3300.54  &  3305.45   &  3353.07   
 &  3363.33   &  3470.73   &  3473.02   
 &  3475.75   &  3496.87   &  3537.43
 &  3540.23  \\
\end{tabular}
\end{center}

\begin{figure}[htb]
\includegraphics[width=0.65\textwidth]{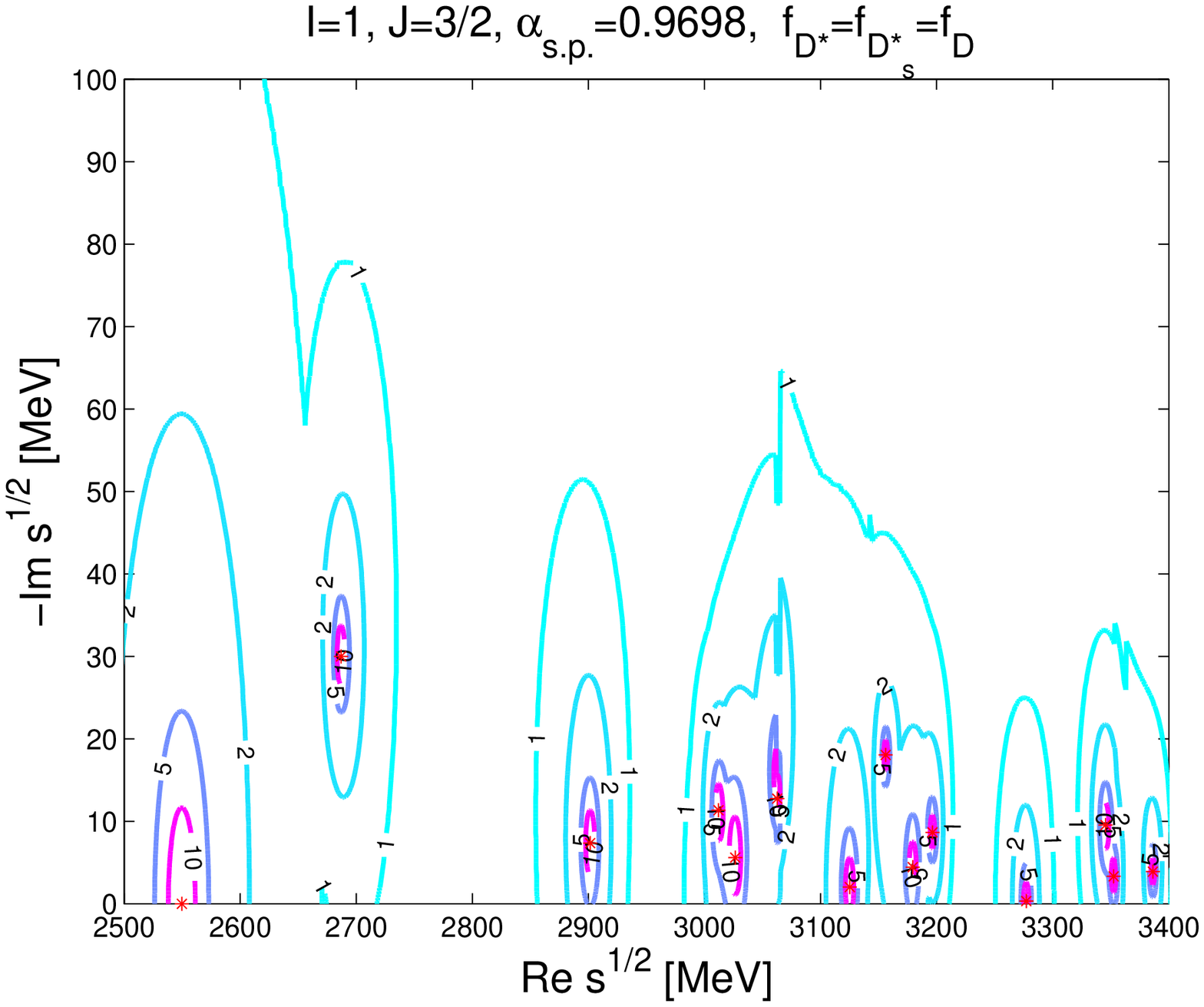}
 \caption{\label{fig:I1J32}
The same as Fig.~\protect\ref{fig:I0J12} for the $I=1$, $J=3/2$
sector.  }
\end{figure}

\begin{table}[hbt]
\caption{The same as
Table~\protect\ref{table:I0J12}, but for the $I=1$, $J=3/2$ sector.}
\medskip
\begin{tabular}{|c|c|l|}
  \hline
\multicolumn{3}{|c|}{$I=1$, $J=3/2$} \\
  \hline
  $M_{R}$     &  $\Gamma_{R}$  & Couplings to main channels  \\
  \hline
2549.8  & 0.0  &  $g_{ND^*}=2.52$,  $g_{\Sigma D_s^*}=2.22$, $g_{\Delta D}=4.23$, $g_{\Sigma^* D_s}=1.48$, $g_{\Delta D^*}=5.28$, $g_{\Sigma^* D_s^*}=2.29$\\
 \hline
2686.9  & 60.0  &   {\boldmath $g_{\Sigma_c^* \pi}=1.91$}, $g_{ND^*}=2.68$, $g_{\Sigma D_s^*}=0.96$, $g_{\Lambda_c \rho}=1.02$,  $g_{\Sigma^* D_s^*}=0.97$\\
 \hline
2901.6  & 14.7 &   {\boldmath $g_{\Sigma_c^* \pi}=0.58$}, $g_{ND^*}=2.77$, $g_{\Lambda_c \rho}=1.52$, $g_{\Delta D}=1.26$, $g_{\Delta D^*}=1.05$, $g_{\Sigma_c^* \rho}=1.20$ \\
\hline
3012.1  & 22.7 & {\boldmath $g_{ND^*}=0.90$},  $g_{\Sigma D_s^*}=2.05$, $g_{\Sigma_c^* \eta}=2.04$\\
\hline
3026.5  & 11.2 &  {\boldmath $g_{ND^*}=0.72$}, $g_{\Sigma_c \omega}=2.10$,  $g_{\Delta D}=2.40$, $g_{\Delta D^*}=2.83$\\
 \hline
3062.6  & 25.5 &  {\boldmath $g_{\Sigma_c^* \eta}=0.61$}, $g_{\Sigma D_s^*}=1.67$, $g_{\Delta D^*}=1.56$,  $g_{\Sigma^* D_s^*}=1.69$,  $g_{\Sigma_c^* \omega}=1.97$\\
 \hline
3125.1  & 4.05  &   {\boldmath $g_{\Sigma_c^* \pi}=0.17$}, {\boldmath $g_{\Sigma_c^* \eta}=0.21$},  $g_{\Sigma D_s^*}=1.39$,  $g_{\Sigma^* D_s}=1.67$,  $g_{\Xi_c^* K}=1.33$, $g_{\Sigma^* D_s^*}=2.90$, $g_{\Sigma_c^* \omega}=0.93$\\
 \hline
3156.1  & 36.1  &  {\boldmath $g_{\Lambda_c \rho}=0.9$},  $g_{\Sigma D_s^*}=1.45$, $g_{\Sigma_c \rho}=1.50$, 
$g_{\Sigma_c \omega}=1.27$, $g_{\Sigma_c^* \rho}=1.41$, $g_{\Sigma_c^* \omega}=1.39$\\
 \hline
3179.5  & 8.89  &   {\boldmath $g_{\Sigma_c^* \eta}=0.37$}, $g_{\Sigma D_s^*}=2.65$, $g_{\Sigma_c \rho}=1.40$, $g_{\Xi_c K^*}=1.71$, $g_{\Sigma_c^* \omega}=0.85$ \\
 \hline
3196.4 & 17.3 &{\boldmath $g_{\Lambda_c \rho}=0.56$}, $g_{\Sigma D_s^*}=0.96$, $g_{\Sigma_c \rho}=1.17$, $g_{\Sigma_c \omega}=0.87$, $g_{\Xi_c K^*}=0.87$,  $g_{\Sigma_c^* \rho}=2.16$, $g_{\Sigma_c^* \omega}=0.92$ \\
 \hline
3277.2  & 0.62  &  {\boldmath $g_{\Xi_c^* K}=0.11$},  $g_{\Sigma_c \phi}=0.91$,  $g_{{\Xi}'_c K^*}=1.68$,  $g_{\Sigma^* D_s}=2.61$, $g_{\Sigma^* D_s^*}=2.52$\\
 \hline
3345.4  & 19.1  &     {\boldmath $g_{\Xi_c^* K}=0.54$},   $g_{{\Xi}'_c K^*}=1.18$, $g_{\Sigma^* D_s^*}=1.07$,  $g_{\Sigma_c^* \phi}=1.38$,  $g_{\Xi_c^* K^*}=2.57$\\
\hline
3352.6  & 6.65  &  {\boldmath $g_{\Sigma D^*_s}=0.40$},  {\boldmath $g_{\Xi_c^* K}=0.32$}, 
 $g_{\Sigma_c \phi}=2.26$,  $g_{{\Xi}'_c K^*}=1.10$, $g_{\Xi_c K^*}=0.91$, $g_{\Sigma_c^* \phi}=1.72$,  $g_{\Xi_c^* K^*}=0.99$\\
 \hline
3386.3  & 7.79  &    {\boldmath $g_{\Xi_c K^*}=0.48$},  $g_{\Sigma_c \phi}=1.40$,  $g_{{\Xi}'_c K^*}=1.00$, 
$g_{\Sigma_c^* \phi}=2.20$,  $g_{\Xi_c^* K^*}=1.18$ \\
 \hline
 \end{tabular}
 \label{table:I1J32}
 \end{table}

Our results are presented in Fig.~\ref{fig:I1J32} and
Table~\ref{table:I1J32}. We obtain many resonances and they
all couple strongly, without exception, to channels with vector mesons.
A bound state at 2550 MeV, 
whose main baryon-meson components  contain
a charmed meson, 
lies  below the threshold of any possible decay channel. It would develop a
narrow decay width to $\Lambda_c \pi$ states if we incorporated $d$-wave
interactions. This resonance is the charm sector counterpart of the
hyperonic $\Sigma(1670)$ resonance, which has also been seen in chiral unitary models
\cite{Kolomeitsev:2003kt,Sarkar:2004jh,Sarkar:2005ap} coupling
strongly to the $\Delta \bar K$ channel. Indeed, the charmed bound state found
here couples strongly to the corresponding $\Delta D$ state in the charm sector, and even more strongly
 to its  vector partner:  $\Delta D^*$. 
 This is precisely the reason for the large binding of this resonance, compared
 to that of its strange sector counterpart. 

Since the observed $I=1$ charmed baryon resonances do not have spin nor parity
assignments, most of the arguments discussed in the $I=1$, $J= 1/2$ section,
can also be made here.
We might assign the resonance at 2687 MeV, decaying 
primarily  to $\Sigma_c^*\pi$,  to
the bump at 2760 MeV observed  from $\Lambda_c\pi\pi$ states
\cite{2880-Artuso:2000xy}. However,  
the corresponding width estimated by scaling our result: 
$\Gamma_{2760} =
(q_{2760}/q_{2690})\Gamma_{2690}$,  would turn out to be around 120 MeV, 
much larger than the experimental value. 
Then, similarly to our discussion in the $I=1$, $J=1/2$ sector,  
the next state at 
2901 MeV, decaying into $\Sigma_c^*\pi$ s-wave pairs, could be a candidate for the $\Sigma_c(2800)$,  
seen in $\Lambda_c\pi$
pairs \cite{Mizuk:2004yu}, if this resonance were also seen in $\Lambda_c\pi\pi$ states.
The little enhancement in the $\Lambda_c\pi\pi$ histogram \cite{2880-Artuso:2000xy}
around $M(\Lambda_c^+\pi^+\pi^-)-M(\Lambda_c^+)\sim 500$ MeV, 
which is obviously distorted by the excitation of the lower
mass resonance at 2760 MeV,  
might well correspond roughly to the mass of the $\Sigma_c(2800)$.

Our model predicts a few narrow resonances above 3 GeV. Of particular interest
are those coupling strongly to $N D^*$, since they could be observed in
$D\pi N$ invariant mass distributions and, if sufficiently narrow, these states
should be reflected into the $D N$
invariant mass spectra Although the discussion in
Ref.~\cite{2940-Aubert:2006sp} mentions explicitly the absence of signal in the
$D^*(2010)^+ p$ or $D^*(2007)^0 p$ mass distributions, the invariant mass
distribution of $D^+ p$ pairs shows some structures which, unfortunately,
are embedded into the statistical error fluctuations. Clearly, more data needs
to be collected in order to see whether some of these structures acquire
statistical significance to become new charmed isotriplet baryons.

\subsection{$I=2$, $J=3/2$}
\label{ch5}

Here we present our results in the $I=2$, $J=3/2$ sector, where
there are only five channels:
\begin{center}
\begin{tabular}{@{\extracolsep{1mm}}cccccccccccccccccccc}
 $ \SigmacS \pion $ &  $ \Delta \D 
  $ &  $ \Delta \DS $ 
 &  $ \Sigmac \rho $ &  $ \SigmacS \rho 
  $\\
 2656.01  &  3077.23   &  3218.35   
 &  3229.05   &  3293.46  \\
\end{tabular}
\end{center}

\begin{figure}[htb]
\includegraphics[width=0.65\textwidth]{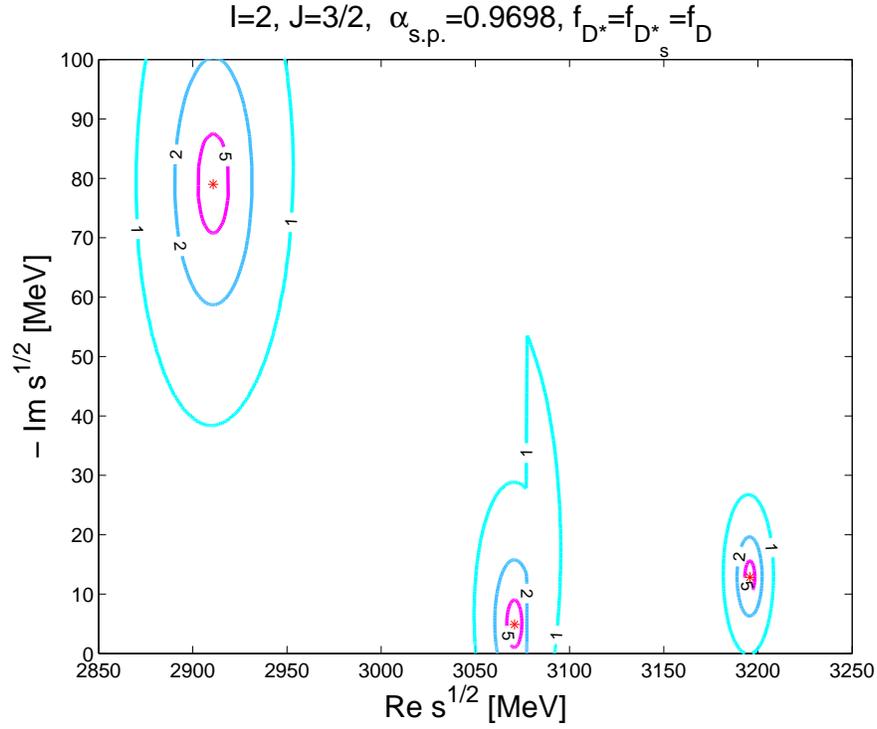}
 \caption{\label{fig:I2J32}
The same as Fig.~\protect\ref{fig:I0J12} for the $I=2$, $J=3/2$
sector. }
\end{figure}

\begin{table}[hbt]
\caption{The same as
Table~\protect\ref{table:I0J12}, but for the $I=2$, $J=3/2$ sector.}
\medskip
\begin{tabular}{|c|c|l|}
  \hline
\multicolumn{3}{|c|}{$I=2$, $J=3/2$} \\
  \hline
  $M_{R}$     &  $\Gamma_{R}$  & Couplings to all channels  \\
  \hline
  2910.8   &158.0  &  {\boldmath $g_{\Sigma_c^* \pi}=1.67$},  $g_{\Delta D}=3.52$, $g_{\Delta D^*}=3.95$,  $g_{\Sigma_c \rho}=0.72 $, $g_{\Sigma_c^* \rho}=0.68$ \\
 \hline
   3070.8  &9.79  &  {\boldmath $g_{\Sigma_c^* \pi}=0.40$},  $g_{\Delta D}=1.48$, $g_{\Delta D^*}=2.73$,  $g_{\Sigma_c \rho}=1.35 $, $g_{\Sigma_c^* \rho}=1.18$ \\
 \hline
     3195.8   &25.7  &  {\boldmath $g_{\Sigma_c^* \pi}=0.50$}, {\boldmath $g_{\Delta D}=0.51$}, $g_{\Delta D^*}=1.07$,  $g_{\Sigma_c \rho}=0.97 $, $g_{\Sigma_c^* \rho}=2.38$ \\
  \hline
  \end{tabular}
 \label{table:I2J32}
 \end{table}

Our results, presented in Fig.~\ref{fig:I2J32} and
Table~\ref{table:I2J32}, indicate that the model generates a
state at 2911 MeV, with mainly $N D$ and $N D^*$ components, which is very 
wide because it lies above the threshold of $\Sigma_c^*\pi$ states
to which it also couples significantly.  We also find a narrow resonance at 
3071 MeV, which lies
only a few MeV below the $\Delta D$ threshold and coupling more strongly to
$\Delta D^*$ states, and a wider resonance at
3196 MeV which qualifies as a $\Sigma^*_c \rho$ quasi-bound state.

\subsection{$J=5/2$ $(I=0,1$ and $2)$}
\label{ch6}

We finally present our results for the isospin sectors $I=0$, 1 and 2 
with maximum value of spin $J=5/2$. The states building up the coupled-channel
space in this $J=5/2$ case are composed of a $3/2^+$ baryon and a vector meson. Note 
that, as seen from the coefficients in Tables~\ref{tab:i0j52s0c1},
\ref{tab:i1j52s0c1} and \ref{tab:i2j52s0c1}, all these cases
present the peculiarity that
channels without strange quarks (as $\Sigma_c^* \rho$ or $\Delta D^*$)
do not mix with channels having an associated
$s\bar{s}$ pair (as $\Sigma^*_c \phi$ or $\Xi_c^* K^*$). Therefore, in the case
$I=0$, $J=5/2$, since the interaction is strongly attractive in the two channels involved,
namely
\begin{center}
\begin{tabular}{@{\extracolsep{1mm}}ccccc}
$  \SigmacS \rho  $
 &  $  \CascadacS \kaonS$ 
\\
 3293.46  &  3540.23  \ ,
\end{tabular}
\end{center}
we obtain two states at 3258 MeV and 3440 MeV 
that are, respectively, bound $ \SigmacS \rho$ and  $\CascadacS \kaonS$ systems,
as seen in Table~\ref{table:I0J52}.
Since we do not take into account
three-particle channels, these resonances appear as stable bound
states.

\begin{table}[hbt]
\caption{The same as
Table~\protect\ref{table:I0J12}, but for the $I=0$, $J=5/2$ sector.}
\medskip
\begin{tabular}{|c|c|l|}
  \hline
\multicolumn{3}{|c|}{$I=0$, $J=5/2$} \\
  \hline
  $M_{R}$     &  $\Gamma_{R}$  & Couplings to main channels  \\
  \hline
 3258.0   & $0.0$  & $g_{\SigmacS \rho}=2.24 $ \\
  \hline
 3440.2   & $0.0$  & $g_{ \CascadacS \kaonS}=2.87$ \\
  \hline
  \end{tabular}
 \label{table:I0J52}
 \end{table}
 
 In the $ I=1$, $ J=5/2$ sector there are 6 channels involved:
\begin{center}
\begin{tabular}{@{\extracolsep{1mm}}cccccccccccccccccccc}
 $ \Delta \DS $ &  $ \SigmacS \rho 
  $ &  $ \SigmacS \omega $ 
 &  $ \SigmaS \DsS $ &  $ \SigmacS \phi 
  $ &  $ \CascadacS \kaonS $\\
 3218.35  &  3293.46   &  3300.54   
 &  3496.87   &  3537.43  
 &  3540.23  \\
\end{tabular}
\end{center}

\begin{table}[hbt]
\caption{The same as
Table~\protect\ref{table:I0J12}, but for the $I=1$, $J=5/2$ sector.}
\medskip
\begin{tabular}{|c|c|l|}
  \hline
\multicolumn{3}{|c|}{$I=1$, $J=5/2$} \\
  \hline
  $M_{R}$     &  $\Gamma_{R}$  & Couplings to main channels  \\
  \hline
  3084.5   &0.0  &  $g_{\Delta D^*}=3.99$,   $g_{\Sigma_c^* \rho}=0.98$,   $g_{\Sigma_c^* \omega}=1.92$ \\
 \hline
   3233.8   &0.17  &   {\boldmath $g_{\Delta D^*}=0.14$},   $g_{\Sigma_c^* \rho}=2.14$,   $g_{\Sigma_c^* \omega}=3.14$ \\
 \hline
    3276.7   &0.0  &  $g_{\Sigma^* D_s^*}=4.32 $, $g_{\Sigma_c^* \phi}=1.17$, $g_{\Xi_c^* K^*}=1.95$ \\
 \hline
     3421.9   &0.0  & $g_{\Sigma^* D_s^*}=0.15 $, $g_{\Sigma_c^* \phi}=2.52$, $g_{\Xi_c^* K^*}=1.65$ \\
  \hline
  \end{tabular}
 \label{table:I1J52}
 \end{table}
 
Table~\ref{table:I1J52} shows that we generate 3 bound states and a narrow
resonance. The lower two states couple to the non-strange baryon-meson
channels, while the upper two belong to the associated strangeness sector.
 
Finally, we present in Table~\ref{table:I2J52} the results for the 
$I=2$, $J=5/2$ sector, which involves two channels:
\begin{center}
\begin{tabular}{@{\extracolsep{1mm}}cccccccccccccccccccc}
 $ \Delta \DS $ &  $ \SigmacS \rho 
  $\\
 3218.35  &  3293.46  \\
\end{tabular}
\end{center}

\begin{table}[hbt]
\caption{The same as
Table~\protect\ref{table:I0J12}, but for the $I=2$, $J=5/2$ sector.}
\medskip
\begin{tabular}{|c|c|l|}
  \hline
\multicolumn{3}{|c|}{$I=2$, $J=5/2$} \\
  \hline
  $M_{R}$     &  $\Gamma_{R}$  & Couplings to all channels  \\
  \hline
  3126.6   &0.0  & $g_{\Delta D^*}=3.62$,   $g_{\Sigma_c^* \rho}=2.0$ \\
  \hline
  \end{tabular}
 \label{table:I2J52}
 \end{table}

As seen from the coefficients in Table~\ref{tab:i2j52s0c1}, the interaction is
zero for $\Delta D^*$ states and repulsive for $\Sigma_c^* \rho$ states. In
spite of this, our model generates a bound state at 3127 MeV, about 90 MeV below
the threshold of the $\Delta D^*$ channel and coupling strongly to it. This
bound state 
is entirely due to the coupling between $\Delta D^*$ and $\Sigma_c^* \rho$ states,
facilitated by the significant value of the mixing coefficient and by the
vicinity of the channels involved, whose thresholds are only 75 MeV apart.
We have checked that the bound state disappears if the mixing coefficient is
reduced to 2/3 its actual value.

\section{Comparison with SU(4) models}
\label{sec:su4results}

In this section we would like to compare our results with those obtained
within a standard SU(4) model denoted as TVME in the present work.  We will follow the TVME approach, first 
developed in Ref.~\cite{Hofmann:2005sw}, but with modifications implemented 
in Ref.~\cite{Mizutani:2006vq}, which we describe briefly in the 
following.
Firstly, the last term of the interaction kernel used in 
Ref.~\cite{Hofmann:2005sw}, namely
\begin{eqnarray}
V^{(I,S,C)}(\sqrt{s}\,) = \!\!\sum_{V\in [16]}\frac{C^{(I,S,C)}_V}{8\,m_V^2}\, \Big(
2\,\sqrt{s}-M-\bar M +(\bar M-M)\,\frac{\bar m^2 -m^2}{m_V^2} \Big) \ ,
\label{VWT}
\end{eqnarray}
is dropped. This was shown in \cite{Mizutani:2006vq} to be
more consistent with the zero range ($t\to 0$) limit
of the TVME employed in Ref.~\cite{Hofmann:2005sw}.
In the second place, a common heavy vector meson mass $m_H$
and a common light vector meson mass $m_V$ is used in the kernel,
with the relation $m_H\sim 2 m_V$. In this way,
the resulting interaction is of the same form as that shown in 
Eq.~(\ref{eq:vsu8break}), namely
\begin{equation}
V^{IJSC}_{ab}(\sqrt{s})= \chi_c\, D^{IJSC}_{ab}
\frac{2\sqrt{s}-M_a-M_b}{4\,f^2} \sqrt{\frac{E_a+M_a}{2M_a}}
\sqrt{\frac{E_b+M_b}{2M_b}} \ , \label{eq:mizutani}
\end{equation}
but with an additional factor $\chi_c$, which is $\chi_c=1/4$ 
for charm-exchange transitions and $\chi_c=1$ otherwise. This factor accounts
approximately for the ratio $(m_V/m_H)^2$ between the squared masses
of uncharmed and charmed vector mesons, such as the $\rho$ and $D^*$,
respectively. The additional scalar-isoscalar attractive interaction
characterized by a $\Sigma_{DN}$ term, implemented in
Ref.~\cite{Mizutani:2006vq}, is not considered here,  since it
would not affect the qualitative aspects of our results.


The main difference between the SU(4) models
\cite{Lutz:2003jw,Hofmann:2005sw,Hofmann:2006qx,Mizutani:2006vq} and
our SU(8) model, which is a must to be consistent with heavy-quark
symmetry, is that the latter includes the vector meson-baryon
channels.  Another essential difference lies in the transition
amplitudes between states containing heavy mesons because, according
to Eq.~(\ref{eq:vsu8break}), they scale with the inverse of a
heavy-meson decay constant for each heavy meson involved, whereas in
the SU(4) models the decay constant is kept fixed for all
transitions. Therefore, as discussed at the end of
Sect.~\ref{subsec:su8-break}, amplitudes such as $N D\to N D $ or
$\Lambda D_s \to N D$ are typically a factor 3 smaller than those of
the SU(4) models.

\begin{table}[htb]
\caption{The same as
Table~\protect\ref{table:I0J12}, but for the SU(4) model, see more details in the text. }
\medskip
\begin{tabular}{|c|c|l|}
 \hline
  $M_{R}$     &  $\Gamma_{R}$  & Couplings to all SU(4) channels  \\
  \hline
  \hline
\multicolumn{3}{|c|}{$I=0$, $J=1/2$} \\
  \hline
 2595.4  & 2.01  & {\boldmath $g_{\Sigma_c \pi}=0.67$}, $g_{ND}=6.03$, $g_{\Lambda_c \eta}=0.12$,  $g_{\Xi_c K}= 0.07$,   $g_{{\Xi}'_c K}= 0.17$,   
$g_{\Lambda D_s}=3.08$, $g_{\Lambda_c \eta'}=0.29$ \\
  \hline
 2625.4   & 103.0  & {\boldmath  $g_{\Sigma_c \pi}=2.30$}, $g_{ND}=1.55$, $g_{\Lambda_c \eta}=0.04$,  $g_{\Xi_c K}= 0.03$,  $g_{{\Xi}'_c K}= 0.67$, 
$g_{\Lambda D_s}=1.05$, $g_{\Lambda_c \eta'}=0.1$\\
  \hline
 2799.5  & 0.0   & {\boldmath $g_{\Sigma_c \pi}=0.35 \cdot 10^{-2}$}, $g_{ND}=0.05$, $g_{\Lambda_c \eta}=1.47$,  $g_{\Xi_c K}= 2.57$, $g_{{\Xi}'_c K}= 0.02$,
$g_{\Lambda D_s}=0.26$, $g_{\Lambda_c \eta'}=0.02$\\
   \hline
 3024.8 &  31.3 & {\boldmath $g_{\Sigma_c \pi}=0.59$, $g_{ND}=0.50$, $g_{\Lambda_c \eta}=0.16$,  $g_{\Xi_c K}= 0.11$},  $g_{{\Xi}'_c K}= 2.22$,  
$g_{\Lambda D_s}=1.48$, $g_{\Lambda_c \eta'}=0.02$  \\
  \hline
  \hline
\multicolumn{3}{|c|}{$I=1$, $J=1/2$} \\
  \hline
2661.2  & 36.5 &  {\boldmath $g_{\Lambda_c \pi}=0.65$},  {\boldmath $g_{\Sigma_c \pi}=1.1$}, $g_{ND}=4.87$, $g_{\Sigma_c \eta}=0.04$, $g_{\Xi_c K}=0.48$,
  $g_{{\Xi}'_c K}=0.45$, $g_{\Sigma D_s}=3.99$, $g_{\Sigma_c \eta'}=0.35$\\
 \hline
2694.7 & 153.0  & {\boldmath $g_{\Lambda_c \pi}=0.34$},   {\boldmath $g_{\Sigma_c \pi}=2.35$}, $g_{ND}=1.95$, $g_{\Sigma_c \eta}=0.21$, 
$g_{\Xi_c K}=0.31$, $g_{{\Xi}'_c K}=1.19$, $g_{\Sigma D_s}=2.03$, $g_{\Sigma_c \eta'}=0.18$\\
 \hline
2938.0  &  291.0  & {\boldmath $g_{\Lambda_c \pi}=1.87$},  {\boldmath $g_{\Sigma_c \pi}=0.09$},  {\boldmath $g_{ND}=0.32$}, $g_{\Sigma_c \eta}=0.13$, 
$g_{\Xi_c K}=2.46$, $g_{{\Xi}'_c K}=0.07$,  $g_{\Sigma D_s}=1.35$, $g_{\Sigma_c \eta'}=0.19$\\
  \hline
2956.5  &29.5 &  {\boldmath $g_{\Lambda_c \pi}=0.05$},  {\boldmath $g_{\Sigma_c \pi}=0.63$}, {\boldmath $g_{ND}=0.55$}, $g_{\Sigma_c \eta}=1.85$, 
$g_{\Xi_c K}=0.26$, $g_{{\Xi}'_c K}=1.60$, $g_{\Sigma D_s}=1.55$, $g_{\Sigma_c \eta'}=0.03$\\
 \hline
  \hline
\multicolumn{3}{|c|}{$I=0$, $J=3/2$} \\
  \hline
2678.0  & 82.7  &   {\boldmath $g_{\Sigma_c^* \pi}=2.23$}, $g_{\Xi_c^* K}=0.61$\\
 \hline
3118.7 & 29.6  &  {\boldmath $g_{\Sigma_c^* \pi}=0.67$}, $g_{\Xi_c^* K}=2.06$\\
 \hline
  \hline
\multicolumn{3}{|c|}{$I=1$, $J=3/2$} \\
  \hline
2543.5  & 0.0  &   $g_{\Sigma_c^* \pi}=0.22$, $g_{\Sigma_c^* \eta}=0.12$, $g_{\Delta D}=7.21$, $g_{\Xi_c^* K}=0.07$, $g_{\Sigma^* D_s}=3.27$, $g_{\Sigma_c^* \eta'}=0.28$ \\
 \hline
2745.5  & 155.0  &   {\boldmath $g_{\Sigma_c^* \pi}=2.2$}, $g_{\Sigma_c^* \eta}=0.22$, $g_{\Delta D}=0.45$, $g_{\Xi_c^* K}=1.09$, $g_{\Sigma^* D_s}=0.52$, $g_{\Sigma_c^* \eta'}=0.06$\\
 \hline
3042.4  & 31.2 &   {\boldmath $g_{\Sigma_c^* \pi}=0.74$}, $g_{\Sigma_c^* \eta}=1.75$, $g_{\Delta D}=0.04$, $g_{\Xi_c^* K}=1.52$, $g_{\Sigma^* D_s}=0.07$, $g_{\Sigma_c^* \eta'}=0.02$ \\
 \hline
\end{tabular}
\label{tab:su4}
\end{table}

The results obtained within the SU(4)  approach, Eq. (\ref{eq:mizutani}), are collected in Table~\ref{tab:su4}, only for those $IJ$ combinations which generate resonances.
That excludes the $I=2,\ J=1/2$ sector, which has a repulsive interaction in the only allowed channel, $\Sigma_c \pi$; as well as the two-channel ($\Sigma_c^* \pi$, $\Delta D$) $I=2,\ J=3/2$ sector, where the mixing coefficient is not strong enough to overcome the repulsive  $\Sigma_c^* \pi$ interaction. 
Note that we have slightly modified the
subtraction point with a parameter $\alpha=0.959$, see Eq.~(\ref{eq:sp}), in
order to place the $\Lambda(2595)$ resonance on the right energy. The
generation of the resonance would have been facilitated if an attractive
scalar-isoscalar $\Sigma_{DN}$ term were included, in which case the
subtraction point would have to be modified by a value of $\alpha=0.979$, 
closer to one.

The main observation in the $I=0$, $J=1/2$ sector is that the model is able to
generate a narrow
resonance around 2600 MeV, which can be readily identified
with the $\Lambda_c(2595)$. It couples strongly to $ND$ and its
nature is therefore very different from the one obtained in the 
SU(8) model, which qualifies as being mainly a $N D^*$ quasibound state. The
reason for the SU(4) model being able to generate the $\Lambda_c(2595)$ 
is that the lack of baryon-meson states in this sector (namely, those involving 
a vector meson and/or a decuplet baryon)
is somehow compensated by the enhanced strength in the baryon-meson channels
involving a charmed meson. In other words, a resonance
that has a strong $N D^*$ component  in the SU(8) approach may reappear 
as a $ND$ state  in the reduced SU(4) model space  due to the enhanced interaction 
kernel. We also find relatively narrow
resonances at 2799 and 3025 MeV that couple, respectively, mostly to
$\Xi_c K$ and $\Xi_c^\prime K$ states involving uncharmed
meson channels and, therefore, are the analogous states as those
obtained within the SU(8) model at 2821 and 3053 MeV respectively [see
Table~\ref{table:I0J12}], although the last state in the SU(4) model
 couples appreciably to
$\Sigma_c\pi$ and $ND$, and hence is substantially wider.
Logically the SU(4)
model does not produce the resonances that   
 couple strongly to states with light vector mesons in the SU(8) approach. 

In the $I=1$, $J=1/2$ sector, we observe that the SU(4) model obtains a narrow
resonance at 2661 MeV which couples strongly to $ND$. Since the strength of
this amplitude is reduced, the SU(8) model does not find such  a type of resonance, 
but  instead, a very narrow one at 2554 MeV that couples strongly to $\Delta
D^*$ and more moderately to $\Sigma^* D^*_s$ and $ND$, see
Table~\ref{table:I1J12}.  So again, while the
two resonances do not differ that much in energy, their very nature is
substantially different. The other relatively narrow resonance at 2957 MeV,
coupling mainly to $\Sigma_c\eta$ and $\Xi_c^\prime K$,
finds its analogue with the one at 2974 MeV obtained in the SU(8) model, although
this one has also vector-meson components of comparable size.
The narrow states at  3158 MeV and 3346 MeV, which couple mainly to 
vector meson components and decay into $\Sigma_c \eta$ and $\Sigma\D_s$ pairs,
respectively, are not obtained
in the SU(4) approach.

In the $I=0$, $J=3/2$ sector, the SU(4) model produces
resonances at 2678 MeV and 3119 MeV that
couple mainly to $\Sigma_c^* \pi$ and $\Xi^*_c K$, respectively. 
The first one is equivalent to that found at 2659 MeV in the SU(8) model, see
Table.~\ref{table:I0J32}.
The second one could be identified basically with that appearing at 3094 MeV 
in the SU(8) model, that also has stronger
vector-meson components.
It has been argued
in previous works \cite{Lutz:2003jw,Hofmann:2006qx} and in the present one, 
that the lowest-lying resonance could correspond to the observed three star
$\Lambda_c(2625)$ of $J^P=3/2^-$, which is quoted to be the analogous in the
charm sector of the $\Lambda(1520)$ \cite{pdg06}.
The SU(8) model produces, apart from the state at 2940 MeV that could be
identified with the $\Lambda_c(2940)$, a few other narrow resonances beyond
3 GeV that couple strongly to states containing vector mesons and, therefore,
cannot be seen in the SU(4) model.

In the $I=1$, $J=3/2$ sector, there are three resonances that couple
 mainly to $\Delta D$, $\Sigma_c^* \pi$ and $\Sigma_c^* \eta$, respectively, and
are the analogies of some of the lower energy states obtained in the SU(8) 
model, namely those at 2550 MeV, 2687 MeV and 3012 MeV. Note,
however, that the resonance at 2550 MeV, which is the counterpart
in the charm sector of the
hyperonic $\Sigma(1670)$ resonance having mainly $\Delta \bar K$ components
\cite{Kolomeitsev:2003kt,Sarkar:2004jh,Sarkar:2005ap}, 
becomes a $\Delta D^*$
resonance in the SU(8) model, although its  coupling  to $\Delta D$, $N
D^*$, $\Sigma D^*_s$, and $\Sigma^* D^*_s$ is also significant.
The three resonances obtained here with the simulated SU(4) model are the same as
those found in Ref.~\cite{Hofmann:2006qx}, however with a notable exception
for the width of the state around 3 GeV which is quoted to be very narrow, while
we find it to be around 30 MeV, due basically to a quite sizable coupling
to $\Sigma_c^* \pi$ states. Note that, in this sector, the WT coefficient is
$-2$ for the $\Sigma^*_c \pi$ transition and $0$ for the
 $\Sigma^*_c\eta$, so the sizable coupling of this resonance
 to this latter channel is necessarily due to the non-perturbative
 coupled channels processes  involving non-diagonal transitions.

We end this section by noting that the present simulation of the SU(4)
model of Ref.~\cite{Lutz:2003jw} reproduces all the resonances given
in Refs.~\cite{Lutz:2003jw,Hofmann:2006qx} at approximately the same
positions.  Only the widths of some resonances appear to be wider
here.  This is because we are using an interaction kernel that ignores
the tensor term of the t-channel vector meson propagator, more
consistent with the limit $t\to 0$, which is implicit in the contact
WT interaction \cite{Mizutani:2006vq}.

\section{Conclusions}
\label{sec:conclusions}

In the present work, we have studied charmed baryon resonances within
a coupled channels unitary approach that implements, for the first
time, the characteristic features of heavy-quark symmetry, as for
instance the fact that $D$ and $D^*$ mesons have to be treated on an
equal footing. This is accomplished by extending the t-channel
vector-exchange SU(4) models (TVME) used in
Refs. \cite{Lutz:2003jw,Hofmann:2005sw,Hofmann:2006qx,Mizutani:2006vq}
to SU(8) spin-flavor symmetry, then by implementing a somewhat
different way of breaking the flavor symmetry.  More concretely our
tree-level $s$-wave WT amplitudes are obtained not only by adopting
the physical hadron masses, but also by introducing the physical weak
decay constants of the mesons involved in the transitions.  This
procedure reduces considerably the strength of diagonal amplitudes
involving a charmed meson, as comparing to SU(4) models.

The  present SU(8) model generates dynamically the resonances with negative parity in all the
isospin-spin sectors that one can form from a $s$-wave interaction
between the mesons of the $0^-,1^-$ multiplets with  the $1/2^+,3/2^+$ baryons. 
In this work we have focused only on the strangeness and charm quantum numbers appropriate for the $N D$ interaction, namely $S=0$ and $C=1$.
In the $I=0$, $J=1/2$ sector, the lowest narrow state obtained can be readily
identified with the $\Lambda_c(2595)$ which in our approach may be interpreted as
 being primarily a  $N D^*$ bound state. We may  also identify our state
at 2660 MeV in the $I=0$, $J=3/2$ sector, which couples strongly to the
$\Sigma^*_c\pi$ channel, to be  the three star  $\Lambda_c(2625)$. The latter is
usually referred to as being the charm counterpart of the $\Lambda(1520)$, 
which shows a strong coupling to $\Sigma^*\pi$ in dynamical models. 
Similarly, in the $I=1$, $J=3/2$ sector we find a resonance at 2550 MeV
that couples strongly to $\Delta D$ and $\Delta D^*$ states.  This appears to be
 the counterpart of  the $\Sigma(1670)$ hyperon which 
couples to $\Delta \bar{K}$ in dynamical models. We also find in this sector a state at 
2901 MeV that could be a candidate for the $\Sigma_c(2800)$.

Our results have been compared with those obtained in the SU(4) models
\cite{Lutz:2003jw,Hofmann:2005sw,Hofmann:2006qx,Mizutani:2006vq}.  It
appears that our SU(8) approach reproduces all the resonances
generated in the SU(4) models which couple strongly to the channels
consisting of a pseudoscalar octet meson and a charmed baryon. They
may be identified as chirally excited charmed baryons.  On the other
hand, due to the different pattern of flavor symmetry breaking,
resonances in the SU(4) model that couple strongly to baryon-meson
states containing a charmed meson and an un-charmed baryon (such as $N
D$ and $Y D_s$) are, in principle, not necessarily reproduced within
our SU(8) approach which has adopted the physical decay constants,
resulting in a reduced interaction in the corresponding transition
amplitudes.  However, the additional states implemented by the
enlarged model space in SU(8) due to the requirement of heavy-quark
symmetry, such as $N D^*$  and $\Delta D^*$,
compensate largely for the reduced interaction, producing resonances
in the same region or even more bound, than those obtained in the
SU(4) model, but with a quite different composition.  A prime example
is the $\Lambda_c(2595)$ in the $I=0,\ J=1/2$ sector. Within the SU(4)
models this resonance is dynamically generated mainly from $N D$
states and it is then interpreted as a $N D$ quasi-bound state, with a
mixture of $\Lambda D_s$ components, which lies very close to the
threshold of its only decaying channel, $\Sigma_c\pi$. Instead, the
SU(8) model interprets this resonance as being mainly a $ND^*$
quasi-bound state, although with sizable $N D$, $\Lambda D_s$ and
$\Lambda D^*_s$ components.

Our SU(8) approach predicts more states than those obtained within the
SU(4)
model. Not all of them find a direct identification with one of the
observed
resonances in the $C=1$, $S=0$ sector. We note, however, that many of
our
states couple weakly to the baryon-meson pairs from which the resonances
are
measured. It should also be pointed that some of our states would become
wider
or even disappear in a more realistic model containing also three-body
channels
and higher multipolarity interactions. We finally note that the
experimental
spectra show a limited amount of counts and, in order to disentangle new
resonant structures from data, more statistics is clearly needed.

\section*{Acknowledgments}
This work is partly supported by the EU contract
FLAVIAnet MRTN-CT-2006-035482,
by the contracts FIS2005-03142 and 
FIS2005-00810
from MEC (Spain) and
FEDER,  by the Generalitat de
Catalunya contract 2005SGR-00343,
and the Junta de Andaluc{\'\i}a
grants FQM225, FQM481 and P06-FQM-01735.
This research is part of the EU
Integrated Infrastructure Initiative Hadron Physics Project under
contract number RII3-CT-2004-506078.  T.M. wishes to express his appreciation
for the support of his visit to Barcelona through the research grants program
of the Facultat de F\'{\i}sica (University of Barcelona).

\appendix
\section{Hadron spin-flavor wave functions and tensor method}

\label{sec:spwf}

 This Appendix is aimed for a detailed account of how to calculate the 
 matrix element $D^{IJSC}_{ab}$  in the expression for the WT interaction, 
 Eq.~(13).  It consists of defining the hadronic wave functions from the quark
 model and  the tensor representation of the SU(8) operators and wave
 functons.

\subsection{Wavefunctions for the ground state hadrons}
\label{subsec:a.1}

Pseudoscalar mesons:
\begin{eqnarray}
| \pion \rangle &=&-\frac{1}{\sqrt{2}}\left(
\Qur\Qdbj - \Quj\Qdbr
\right)|0\rangle \,,
\nonumber  \\
| \kaon \rangle &=&\frac{1}{\sqrt{2}}\left(
\Qur\Qsbj - \Quj\Qsbr
\right)|0\rangle \,,
\nonumber  \\
|\kaonb \rangle &=& -\frac{1}{\sqrt{2}}\left(
\Qsr\Qdbj - \Qsj\Qdbr
\right)|0\rangle \,,
\nonumber  \\
|\D \rangle &=& -\frac{1}{\sqrt{2}}\left(
\Qcr\Qdbj - \Qcj\Qdbr
\right)|0\rangle \,,
\nonumber  \\
|\Ds \rangle &=& \frac{1}{\sqrt{2}}\left(
\Qcr\Qsbj - \Qcj\Qsbr
\right)|0\rangle \,,
\nonumber  \\
|\Db \rangle &=& -\frac{1}{\sqrt{2}}\left(
\Qur\Qcbj - \Quj\Qcbr
\right)|0\rangle \,,
\nonumber  \\
|\Dsb \rangle &=& -\frac{1}{\sqrt{2}}\left(
\Qsr\Qcbj - \Qsj\Qcbr
\right)|0\rangle \,,
\nonumber  \\
|\eta \rangle &=& \frac{1}{\sqrt{12}}\left(
\Qur\Qubj - \Quj\Qubr
+\Qdr\Qdbj - \Qdj\Qdbr
-2\Qsr\Qsbj +2 \Qsj\Qsbr
\right)|0\rangle \,,
\nonumber  \\
|\etap \rangle &=& \frac{1}{\sqrt{6}}\left(
\Qur\Qubj - \Quj\Qubr
+\Qdr\Qdbj - \Qdj\Qdbr
+\Qsr\Qsbj - \Qsj\Qsbr
\right)|0\rangle \,,
\nonumber  \\
|\etac \rangle &=& \frac{1}{\sqrt{2}}\left(
\Qcr\Qcbj - \Qcj\Qcbr
\right)|0\rangle \,.
\end{eqnarray}

Vector mesons:
\begin{eqnarray}
| \rho \rangle &=& -\Qur\Qdbr |0\rangle \,,
\nonumber  \\
| \kaonS \rangle &=& \Qur\Qsbr |0\rangle \,,
\nonumber  \\
| \kaonSb \rangle &=& -\Qsr\Qdbr |0\rangle \,,
\nonumber  \\
| \DS \rangle &=& -\Qcr\Qdbr |0\rangle \,,
\nonumber  \\
| \DsS \rangle &=& \Qcr\Qsbr |0\rangle \,,
\nonumber  \\
| \DSb \rangle &=& -\Qur\Qcbr |0\rangle \,,
\nonumber  \\
| \DsSb \rangle &=& -\Qsr\Qcbr |0\rangle \,,
\nonumber  \\
|\omega \rangle &=& \frac{1}{\sqrt{2}}\left(
\Qur\Qubr +\Qdr\Qdbr
\right)|0\rangle \,,
\nonumber  \\
|\phi \rangle &=& - \Qsr\Qsbr |0\rangle \,,
\nonumber  \\
|\Jpsi \rangle &=&  \Qcr\Qcbr |0\rangle \,.
\end{eqnarray}

Spin 1/2 baryons:
\begin{eqnarray}
| \Lambda \rangle &=& \frac{1}{\sqrt{2}}\left(
\Qur\Qdj\Qsr - \Quj\Qdr\Qsr
\right)|0\rangle \,,
\nonumber  \\
| \Nucleon \rangle &=& \frac{1}{\sqrt{3}}\left(
\Qur\Qur\Qdj - \Qur\Quj\Qdr
\right)|0\rangle \,,
\nonumber  \\
| \Sigma \rangle &=& \frac{1}{\sqrt{3}}\left(
\Qur\Qur\Qsj - \Qur\Quj\Qsr
\right)|0\rangle \,,
\nonumber  \\
| \Cascada \rangle &=& \frac{1}{\sqrt{3}}\left(
\Qur\Qsj\Qsr - \Quj\Qsr\Qsr
\right)|0\rangle \,,
\nonumber  \\
| \Sigma_c \rangle &=& \frac{1}{\sqrt{3}}\left(
\Qur\Qur\Qcj - \Qur\Quj\Qcr
\right)|0\rangle \,,
\nonumber  \\
| \Cascadacp \rangle &=& \frac{1}{\sqrt{6}}\left(
2\Qur\Qsr\Qcj - \Qur\Qsj\Qcr - \Quj\Qsr\Qcr
\right)|0\rangle \,,
\nonumber  \\
| \Omegac \rangle &=& \frac{1}{\sqrt{3}}\left(
\Qsr\Qsr\Qcj - \Qsr\Qsj\Qcr
\right)|0\rangle \,,
\nonumber  \\
| \Cascadac \rangle &=& \frac{1}{\sqrt{2}}\left(
\Qur\Qsj\Qcr - \Quj\Qsr\Qcr
\right)|0\rangle \,,
\nonumber  \\
| \Lambdac \rangle &=& \frac{1}{\sqrt{2}}\left(
\Qur\Qdj\Qcr - \Quj\Qdr\Qcr
\right)|0\rangle \,,
\nonumber  \\
| \Cascadacc \rangle &=& \frac{1}{\sqrt{3}}\left(
\Qur\Qcj\Qcr - \Quj\Qcr\Qcr
\right)|0\rangle \,,
\nonumber  \\
| \Omegacc \rangle &=& \frac{1}{\sqrt{3}}\left(
\Qsr\Qcj\Qcr - \Qsj\Qcr\Qcr
\right)|0\rangle \,.
\end{eqnarray}

Spin 3/2 baryons:
\begin{eqnarray}
| \Delta \rangle &=& \frac{1}{\sqrt{6}} \Qur\Qur\Qur |0\rangle \,,
\nonumber  \\
| \SigmaS \rangle &=& \frac{1}{\sqrt{2}} \Qur\Qur\Qsr |0\rangle \,,
\nonumber  \\
| \CascadaS \rangle &=& \frac{1}{\sqrt{2}} \Qur\Qsr\Qsr |0\rangle \,,
\nonumber  \\
| \Omega \rangle &=& \frac{1}{\sqrt{6}} \Qsr\Qsr\Qsr |0\rangle \,,
\nonumber  \\
| \SigmacS \rangle &=& \frac{1}{\sqrt{2}} \Qur\Qur\Qcr |0\rangle \,,
\nonumber  \\
| \CascadacS \rangle &=&  \Qur\Qsr\Qcr |0\rangle \,,
\nonumber  \\
| \OmegacS \rangle &=& \frac{1}{\sqrt{2}} \Qsr\Qsr\Qcr |0\rangle \,,
\nonumber  \\
| \CascadaccS \rangle &=& \frac{1}{\sqrt{2}} \Qur\Qcr\Qcr |0\rangle \,,
\nonumber  \\
| \OmegaccS \rangle &=& \frac{1}{\sqrt{2}} \Qsr\Qcr\Qcr |0\rangle \,,
\nonumber  \\
| \Omegaccc \rangle &=& \frac{1}{\sqrt{6}} \Qcr\Qcr\Qcr |0\rangle \,.
\end{eqnarray}

$\left [{\rm Notes}\right ]$:
\begin{itemize}
\item Each state represents  a member of a  spin and isospin multiplet. The
wavefunctions made explicit are those with the highest value of $I_3$ and
$J_3$ in the multiplet.

\item $\Qur$ creates a $u$ quark with spin up, $\Qdbj$ creates a quark
$\bar d$ with spin down, and so on. As usual, color takes care of the
fermionic statistics,  so the quark operators are bosonic.  The basis
$\{|\uparrow\rangle$, $|\downarrow\rangle\}$ is  of  the standard SU(2)
 by which step operators $J_\pm$  have non-negative matrix elements.  Also
standard are $\{|u\rangle$, $|d\rangle\}$, $\{|d\rangle$,
$|s\rangle\}$, $\{|s\rangle$, $|c\rangle\}$, and $\{-|\bar{d}\rangle$,
$|\bar{u}\rangle\}$, $\{-|\bar{s}\rangle$, $|\bar{d}\rangle\}$,
$\{-|\bar{c}\rangle$, $|\bar{s}\rangle\}$.\footnote{This is the
convention used in \cite{Baird:1964zm,Rabl:1975zy} for arbitrary
SU($n$). Note that the convention in \cite{deSwart:1963gc} for SU(3)
is, instead, $\{|u\rangle$, $|d\rangle\}$, $\{|u\rangle$,
$|s\rangle\}$, $\{-|\bar{d}\rangle$, $|\bar{u}\rangle\}$,
$\{-|\bar{s}\rangle$, $|\bar{u}\rangle\}$.  } This means that the
flavor-SU(4) step operators act as
\begin{eqnarray}
E_{12}|u\rangle =|d\rangle,\quad
E_{23}|d\rangle =|s\rangle,\quad
E_{34}|s\rangle =|c\rangle,\quad
\nonumber \\
E_{12}|\bar{d}\rangle =-|\bar{u}\rangle,\quad
E_{23}|\bar{s}\rangle =-|\bar{d}\rangle,\quad
E_{34}|\bar{c}\rangle =-|\bar{s}\rangle.
\end{eqnarray}

\item No special choice of relative phases has been made between different
SU(4) multiplets. The relative phases of the hadronic states in an
SU(4) multiplet are taken such that $E_{12}$, $E_{23}$ and $E_{34}$ have 
 non-negative matrix elements. Exceptions are the neutral mesons $\eta$, $\etap$,
$\etac$, $\Jpsi$, $\omega$ and $\phi$.

\item The pseudoscalar meson $\eta$ is (up to a sign) the isoscalar meson of
the SU(3) octet and $\etap$ the SU(3) singlet. $\etac$ is purely
$c\bar{c}$. In the spin 1 sector, $\Jpsi$ is purely $c\bar{c}$ whereas
$\phi$ and $\omega$ have been rotated in the SU(3) sector so that
$\phi$ is purely $s\bar{s}$ (ideal mixing).
\end{itemize}

\subsection{Tensor representation for matrix elements}

Having  identified  the group structure of the extended WT
interaction, the calculation of matrix elements can be done using
standard group theoretical techniques. Methods based on computing
SU($n$) Clebsch-Gordan coefficients (e.g. \cite{Carter:1965}),
extracting isoscalar factors (in SU(3) \cite{deSwart:1963gc}, and then
SU(3)-scalar factors in SU(4), etc) are useful for  an explicit hand
calculations for small groups but become more involved as the groups
get larger. Instead, we follow a different route here. A tensor representation
spanned by quark operators is used. This method only requires to know
Kronecker deltas and Wick contractions. Although it may not be  practical
for  hand calculations,  it is conceptually simple so that it is
easy enough to automatize the calculation for larger groups.

Following \cite{GarciaRecio:2006wb} the WT Hamiltonian
with spin-flavor symmetry takes the form
\begin{eqnarray}
-{\cal L}_{{\rm WT}}={\cal H}_{{\rm WT}}
&=& -\frac{{\rm i}}{4f^2}:[\Phi, \partial_0 \Phi]^i{}_j
 {\cal B}^\dagger_{i\ell m} {\cal B}^{j\ell m}:
\end{eqnarray}
where ${\cal B}^{ijk}$ is the baryon field, $\Phi{}^i{}_j$ the meson
field, and the  SU(8) way of labeling the spin-flavor indices goes for $i$, $j$, etc.
from $1$ to $8$ for four
flavors. ${\cal B}^{ijk}$ is completely symmetric under permutation of
indices.  There is no universally accepted convention for the
normalization.   Here we choose that  ${\cal B}^{ijk}$ is normalized 
in a standard manner for  a 
fermionic field when the indices $i$, $j$, $k$ are all
different. Upon extracting the kinematical factors
$(\sqrt{s}-M)/(2f^2)$,  the spin-flavor dependence is  all contained in the
matrices $D^{IJ}_{ab}$. These are the matrix elements of the operator
(see \cite{GarciaRecio:2006wb})
\begin{equation}
{\cal G}=:(M^i{}_k M^\dagger{}^k{}_j - M^\dagger{}^i{}_k M^k{}_j )
B^\dagger{}_{i\ell m} B^{j\ell m}:
\end{equation}
where $B^{ijk}$ and $B^\dagger_{ijk}$ denote the baryon annihilation and creation
operators, normalized as
\begin{equation}
[B^{ijk},B^\dagger_{i^\prime j^\prime k^\prime}]
= \delta^i_{i^\prime} \delta^j_{j^\prime} \delta^k_{k^\prime}
+
\delta^i_{j^\prime} \delta^j_{i^\prime} \delta^k_{k^\prime}
+ \cdots
\qquad \text{($3!$ permutations)}
\,.
\label{eq:A.8}
\end{equation}
Likewise $M^i{}_j$ and $M^\dagger{}^j{}_i=(M^i{}_j)^\dagger$ denote
the meson annihilation and creation operators, normalized as
\begin{equation}
[M^i{}_j,M^\dagger{}^{j^\prime}{}_{i^\prime}]
= \delta^i_{i^\prime} \delta^j_{j^\prime}
\,.
\label{eq:A.9}
\end{equation}

These characterizations automatically guarantee that the normalization and spin-flavor
transformation properties of the states $\langle 0|B^{ijk}$ and
$\langle 0|M^i{}_j$ coincide with those of $\langle 0|Q^i Q^j Q^k$ and
$\langle 0|Q^i\bar{Q}_j$, respectively. Here $Q^i$ and $\bar{Q}_i$ are
the SU(8) labeled annihilation operators of quarks and antiquarks,
respectively. They transform as
\begin{equation}
Q^i\to U^i{}_jQ^j,\quad
\bar{Q}_i\to (U^i{}_j)^*\bar{Q}_j=
\bar{Q}_j  (U^{-1})^j{}_i,
\qquad U\in {\rm SU}(8)
\end{equation}
and have standard normalization
\begin{equation}
[Q^i,Q^\dagger{}_j] = \delta^i_j \,,
\qquad
[\bar{Q}_i,\bar{Q}^\dagger{}^j] = \delta^j_i
\,.
\end{equation}

The correspondence between the SU(8) and the explicit spin-SU(4) flavor representations are the following: since  $\bar{Q}_i$ transforms as $Q^\dagger_i$, the conventions in subsection
\ref{subsec:a.1} imply the identifications (up to a global sign)
\begin{eqnarray}
Q^i &=& (Q_{u\uparrow},Q_{d\uparrow},Q_{s\uparrow},Q_{c\uparrow},
Q_{u\downarrow},Q_{d\downarrow},Q_{s\downarrow},Q_{c\downarrow})\,,
\nonumber \\
\bar{Q}_i &=&
(Q_{\bar{u}\downarrow},Q_{\bar{d}\downarrow},
Q_{\bar{s}\downarrow},Q_{\bar{c}\downarrow},
-Q_{\bar{u}\uparrow},-Q_{\bar{d}\uparrow},
-Q_{\bar{s}\uparrow},-Q_{\bar{c}\uparrow})\,.
\end{eqnarray}
That is $Q^1=Q_{u\uparrow}$, $\bar{Q}_6=-Q_{\bar{d}\uparrow}$,
$\bar{Q}^\dagger{}^6=-Q^\dagger_{\bar{d}\uparrow}$, etc. The 
 correct book keeping
between  an explicit flavor-spin index and  the SU(8)  one (running from 1 to 8) needs to be
maintained to apply the Kronecker deltas in (\ref{eq:A.8}) and
(\ref{eq:A.9}). Using these identifications together with
\begin{eqnarray}
Q^\dagger_i Q^\dagger_j Q^\dagger_k &\to& B^\dagger_{ijk} \,,
\nonumber \\
\bar{Q}^\dagger{}^i Q^\dagger_j &\to& M^\dagger{}^i{}_j \,,
\end{eqnarray}
the mesonic and baryonic wavefunctions can be rewritten in terms of
the operators $M^\dagger{}^i{}_j$ and $B^\dagger_{ijk}$. For instance,
\begin{eqnarray}
| \pion \rangle &=&-\frac{1}{\sqrt{2}}\left(
M^\dagger{}^2{}_1+ M^\dagger{}^6{}_5
\right)|0\rangle \,,
\nonumber  \\
| \Nucleon \rangle &=& \frac{1}{\sqrt{3}}\left(
B^\dagger_{116}-B^\dagger_{152}
\right)|0\rangle \,,
\end{eqnarray}
(once again for the highest weights in the spin-isospin multiplet).
It is straightforward now to obtain the matrix elements $\langle M
B|{\cal G}|M^\prime B^\prime\rangle$ using the commutation relations,
or equivalently Wick's theorem. A non-trivial confirmation of the calculation
is to see that the matrices $D_{ab}^{IJSC}$ so obtained have the correct
eigenvalues $6,0,-2,-16,-22$ (the $0$ being the consequence of the
inclusion of the SU(8) singlet among the mesons). Another  important point to check
 is that all the  matrix elements computed are (up to a
sign) square roots of rational numbers. While this is not surprising
for matrix elements of SU($n$) with standard bases, it is found to be
the case also for the neutral vector mesons with ideal mixing.

\section{Coefficients of the $s$-wave tree level amplitudes}
\label{app:tables}

This Appendix gives the coefficients $D_{ab}^{IJSC}$ of the $s$-wave tree level Baryon-Meson\footnote{
We say Baryon-Meson amplitudes to stress that the given $D^{IJSC}$ matrix elements corresponds to the state 
of the two colliding hadrons coupled in this order: Baryon-Meson. Change of phases in the matrix elements can be derived of using other order.} 
amplitudes of Eq.~(\ref{eq:vsu8}) for the various $IJSC$ sectors, in the 
particular case of $S=0$ and $C=1$.

\begin{sidewaystable}
\centering
\caption{
  $ I=0$, $ J=1/2$, $ S=0$, $ C= 1$.
Baryon-Meson states with more
than one $c$ quark have not been included.
}
\begin{ruledtabular}
\begin{tabular}{rrrrrrrrrrrrrrrrrrrrrrrrrrrrr}
 &  $ \Sigmac \pion $ 
 &  $ \Nucleon \D $ &  $ \Lambdac \eta 
  $ &  $ \Nucleon \DS $ 
 &  $ \Cascadac \kaon $ &  $ \Lambdac 
  \omega $ &  $ \Cascadacp \kaon $ 
 &  $ \Lambda \Ds $ &  $ \Lambda \DsS 
  $ &  $ \Sigmac \rho $ 
 &  $ \Lambdac \etap $ &  $ \SigmacS 
  \rho $ &  $ \Lambdac \phi $ 
 &  $ \Cascadac \kaonS $ &  $ \Cascadacp 
  \kaonS $ &  $ \CascadacS \kaonS $\\
$ \Sigmac \pion $& $ -4 $ 
 & $ \sqrt{\frac{ 3 }{ 2 }} 
   $ & $ 0 $ 
 & $ -\sqrt{\frac{ 1 }{ 2 }} 
   $ & $ 0 $ & $ 0 $ 
 & $ -\sqrt{ 3 } $ 
 & $ 0 $ & $ 0 $ 
 & $ -\sqrt{\frac{ 64 }{ 3 }} 
   $ & $ 0 $ 
 & $ \sqrt{\frac{ 32 }{ 3 }} 
   $ & $ 0 $ 
 & $ -\sqrt{ 3 } $ 
 & $ -2 $ 
 & $ \sqrt{ 2 } $ \\
$ \Nucleon \D $
 & $ \sqrt{\frac{ 3 }{ 2 }} 
   $ & $ -3 $ 
 & $ \sqrt{\frac{ 1 }{ 2 }} 
   $ & $ -\sqrt{ 27 } $ 
 & $ 0 $ 
 & $ \sqrt{\frac{ 9 }{ 2 }} 
   $ & $ 0 $ 
 & $ -\sqrt{ 3 } $ 
 & $ -3 $ 
 & $ -\sqrt{\frac{ 1 }{ 2 }} 
   $ & $ 1 $ & $ 2 $ 
 & $ 0 $ & $ 0 $ 
 & $ 0 $ & $ 0 $ \\
$ \Lambdac \eta $& $ 0 $ 
 & $ \sqrt{\frac{ 1 }{ 2 }} 
   $ & $ 0 $ 
 & $ \sqrt{\frac{ 3 }{ 2 }} 
   $ & $ -\sqrt{ 3 } $ 
 & $ 0 $ & $ 0 $ 
 & $ -\sqrt{\frac{ 2 }{ 3 }} 
   $ & $ -\sqrt{ 2 } $ 
 & $ 0 $ & $ 0 $ 
 & $ 0 $ & $ 0 $ 
 & $ 0 $ 
 & $ -\sqrt{ 3 } $ 
 & $ -\sqrt{ 6 } $ \\
$ \Nucleon \DS $
 & $ -\sqrt{\frac{ 1 }{ 2 }} 
   $ & $ -\sqrt{ 27 } $ 
 & $ \sqrt{\frac{ 3 }{ 2 }} 
   $ & $ -9 $ & $ 0 $ 
 & $ -\sqrt{\frac{ 3 }{ 2 }} 
   $ & $ 0 $ & $ -3 $ 
 & $ -\sqrt{ 27 } $ 
 & $ \sqrt{\frac{ 25 }{ 6 }} 
   $ & $ \sqrt{ 3 } $ 
 & $ \sqrt{\frac{ 4 }{ 3 }} 
   $ & $ 0 $ & $ 0 $ 
 & $ 0 $ & $ 0 $ \\
$ \Cascadac \kaon $& $ 0 $ 
 & $ 0 $ 
 & $ -\sqrt{ 3 } $ 
 & $ 0 $ & $ -2 $ 
 & $ 0 $ & $ 0 $ 
 & $ \sqrt{\frac{ 1 }{ 2 }} 
   $ & $ \sqrt{\frac{ 3 }{ 2 
    }} $ & $ -\sqrt{ 3 } 
   $ & $ 0 $ 
 & $ -\sqrt{ 6 } $ 
 & $ 0 $ & $ 0 $ 
 & $ 0 $ & $ 0 $ \\
$ \Lambdac \omega $& $ 0 $ 
 & $ \sqrt{\frac{ 9 }{ 2 }} 
   $ & $ 0 $ 
 & $ -\sqrt{\frac{ 3 }{ 2 }} 
   $ & $ 0 $ & $ 0 $ 
 & $ -1 $ & $ 0 $ 
 & $ 0 $ & $ -4 $ 
 & $ 0 $ & $ \sqrt{ 8 } 
   $ & $ 0 $ & $ -1 $ 
 & $ -\sqrt{\frac{ 4 }{ 3 }} 
   $ & $ \sqrt{\frac{ 2 }{ 3 
    }} $ \\
$ \Cascadacp \kaon $
 & $ -\sqrt{ 3 } $ 
 & $ 0 $ & $ 0 $ 
 & $ 0 $ & $ 0 $ 
 & $ -1 $ & $ -2 $ 
 & $ \sqrt{\frac{ 3 }{ 2 }} 
   $ & $ -\sqrt{\frac{ 1 }{ 2 
     }} $ & $ -2 $ 
 & $ 0 $ & $ \sqrt{ 2 } 
   $ & $ -\sqrt{ 2 } $ 
 & $ 0 $ 
 & $ -\sqrt{\frac{ 16 }{ 3 }} 
   $ & $ \sqrt{\frac{ 8 }{ 3 
    }} $ \\
$ \Lambda \Ds $& $ 0 $ 
 & $ -\sqrt{ 3 } $ 
 & $ -\sqrt{\frac{ 2 }{ 3 }} 
   $ & $ -3 $ 
 & $ \sqrt{\frac{ 1 }{ 2 }} 
   $ & $ 0 $ 
 & $ \sqrt{\frac{ 3 }{ 2 }} 
   $ & $ -1 $ 
 & $ -\sqrt{ 3 } $ 
 & $ 0 $ 
 & $ \sqrt{\frac{ 1 }{ 3 }} 
   $ & $ 0 $ 
 & $ -\sqrt{ 3 } $ 
 & $ \sqrt{\frac{ 3 }{ 2 }} 
   $ & $ -\sqrt{\frac{ 1 }{ 2 
     }} $ & $ 2 $ \\
$ \Lambda \DsS $& $ 0 $ 
 & $ -3 $ 
 & $ -\sqrt{ 2 } $ 
 & $ -\sqrt{ 27 } $ 
 & $ \sqrt{\frac{ 3 }{ 2 }} 
   $ & $ 0 $ 
 & $ -\sqrt{\frac{ 1 }{ 2 }} 
   $ & $ -\sqrt{ 3 } $ 
 & $ -3 $ & $ 0 $ 
 & $ 1 $ & $ 0 $ 
 & $ 1 $ 
 & $ -\sqrt{\frac{ 1 }{ 2 }} 
   $ & $ \sqrt{\frac{ 25 }{ 6 
    }} $ & $ \sqrt{\frac{ 4 
    }{ 3 }} $ \\
$ \Sigmac \rho $
 & $ -\sqrt{\frac{ 64 }{ 3 }} 
   $ & $ -\sqrt{\frac{ 1 }{ 2 
     }} $ & $ 0 $ 
 & $ \sqrt{\frac{ 25 }{ 6 }} 
   $ & $ -\sqrt{ 3 } $ 
 & $ -4 $ & $ -2 $ 
 & $ 0 $ & $ 0 $ 
 & $ -\frac{ 20 }{ 3  } $ 
 & $ 0 $ 
 & $ -\sqrt{\frac{ 8 }{ 9 }} 
   $ & $ 0 $ & $ -2 $ 
 & $ -\sqrt{\frac{ 49 }{ 3 }} 
   $ & $ -\sqrt{\frac{ 2 }{ 3 
     }} $ \\
$ \Lambdac \etap $& $ 0 $ 
 & $ 1 $ & $ 0 $ 
 & $ \sqrt{ 3 } $ 
 & $ 0 $ & $ 0 $ 
 & $ 0 $ 
 & $ \sqrt{\frac{ 1 }{ 3 }} 
   $ & $ 1 $ & $ 0 $ 
 & $ 0 $ & $ 0 $ 
 & $ 0 $ & $ 0 $ 
 & $ 0 $ & $ 0 $ \\
$ \SigmacS \rho $
 & $ \sqrt{\frac{ 32 }{ 3 }} 
   $ & $ 2 $ & $ 0 $ 
 & $ \sqrt{\frac{ 4 }{ 3 }} 
   $ & $ -\sqrt{ 6 } $ 
 & $ \sqrt{ 8 } $ 
 & $ \sqrt{ 2 } $ 
 & $ 0 $ & $ 0 $ 
 & $ -\sqrt{\frac{ 8 }{ 9 }} 
   $ & $ 0 $ 
 & $ -\frac{ 22 }{ 3  } $ 
 & $ 0 $ & $ \sqrt{ 2 } 
   $ & $ -\sqrt{\frac{ 2 }{ 3 
     }} $ 
 & $ -\sqrt{\frac{ 64 }{ 3 }} 
   $ \\
$ \Lambdac \phi $& $ 0 $ 
 & $ 0 $ & $ 0 $ 
 & $ 0 $ & $ 0 $ 
 & $ 0 $ 
 & $ -\sqrt{ 2 } $ 
 & $ -\sqrt{ 3 } $ 
 & $ 1 $ & $ 0 $ 
 & $ 0 $ & $ 0 $ 
 & $ 0 $ 
 & $ -\sqrt{ 2 } $ 
 & $ \sqrt{\frac{ 8 }{ 3 }} 
   $ & $ -\sqrt{\frac{ 4 }{ 3 
     }} $ \\
$ \Cascadac \kaonS $
 & $ -\sqrt{ 3 } $ 
 & $ 0 $ & $ 0 $ 
 & $ 0 $ & $ 0 $ 
 & $ -1 $ & $ 0 $ 
 & $ \sqrt{\frac{ 3 }{ 2 }} 
   $ & $ -\sqrt{\frac{ 1 }{ 2 
     }} $ & $ -2 $ 
 & $ 0 $ & $ \sqrt{ 2 } 
   $ & $ -\sqrt{ 2 } $ 
 & $ -2 $ 
 & $ -\sqrt{\frac{ 16 }{ 3 }} 
   $ & $ \sqrt{\frac{ 8 }{ 3 
    }} $ \\
$ \Cascadacp \kaonS $& $ -2 $ 
 & $ 0 $ 
 & $ -\sqrt{ 3 } $ 
 & $ 0 $ & $ 0 $ 
 & $ -\sqrt{\frac{ 4 }{ 3 }} 
   $ & $ -\sqrt{\frac{ 16 }{ 3 
     }} $ 
 & $ -\sqrt{\frac{ 1 }{ 2 }} 
   $ & $ \sqrt{\frac{ 25 }{ 6 
    }} $ 
 & $ -\sqrt{\frac{ 49 }{ 3 }} 
   $ & $ 0 $ 
 & $ -\sqrt{\frac{ 2 }{ 3 }} 
   $ & $ \sqrt{\frac{ 8 }{ 3 
    }} $ 
 & $ -\sqrt{\frac{ 16 }{ 3 }} 
   $ & $ -2 $ & $ 0 $ \\
$ \CascadacS \kaonS $
 & $ \sqrt{ 2 } $ 
 & $ 0 $ 
 & $ -\sqrt{ 6 } $ 
 & $ 0 $ & $ 0 $ 
 & $ \sqrt{\frac{ 2 }{ 3 }} 
   $ & $ \sqrt{\frac{ 8 }{ 3 
    }} $ & $ 2 $ 
 & $ \sqrt{\frac{ 4 }{ 3 }} 
   $ & $ -\sqrt{\frac{ 2 }{ 3 
     }} $ & $ 0 $ 
 & $ -\sqrt{\frac{ 64 }{ 3 }} 
   $ & $ -\sqrt{\frac{ 4 }{ 3 
     }} $ & $ \sqrt{\frac{ 8 
    }{ 3 }} $ & $ 0 $ 
 & $ -2 $ \\
\end{tabular}
\end{ruledtabular}
\label{tab:i0j12s0c1}
\end{sidewaystable}

\begin{sidewaystable}
\centering
\caption{
  $ I=1$, $ J=1/2$, $ S=0$, $ C= 1$.
Baryon-Meson states with more
than one $c$ quark have not been included.
}
\begin{ruledtabular}
\begin{tabular}{rrrrrrrrrrrrrrrrrrrrrrrrrrrrr}
 &  $ \Lambdac \pion $ 
 &  $ \Sigmac \pion $ &  $ \Nucleon \D 
  $ &  $ \Nucleon \DS $ 
 &  $ \Cascadac \kaon $ &  $ \Sigmac 
  \eta $ &  $ \Lambdac \rho $ 
 &  $ \Cascadacp \kaon $ 
 &  $ \Sigma \Ds $ &  $ \Delta \DS 
  $ &  $ \Sigmac \rho $ 
 &  $ \Sigmac \omega $ &  $ \SigmacS 
  \rho $ &  $ \SigmacS \omega $ 
 &  $ \Sigma \DsS $ &  $ \Cascadac 
  \kaonS $ &  $ \Sigmac \etap $ 
 &  $ \Cascadacp \kaonS $ 
 &  $ \Sigmac \phi $ &  $ \SigmaS \DsS 
  $ &  $ \SigmacS \phi $ 
 &  $ \CascadacS \kaonS $\\
$ \Lambdac \pion $& $ 0 $ 
 & $ 0 $ 
 & $ \sqrt{\frac{ 3 }{ 2 }} 
   $ & $ \sqrt{\frac{ 9 }{ 2 
    }} $ & $ 1 $ & $ 0 $ 
 & $ 0 $ & $ 0 $ 
 & $ 0 $ & $ 0 $ 
 & $ \sqrt{ 8 } $ 
 & $ 0 $ & $ 4 $ 
 & $ 0 $ & $ 0 $ 
 & $ 0 $ & $ 0 $ 
 & $ 1 $ & $ 0 $ 
 & $ 0 $ & $ 0 $ 
 & $ \sqrt{ 2 } $ \\
$ \Sigmac \pion $& $ 0 $ 
 & $ -2 $ & $ 1 $ 
 & $ -\sqrt{\frac{ 1 }{ 3 }} 
   $ & $ 0 $ & $ 0 $ 
 & $ \sqrt{ 8 } $ 
 & $ -\sqrt{ 2 } $ 
 & $ 0 $ 
 & $ \sqrt{\frac{ 8 }{ 3 }} 
   $ & $ -\sqrt{\frac{ 16 }{ 3 
     }} $ & $ 0 $ 
 & $ \sqrt{\frac{ 8 }{ 3 }} 
   $ & $ 0 $ & $ 0 $ 
 & $ -\sqrt{ 2 } $ 
 & $ 0 $ 
 & $ -\sqrt{\frac{ 8 }{ 3 }} 
   $ & $ 0 $ & $ 0 $ 
 & $ 0 $ 
 & $ \sqrt{\frac{ 4 }{ 3 }} 
   $ \\
$ \Nucleon \D $
 & $ \sqrt{\frac{ 3 }{ 2 }} 
   $ & $ 1 $ & $ -1 $ 
 & $ \sqrt{\frac{ 1 }{ 3 }} 
   $ & $ 0 $ 
 & $ -\sqrt{\frac{ 1 }{ 6 }} 
   $ & $ \sqrt{\frac{ 9 }{ 2 
    }} $ & $ 0 $ & $ 1 $ 
 & $ \sqrt{\frac{ 32 }{ 3 }} 
   $ & $ -\sqrt{\frac{ 1 }{ 3 
     }} $ & $ \sqrt{\frac{ 1 
    }{ 6 }} $ 
 & $ \sqrt{\frac{ 8 }{ 3 }} 
   $ & $ -\sqrt{\frac{ 4 }{ 3 
     }} $ 
 & $ -\sqrt{\frac{ 1 }{ 3 }} 
   $ & $ 0 $ 
 & $ -\sqrt{\frac{ 1 }{ 3 }} 
   $ & $ 0 $ & $ 0 $ 
 & $ \sqrt{\frac{ 8 }{ 3 }} 
   $ & $ 0 $ & $ 0 $ \\
$ \Nucleon \DS $
 & $ \sqrt{\frac{ 9 }{ 2 }} 
   $ & $ -\sqrt{\frac{ 1 }{ 3 
     }} $ & $ \sqrt{\frac{ 1 
    }{ 3 }} $ 
 & $ -\frac{ 1 }{ 3  } $ 
 & $ 0 $ 
 & $ \sqrt{\frac{ 1 }{ 18 }} 
   $ & $ -\sqrt{\frac{ 3 }{ 2 
     }} $ & $ 0 $ 
 & $ -\sqrt{\frac{ 1 }{ 3 }} 
   $ & $ -\sqrt{\frac{ 32 }{ 9 
     }} $ & $ \frac{ 5 }{ 3 
     } $ 
 & $ -\sqrt{\frac{ 25 }{ 18 }} 
   $ & $ \sqrt{\frac{ 8 }{ 9 
    }} $ 
 & $ -\frac{ 2 }{ 3  } $ 
 & $ \frac{ 1 }{ 3  } $ 
 & $ 0 $ 
 & $ \frac{ 1 }{ 3  } $ 
 & $ 0 $ & $ 0 $ 
 & $ -\sqrt{\frac{ 8 }{ 9 }} 
   $ & $ 0 $ & $ 0 $ \\
$ \Cascadac \kaon $& $ 1 $ 
 & $ 0 $ & $ 0 $ 
 & $ 0 $ & $ 0 $ 
 & $ 0 $ & $ 0 $ 
 & $ 0 $ 
 & $ \sqrt{\frac{ 3 }{ 2 }} 
   $ & $ 0 $ 
 & $ -\sqrt{ 2 } $ 
 & $ -1 $ & $ -2 $ 
 & $ -\sqrt{ 2 } $ 
 & $ \sqrt{\frac{ 9 }{ 2 }} 
   $ & $ 0 $ & $ 0 $ 
 & $ -2 $ 
 & $ -\sqrt{ 2 } $ 
 & $ 0 $ & $ -2 $ 
 & $ -\sqrt{ 8 } $ \\
$ \Sigmac \eta $& $ 0 $ 
 & $ 0 $ 
 & $ -\sqrt{\frac{ 1 }{ 6 }} 
   $ & $ \sqrt{\frac{ 1 }{ 18 
    }} $ & $ 0 $ & $ 0 $ 
 & $ 0 $ 
 & $ -\sqrt{ 3 } $ 
 & $ -\sqrt{\frac{ 2 }{ 3 }} 
   $ & $ \frac{ 4 }{ 3  } 
   $ & $ 0 $ & $ 0 $ 
 & $ 0 $ & $ 0 $ 
 & $ \sqrt{\frac{ 2 }{ 9 }} 
   $ & $ -\sqrt{ 3 } $ 
 & $ 0 $ & $ -2 $ 
 & $ 0 $ 
 & $ -\frac{ 4 }{ 3  } $ 
 & $ 0 $ & $ \sqrt{ 2 } 
   $ \\
$ \Lambdac \rho $& $ 0 $ 
 & $ \sqrt{ 8 } $ 
 & $ \sqrt{\frac{ 9 }{ 2 }} 
   $ & $ -\sqrt{\frac{ 3 }{ 2 
     }} $ & $ 0 $ 
 & $ 0 $ & $ 0 $ 
 & $ 1 $ & $ 0 $ 
 & $ 0 $ & $ 0 $ 
 & $ \sqrt{\frac{ 16 }{ 3 }} 
   $ & $ 0 $ 
 & $ -\sqrt{\frac{ 8 }{ 3 }} 
   $ & $ 0 $ & $ 1 $ 
 & $ 0 $ 
 & $ \sqrt{\frac{ 4 }{ 3 }} 
   $ & $ 0 $ & $ 0 $ 
 & $ 0 $ 
 & $ -\sqrt{\frac{ 2 }{ 3 }} 
   $ \\
$ \Cascadacp \kaon $& $ 0 $ 
 & $ -\sqrt{ 2 } $ 
 & $ 0 $ & $ 0 $ 
 & $ 0 $ 
 & $ -\sqrt{ 3 } $ 
 & $ 1 $ & $ 0 $ 
 & $ -\sqrt{\frac{ 1 }{ 2 }} 
   $ & $ 0 $ 
 & $ -\sqrt{\frac{ 8 }{ 3 }} 
   $ & $ -\sqrt{\frac{ 4 }{ 3 
     }} $ & $ \sqrt{\frac{ 4 
    }{ 3 }} $ 
 & $ \sqrt{\frac{ 2 }{ 3 }} 
   $ & $ \sqrt{\frac{ 1 }{ 6 
    }} $ & $ -2 $ 
 & $ 0 $ & $ 0 $ 
 & $ -\sqrt{\frac{ 8 }{ 3 }} 
   $ & $ \sqrt{\frac{ 16 }{ 3 
    }} $ & $ \sqrt{\frac{ 4 
    }{ 3 }} $ & $ 0 $ \\
$ \Sigma \Ds $& $ 0 $ 
 & $ 0 $ & $ 1 $ 
 & $ -\sqrt{\frac{ 1 }{ 3 }} 
   $ & $ \sqrt{\frac{ 3 }{ 2 
    }} $ 
 & $ -\sqrt{\frac{ 2 }{ 3 }} 
   $ & $ 0 $ 
 & $ -\sqrt{\frac{ 1 }{ 2 }} 
   $ & $ -1 $ 
 & $ -\sqrt{\frac{ 32 }{ 3 }} 
   $ & $ 0 $ & $ 0 $ 
 & $ 0 $ & $ 0 $ 
 & $ \sqrt{\frac{ 1 }{ 3 }} 
   $ & $ \sqrt{\frac{ 9 }{ 2 
    }} $ & $ \sqrt{\frac{ 1 
    }{ 3 }} $ 
 & $ \sqrt{\frac{ 1 }{ 6 }} 
   $ & $ \sqrt{\frac{ 1 }{ 3 
    }} $ 
 & $ -\sqrt{\frac{ 8 }{ 3 }} 
   $ & $ -\sqrt{\frac{ 8 }{ 3 
     }} $ 
 & $ -\sqrt{\frac{ 4 }{ 3 }} 
   $ \\
$ \Delta \DS $& $ 0 $ 
 & $ \sqrt{\frac{ 8 }{ 3 }} 
   $ & $ \sqrt{\frac{ 32 }{ 3 
    }} $ 
 & $ -\sqrt{\frac{ 32 }{ 9 }} 
   $ & $ 0 $ 
 & $ \frac{ 4 }{ 3  } $ 
 & $ 0 $ & $ 0 $ 
 & $ -\sqrt{\frac{ 32 }{ 3 }} 
   $ & $ -\frac{ 32 }{ 3  } 
   $ & $ \sqrt{\frac{ 8 }{ 9 
    }} $ & $ \frac{ 4 }{ 3 
     } $ 
 & $ -\frac{ 2 }{ 3  } $ 
 & $ -\sqrt{\frac{ 8 }{ 9 }} 
   $ & $ \sqrt{\frac{ 32 }{ 9 
    }} $ & $ 0 $ 
 & $ \sqrt{\frac{ 32 }{ 9 }} 
   $ & $ 0 $ & $ 0 $ 
 & $ -\frac{ 16 }{ 3  } $ 
 & $ 0 $ & $ 0 $ \\
$ \Sigmac \rho $
 & $ \sqrt{ 8 } $ 
 & $ -\sqrt{\frac{ 16 }{ 3 }} 
   $ & $ -\sqrt{\frac{ 1 }{ 3 
     }} $ & $ \frac{ 5 }{ 3 
     } $ & $ -\sqrt{ 2 } $ 
 & $ 0 $ & $ 0 $ 
 & $ -\sqrt{\frac{ 8 }{ 3 }} 
   $ & $ 0 $ 
 & $ \sqrt{\frac{ 8 }{ 9 }} 
   $ & $ -\frac{ 14 }{ 3  } 
   $ & $ -\sqrt{\frac{ 128 }{ 9 
     }} $ 
 & $ -\sqrt{\frac{ 8 }{ 9 }} 
   $ & $ -\frac{ 4 }{ 3  } 
   $ & $ 0 $ 
 & $ -\sqrt{\frac{ 8 }{ 3 }} 
   $ & $ 0 $ 
 & $ -\sqrt{\frac{ 98 }{ 9 }} 
   $ & $ 0 $ & $ 0 $ 
 & $ 0 $ 
 & $ -\frac{ 2 }{ 3  } $ \\
$ \Sigmac \omega $& $ 0 $ 
 & $ 0 $ 
 & $ \sqrt{\frac{ 1 }{ 6 }} 
   $ & $ -\sqrt{\frac{ 25 }{ 18 
     }} $ & $ -1 $ 
 & $ 0 $ 
 & $ \sqrt{\frac{ 16 }{ 3 }} 
   $ & $ -\sqrt{\frac{ 4 }{ 3 
     }} $ & $ 0 $ 
 & $ \frac{ 4 }{ 3  } $ 
 & $ -\sqrt{\frac{ 128 }{ 9 }} 
   $ & $ -\frac{ 8 }{ 3  } 
   $ & $ -\frac{ 4 }{ 3  } 
   $ & $ -\sqrt{\frac{ 8 }{ 9 
     }} $ & $ 0 $ 
 & $ -\sqrt{\frac{ 4 }{ 3 }} 
   $ & $ 0 $ 
 & $ -\frac{ 7 }{ 3  } $ 
 & $ 0 $ & $ 0 $ 
 & $ 0 $ 
 & $ -\sqrt{\frac{ 2 }{ 9 }} 
   $ \\
$ \SigmacS \rho $& $ 4 $ 
 & $ \sqrt{\frac{ 8 }{ 3 }} 
   $ & $ \sqrt{\frac{ 8 }{ 3 
    }} $ & $ \sqrt{\frac{ 8 
    }{ 9 }} $ & $ -2 $ 
 & $ 0 $ & $ 0 $ 
 & $ \sqrt{\frac{ 4 }{ 3 }} 
   $ & $ 0 $ 
 & $ -\frac{ 2 }{ 3  } $ 
 & $ -\sqrt{\frac{ 8 }{ 9 }} 
   $ & $ -\frac{ 4 }{ 3  } 
   $ & $ -\frac{ 16 }{ 3  } 
   $ & $ -\sqrt{\frac{ 200 }{ 9 
     }} $ & $ 0 $ 
 & $ \sqrt{\frac{ 4 }{ 3 }} 
   $ & $ 0 $ 
 & $ -\frac{ 2 }{ 3  } $ 
 & $ 0 $ & $ 0 $ 
 & $ 0 $ 
 & $ -\sqrt{\frac{ 128 }{ 9 }} 
   $ \\
$ \SigmacS \omega $& $ 0 $ 
 & $ 0 $ 
 & $ -\sqrt{\frac{ 4 }{ 3 }} 
   $ & $ -\frac{ 2 }{ 3  } 
   $ & $ -\sqrt{ 2 } $ 
 & $ 0 $ 
 & $ -\sqrt{\frac{ 8 }{ 3 }} 
   $ & $ \sqrt{\frac{ 2 }{ 3 
    }} $ & $ 0 $ 
 & $ -\sqrt{\frac{ 8 }{ 9 }} 
   $ & $ -\frac{ 4 }{ 3  } 
   $ & $ -\sqrt{\frac{ 8 }{ 9 
     }} $ 
 & $ -\sqrt{\frac{ 200 }{ 9 }} 
   $ & $ -\frac{ 10 }{ 3  } 
   $ & $ 0 $ 
 & $ \sqrt{\frac{ 2 }{ 3 }} 
   $ & $ 0 $ 
 & $ -\sqrt{\frac{ 2 }{ 9 }} 
   $ & $ 0 $ & $ 0 $ 
 & $ 0 $ 
 & $ -\frac{ 8 }{ 3  } $ \\
$ \Sigma \DsS $& $ 0 $ 
 & $ 0 $ 
 & $ -\sqrt{\frac{ 1 }{ 3 }} 
   $ & $ \frac{ 1 }{ 3  } 
   $ & $ \sqrt{\frac{ 9 }{ 2 
    }} $ & $ \sqrt{\frac{ 2 
    }{ 9 }} $ & $ 0 $ 
 & $ \sqrt{\frac{ 1 }{ 6 }} 
   $ & $ \sqrt{\frac{ 1 }{ 3 
    }} $ & $ \sqrt{\frac{ 32 
    }{ 9 }} $ & $ 0 $ 
 & $ 0 $ & $ 0 $ 
 & $ 0 $ 
 & $ -\frac{ 1 }{ 3  } $ 
 & $ -\sqrt{\frac{ 3 }{ 2 }} 
   $ & $ -\frac{ 1 }{ 3  } 
   $ & $ -\sqrt{\frac{ 25 }{ 18 
     }} $ 
 & $ -\frac{ 5 }{ 3  } $ 
 & $ \sqrt{\frac{ 8 }{ 9 }} 
   $ & $ -\sqrt{\frac{ 8 }{ 9 
     }} $ 
 & $ -\frac{ 2 }{ 3  } $ \\
$ \Cascadac \kaonS $& $ 0 $ 
 & $ -\sqrt{ 2 } $ 
 & $ 0 $ & $ 0 $ 
 & $ 0 $ 
 & $ -\sqrt{ 3 } $ 
 & $ 1 $ & $ -2 $ 
 & $ \sqrt{\frac{ 9 }{ 2 }} 
   $ & $ 0 $ 
 & $ -\sqrt{\frac{ 8 }{ 3 }} 
   $ & $ -\sqrt{\frac{ 4 }{ 3 
     }} $ & $ \sqrt{\frac{ 4 
    }{ 3 }} $ 
 & $ \sqrt{\frac{ 2 }{ 3 }} 
   $ & $ -\sqrt{\frac{ 3 }{ 2 
     }} $ & $ 0 $ 
 & $ 0 $ & $ 0 $ 
 & $ \sqrt{\frac{ 8 }{ 3 }} 
   $ & $ 0 $ 
 & $ -\sqrt{\frac{ 4 }{ 3 }} 
   $ & $ 0 $ \\
$ \Sigmac \etap $& $ 0 $ 
 & $ 0 $ 
 & $ -\sqrt{\frac{ 1 }{ 3 }} 
   $ & $ \frac{ 1 }{ 3  } 
   $ & $ 0 $ & $ 0 $ 
 & $ 0 $ & $ 0 $ 
 & $ \sqrt{\frac{ 1 }{ 3 }} 
   $ & $ \sqrt{\frac{ 32 }{ 9 
    }} $ & $ 0 $ & $ 0 $ 
 & $ 0 $ & $ 0 $ 
 & $ -\frac{ 1 }{ 3  } $ 
 & $ 0 $ & $ 0 $ 
 & $ 0 $ & $ 0 $ 
 & $ \sqrt{\frac{ 8 }{ 9 }} 
   $ & $ 0 $ & $ 0 $ \\
$ \Cascadacp \kaonS $& $ 1 $ 
 & $ -\sqrt{\frac{ 8 }{ 3 }} 
   $ & $ 0 $ & $ 0 $ 
 & $ -2 $ & $ -2 $ 
 & $ \sqrt{\frac{ 4 }{ 3 }} 
   $ & $ 0 $ 
 & $ \sqrt{\frac{ 1 }{ 6 }} 
   $ & $ 0 $ 
 & $ -\sqrt{\frac{ 98 }{ 9 }} 
   $ & $ -\frac{ 7 }{ 3  } 
   $ & $ -\frac{ 2 }{ 3  } 
   $ & $ -\sqrt{\frac{ 2 }{ 9 
     }} $ 
 & $ -\sqrt{\frac{ 25 }{ 18 }} 
   $ & $ 0 $ & $ 0 $ 
 & $ -\frac{ 8 }{ 3  } $ 
 & $ \sqrt{\frac{ 2 }{ 9 }} 
   $ & $ \frac{ 4 }{ 3  } 
   $ & $ \frac{ 2 }{ 3  } 
   $ & $ -\sqrt{\frac{ 8 }{ 9 
     }} $ \\
$ \Sigmac \phi $& $ 0 $ 
 & $ 0 $ & $ 0 $ 
 & $ 0 $ 
 & $ -\sqrt{ 2 } $ 
 & $ 0 $ & $ 0 $ 
 & $ -\sqrt{\frac{ 8 }{ 3 }} 
   $ & $ \sqrt{\frac{ 1 }{ 3 
    }} $ & $ 0 $ & $ 0 $ 
 & $ 0 $ & $ 0 $ 
 & $ 0 $ 
 & $ -\frac{ 5 }{ 3  } $ 
 & $ \sqrt{\frac{ 8 }{ 3 }} 
   $ & $ 0 $ 
 & $ \sqrt{\frac{ 2 }{ 9 }} 
   $ & $ 0 $ 
 & $ -\sqrt{\frac{ 8 }{ 9 }} 
   $ & $ 0 $ 
 & $ \frac{ 2 }{ 3  } $ \\
$ \SigmaS \DsS $& $ 0 $ 
 & $ 0 $ 
 & $ \sqrt{\frac{ 8 }{ 3 }} 
   $ & $ -\sqrt{\frac{ 8 }{ 9 
     }} $ & $ 0 $ 
 & $ -\frac{ 4 }{ 3  } $ 
 & $ 0 $ 
 & $ \sqrt{\frac{ 16 }{ 3 }} 
   $ & $ -\sqrt{\frac{ 8 }{ 3 
     }} $ 
 & $ -\frac{ 16 }{ 3  } $ 
 & $ 0 $ & $ 0 $ 
 & $ 0 $ & $ 0 $ 
 & $ \sqrt{\frac{ 8 }{ 9 }} 
   $ & $ 0 $ 
 & $ \sqrt{\frac{ 8 }{ 9 }} 
   $ & $ \frac{ 4 }{ 3  } 
   $ & $ -\sqrt{\frac{ 8 }{ 9 
     }} $ 
 & $ -\frac{ 8 }{ 3  } $ 
 & $ \frac{ 2 }{ 3  } $ 
 & $ -\sqrt{\frac{ 8 }{ 9 }} 
   $ \\
$ \SigmacS \phi $& $ 0 $ 
 & $ 0 $ & $ 0 $ 
 & $ 0 $ & $ -2 $ 
 & $ 0 $ & $ 0 $ 
 & $ \sqrt{\frac{ 4 }{ 3 }} 
   $ & $ -\sqrt{\frac{ 8 }{ 3 
     }} $ & $ 0 $ 
 & $ 0 $ & $ 0 $ 
 & $ 0 $ & $ 0 $ 
 & $ -\sqrt{\frac{ 8 }{ 9 }} 
   $ & $ -\sqrt{\frac{ 4 }{ 3 
     }} $ & $ 0 $ 
 & $ \frac{ 2 }{ 3  } $ 
 & $ 0 $ 
 & $ \frac{ 2 }{ 3  } $ 
 & $ 0 $ 
 & $ \sqrt{\frac{ 8 }{ 9 }} 
   $ \\
$ \CascadacS \kaonS $
 & $ \sqrt{ 2 } $ 
 & $ \sqrt{\frac{ 4 }{ 3 }} 
   $ & $ 0 $ & $ 0 $ 
 & $ -\sqrt{ 8 } $ 
 & $ \sqrt{ 2 } $ 
 & $ -\sqrt{\frac{ 2 }{ 3 }} 
   $ & $ 0 $ 
 & $ -\sqrt{\frac{ 4 }{ 3 }} 
   $ & $ 0 $ 
 & $ -\frac{ 2 }{ 3  } $ 
 & $ -\sqrt{\frac{ 2 }{ 9 }} 
   $ & $ -\sqrt{\frac{ 128 }{ 9 
     }} $ 
 & $ -\frac{ 8 }{ 3  } $ 
 & $ -\frac{ 2 }{ 3  } $ 
 & $ 0 $ & $ 0 $ 
 & $ -\sqrt{\frac{ 8 }{ 9 }} 
   $ & $ \frac{ 2 }{ 3  } 
   $ & $ -\sqrt{\frac{ 8 }{ 9 
     }} $ & $ \sqrt{\frac{ 8 
    }{ 9 }} $ 
 & $ -\frac{ 10 }{ 3  } 
   $ \\
\end{tabular}
\end{ruledtabular}
\label{tab:i1j12s0c1}
\end{sidewaystable}

\begin{sidewaystable}
\centering
\caption{
  $ I=2$, $ J=1/2$, $ S=0$, $ C= 1$.
Baryon-Meson states with more
than one $c$ quark have not been included.
}
\begin{ruledtabular}
\begin{tabular}{rrrrrrrrrrrrrrrrrrrrrrrrrrrrr}
 &  $ \Sigmac \pion $ 
 &  $ \Delta \DS $ &  $ \Sigmac \rho 
  $ &  $ \SigmacS \rho $\\
$ \Sigmac \pion $& $ 2 $ 
 & $ \sqrt{ 8 } $ 
 & $ \sqrt{\frac{ 16 }{ 3 }} 
   $ & $ -\sqrt{\frac{ 8 }{ 3 
     }} $ \\
$ \Delta \DS $
 & $ \sqrt{ 8 } $ 
 & $ 0 $ 
 & $ \sqrt{\frac{ 8 }{ 3 }} 
   $ & $ -\sqrt{\frac{ 4 }{ 3 
     }} $ \\
$ \Sigmac \rho $
 & $ \sqrt{\frac{ 16 }{ 3 }} 
   $ & $ \sqrt{\frac{ 8 }{ 3 
    }} $ 
 & $ -\frac{ 2 }{ 3  } $ 
 & $ -\sqrt{\frac{ 8 }{ 9 }} 
   $ \\
$ \SigmacS \rho $
 & $ -\sqrt{\frac{ 8 }{ 3 }} 
   $ & $ -\sqrt{\frac{ 4 }{ 3 
     }} $ 
 & $ -\sqrt{\frac{ 8 }{ 9 }} 
   $ & $ -\frac{ 4 }{ 3  } 
   $ \\
\end{tabular}
\end{ruledtabular}
\label{tab:i2j12s0c1}
\end{sidewaystable}

\begin{sidewaystable}
\centering
\caption{
  $ I=0$, $ J=3/2$, $ S=0$, $ C= 1$.
Baryon-Meson states with more
than one $c$ quark have not been included.
}
\begin{ruledtabular}
\begin{tabular}{rrrrrrrrrrrrrrrrrrrrrrrrrrrrr}
 &  $ \SigmacS \pion $ 
 &  $ \Nucleon \DS $ &  $ \Lambdac 
  \omega $ &  $ \CascadacS \kaon $ 
 &  $ \Lambda \DsS $ &  $ \Sigmac \rho 
  $ &  $ \SigmacS \rho $ 
 &  $ \Lambdac \phi $ &  $ \Cascadac 
  \kaonS $ &  $ \Cascadacp \kaonS $ 
 &  $ \CascadacS \kaonS $\\
$ \SigmacS \pion $& $ -4 $ 
 & $ -\sqrt{ 2 } $ 
 & $ 0 $ 
 & $ -\sqrt{ 3 } $ 
 & $ 0 $ 
 & $ -\sqrt{\frac{ 16 }{ 3 }} 
   $ & $ -\sqrt{\frac{ 80 }{ 3 
     }} $ & $ 0 $ 
 & $ \sqrt{ 3 } $ 
 & $ -1 $ 
 & $ -\sqrt{ 5 } $ \\
$ \Nucleon \DS $
 & $ -\sqrt{ 2 } $ 
 & $ 0 $ & $ \sqrt{ 6 } 
   $ & $ 0 $ & $ 0 $ 
 & $ \sqrt{\frac{ 2 }{ 3 }} 
   $ & $ \sqrt{\frac{ 10 }{ 3 
    }} $ & $ 0 $ & $ 0 $ 
 & $ 0 $ & $ 0 $ \\
$ \Lambdac \omega $& $ 0 $ 
 & $ \sqrt{ 6 } $ 
 & $ 0 $ & $ 1 $ 
 & $ 0 $ & $ 2 $ 
 & $ \sqrt{ 20 } $ 
 & $ 0 $ & $ -1 $ 
 & $ \sqrt{\frac{ 1 }{ 3 }} 
   $ & $ \sqrt{\frac{ 5 }{ 3 
    }} $ \\
$ \CascadacS \kaon $
 & $ -\sqrt{ 3 } $ 
 & $ 0 $ & $ 1 $ 
 & $ -2 $ 
 & $ -\sqrt{ 2 } $ 
 & $ -1 $ 
 & $ -\sqrt{ 5 } $ 
 & $ \sqrt{ 2 } $ 
 & $ 0 $ 
 & $ -\sqrt{\frac{ 4 }{ 3 }} 
   $ & $ -\sqrt{\frac{ 20 }{ 3 
     }} $ \\
$ \Lambda \DsS $& $ 0 $ 
 & $ 0 $ & $ 0 $ 
 & $ -\sqrt{ 2 } $ 
 & $ 0 $ & $ 0 $ 
 & $ 0 $ & $ -2 $ 
 & $ \sqrt{ 2 } $ 
 & $ \sqrt{\frac{ 2 }{ 3 }} 
   $ & $ \sqrt{\frac{ 10 }{ 3 
    }} $ \\
$ \Sigmac \rho $
 & $ -\sqrt{\frac{ 16 }{ 3 }} 
   $ & $ \sqrt{\frac{ 2 }{ 3 
    }} $ & $ 2 $ 
 & $ -1 $ & $ 0 $ 
 & $ -\frac{ 8 }{ 3  } $ 
 & $ -\sqrt{\frac{ 20 }{ 9 }} 
   $ & $ 0 $ & $ 1 $ 
 & $ -\sqrt{\frac{ 1 }{ 3 }} 
   $ & $ -\sqrt{\frac{ 5 }{ 3 
     }} $ \\
$ \SigmacS \rho $
 & $ -\sqrt{\frac{ 80 }{ 3 }} 
   $ & $ \sqrt{\frac{ 10 }{ 3 
    }} $ & $ \sqrt{ 20 } 
   $ & $ -\sqrt{ 5 } $ 
 & $ 0 $ 
 & $ -\sqrt{\frac{ 20 }{ 9 }} 
   $ & $ -\frac{ 16 }{ 3  } 
   $ & $ 0 $ 
 & $ \sqrt{ 5 } $ 
 & $ -\sqrt{\frac{ 5 }{ 3 }} 
   $ & $ -\sqrt{\frac{ 25 }{ 3 
     }} $ \\
$ \Lambdac \phi $& $ 0 $ 
 & $ 0 $ & $ 0 $ 
 & $ \sqrt{ 2 } $ 
 & $ -2 $ & $ 0 $ 
 & $ 0 $ & $ 0 $ 
 & $ -\sqrt{ 2 } $ 
 & $ -\sqrt{\frac{ 2 }{ 3 }} 
   $ & $ -\sqrt{\frac{ 10 }{ 3 
     }} $ \\
$ \Cascadac \kaonS $
 & $ \sqrt{ 3 } $ 
 & $ 0 $ & $ -1 $ 
 & $ 0 $ & $ \sqrt{ 2 } 
   $ & $ 1 $ 
 & $ \sqrt{ 5 } $ 
 & $ -\sqrt{ 2 } $ 
 & $ -2 $ 
 & $ \sqrt{\frac{ 4 }{ 3 }} 
   $ & $ \sqrt{\frac{ 20 }{ 3 
    }} $ \\
$ \Cascadacp \kaonS $& $ -1 $ 
 & $ 0 $ 
 & $ \sqrt{\frac{ 1 }{ 3 }} 
   $ & $ -\sqrt{\frac{ 4 }{ 3 
     }} $ & $ \sqrt{\frac{ 2 
    }{ 3 }} $ 
 & $ -\sqrt{\frac{ 1 }{ 3 }} 
   $ & $ -\sqrt{\frac{ 5 }{ 3 
     }} $ 
 & $ -\sqrt{\frac{ 2 }{ 3 }} 
   $ & $ \sqrt{\frac{ 4 }{ 3 
    }} $ & $ -2 $ 
 & $ 0 $ \\
$ \CascadacS \kaonS $
 & $ -\sqrt{ 5 } $ 
 & $ 0 $ 
 & $ \sqrt{\frac{ 5 }{ 3 }} 
   $ & $ -\sqrt{\frac{ 20 }{ 3 
     }} $ & $ \sqrt{\frac{ 10 
    }{ 3 }} $ 
 & $ -\sqrt{\frac{ 5 }{ 3 }} 
   $ & $ -\sqrt{\frac{ 25 }{ 3 
     }} $ 
 & $ -\sqrt{\frac{ 10 }{ 3 }} 
   $ & $ \sqrt{\frac{ 20 }{ 3 
    }} $ & $ 0 $ 
 & $ -2 $ \\
\end{tabular}
\end{ruledtabular}
\label{tab:i0j32s0c1}
\end{sidewaystable}

\begin{sidewaystable}
\centering
\caption{
  $ I=1$, $ J=3/2$, $ S=0$, $ C= 1$.
Baryon-Meson states with more
than one $c$ quark have not been included.
}
\begin{ruledtabular}
\begin{tabular}{rrrrrrrrrrrrrrrrrrrrrrrrrrrrr}
 &  $ \SigmacS \pion $ 
 &  $ \Nucleon \DS $ &  $ \Lambdac \rho 
  $ &  $ \SigmacS \eta $ 
 &  $ \Delta \D $ &  $ \CascadacS \kaon 
  $ &  $ \Delta \DS $ 
 &  $ \Sigmac \rho $ &  $ \Sigmac 
  \omega $ &  $ \SigmacS \rho $ 
 &  $ \SigmacS \omega $ &  $ \Sigma 
  \DsS $ &  $ \SigmaS \Ds $ 
 &  $ \Cascadac \kaonS $ &  $ \Cascadacp 
  \kaonS $ &  $ \Sigmac \phi $ 
 &  $ \SigmacS \etap $ &  $ \SigmaS 
  \DsS $ &  $ \SigmacS \phi $ 
 &  $ \CascadacS \kaonS $\\
$ \SigmacS \pion $& $ -2 $ 
 & $ -\sqrt{\frac{ 4 }{ 3 }} 
   $ & $ -\sqrt{ 8 } $ 
 & $ 0 $ & $ 1 $ 
 & $ -\sqrt{ 2 } $ 
 & $ \sqrt{\frac{ 5 }{ 3 }} 
   $ & $ -\sqrt{\frac{ 4 }{ 3 
     }} $ & $ 0 $ 
 & $ -\sqrt{\frac{ 20 }{ 3 }} 
   $ & $ 0 $ & $ 0 $ 
 & $ 0 $ & $ \sqrt{ 2 } 
   $ & $ -\sqrt{\frac{ 2 }{ 3 
     }} $ & $ 0 $ 
 & $ 0 $ & $ 0 $ 
 & $ 0 $ 
 & $ -\sqrt{\frac{ 10 }{ 3 }} 
   $ \\
$ \Nucleon \DS $
 & $ -\sqrt{\frac{ 4 }{ 3 }} 
   $ & $ -\frac{ 4 }{ 3  } 
   $ & $ \sqrt{ 6 } $ 
 & $ \sqrt{\frac{ 2 }{ 9 }} 
   $ & $ -\sqrt{\frac{ 16 }{ 3 
     }} $ & $ 0 $ 
 & $ -\sqrt{\frac{ 80 }{ 9 }} 
   $ & $ \frac{ 2 }{ 3  } 
   $ & $ -\sqrt{\frac{ 2 }{ 9 
     }} $ & $ \sqrt{\frac{ 20 
    }{ 9 }} $ 
 & $ -\sqrt{\frac{ 10 }{ 9 }} 
   $ & $ \frac{ 4 }{ 3  } 
   $ & $ -\sqrt{\frac{ 4 }{ 3 
     }} $ & $ 0 $ 
 & $ 0 $ & $ 0 $ 
 & $ \frac{ 2 }{ 3  } $ 
 & $ -\sqrt{\frac{ 20 }{ 9 }} 
   $ & $ 0 $ & $ 0 $ \\
$ \Lambdac \rho $
 & $ -\sqrt{ 8 } $ 
 & $ \sqrt{ 6 } $ 
 & $ 0 $ & $ 0 $ 
 & $ 0 $ & $ -1 $ 
 & $ 0 $ & $ 0 $ 
 & $ -\sqrt{\frac{ 4 }{ 3 }} 
   $ & $ 0 $ 
 & $ -\sqrt{\frac{ 20 }{ 3 }} 
   $ & $ 0 $ & $ 0 $ 
 & $ 1 $ 
 & $ -\sqrt{\frac{ 1 }{ 3 }} 
   $ & $ 0 $ & $ 0 $ 
 & $ 0 $ & $ 0 $ 
 & $ -\sqrt{\frac{ 5 }{ 3 }} 
   $ \\
$ \SigmacS \eta $& $ 0 $ 
 & $ \sqrt{\frac{ 2 }{ 9 }} 
   $ & $ 0 $ & $ 0 $ 
 & $ \sqrt{\frac{ 2 }{ 3 }} 
   $ & $ -\sqrt{ 3 } $ 
 & $ \sqrt{\frac{ 10 }{ 9 }} 
   $ & $ 0 $ & $ 0 $ 
 & $ 0 $ & $ 0 $ 
 & $ \sqrt{\frac{ 8 }{ 9 }} 
   $ & $ -\sqrt{\frac{ 2 }{ 3 
     }} $ & $ \sqrt{ 3 } 
   $ & $ -1 $ & $ 0 $ 
 & $ 0 $ 
 & $ -\sqrt{\frac{ 10 }{ 9 }} 
   $ & $ 0 $ 
 & $ -\sqrt{ 5 } $ \\
$ \Delta \D $& $ 1 $ 
 & $ -\sqrt{\frac{ 16 }{ 3 }} 
   $ & $ 0 $ 
 & $ \sqrt{\frac{ 2 }{ 3 }} 
   $ & $ -4 $ & $ 0 $ 
 & $ -\sqrt{\frac{ 80 }{ 3 }} 
   $ & $ -\sqrt{\frac{ 4 }{ 3 
     }} $ 
 & $ -\sqrt{\frac{ 8 }{ 3 }} 
   $ & $ \sqrt{\frac{ 5 }{ 3 
    }} $ & $ \sqrt{\frac{ 10 
    }{ 3 }} $ 
 & $ \sqrt{\frac{ 16 }{ 3 }} 
   $ & $ -2 $ & $ 0 $ 
 & $ 0 $ & $ 0 $ 
 & $ \sqrt{\frac{ 4 }{ 3 }} 
   $ & $ -\sqrt{\frac{ 20 }{ 3 
     }} $ & $ 0 $ 
 & $ 0 $ \\
$ \CascadacS \kaon $
 & $ -\sqrt{ 2 } $ 
 & $ 0 $ & $ -1 $ 
 & $ -\sqrt{ 3 } $ 
 & $ 0 $ & $ 0 $ 
 & $ 0 $ 
 & $ -\sqrt{\frac{ 2 }{ 3 }} 
   $ & $ -\sqrt{\frac{ 1 }{ 3 
     }} $ 
 & $ -\sqrt{\frac{ 10 }{ 3 }} 
   $ & $ -\sqrt{\frac{ 5 }{ 3 
     }} $ & $ \sqrt{\frac{ 2 
    }{ 3 }} $ 
 & $ \sqrt{ 2 } $ 
 & $ 2 $ & $ 0 $ 
 & $ -\sqrt{\frac{ 2 }{ 3 }} 
   $ & $ 0 $ 
 & $ \sqrt{\frac{ 10 }{ 3 }} 
   $ & $ -\sqrt{\frac{ 10 }{ 3 
     }} $ & $ 0 $ \\
$ \Delta \DS $
 & $ \sqrt{\frac{ 5 }{ 3 }} 
   $ & $ -\sqrt{\frac{ 80 }{ 9 
     }} $ & $ 0 $ 
 & $ \sqrt{\frac{ 10 }{ 9 }} 
   $ & $ -\sqrt{\frac{ 80 }{ 3 
     }} $ & $ 0 $ 
 & $ -\frac{ 20 }{ 3  } $ 
 & $ \sqrt{\frac{ 20 }{ 9 }} 
   $ & $ \sqrt{\frac{ 40 }{ 9 
    }} $ & $ \frac{ 1 }{ 3 
     } $ & $ \sqrt{\frac{ 2 
    }{ 9 }} $ 
 & $ \sqrt{\frac{ 80 }{ 9 }} 
   $ & $ -\sqrt{\frac{ 20 }{ 3 
     }} $ & $ 0 $ 
 & $ 0 $ & $ 0 $ 
 & $ \sqrt{\frac{ 20 }{ 9 }} 
   $ & $ -\frac{ 10 }{ 3  } 
   $ & $ 0 $ & $ 0 $ \\
$ \Sigmac \rho $
 & $ -\sqrt{\frac{ 4 }{ 3 }} 
   $ & $ \frac{ 2 }{ 3  } 
   $ & $ 0 $ & $ 0 $ 
 & $ -\sqrt{\frac{ 4 }{ 3 }} 
   $ & $ -\sqrt{\frac{ 2 }{ 3 
     }} $ & $ \sqrt{\frac{ 20 
    }{ 9 }} $ 
 & $ -\frac{ 2 }{ 3  } $ 
 & $ \sqrt{\frac{ 32 }{ 9 }} 
   $ & $ -\sqrt{\frac{ 20 }{ 9 
     }} $ 
 & $ -\sqrt{\frac{ 40 }{ 9 }} 
   $ & $ 0 $ & $ 0 $ 
 & $ \sqrt{\frac{ 2 }{ 3 }} 
   $ & $ -\sqrt{\frac{ 2 }{ 9 
     }} $ & $ 0 $ 
 & $ 0 $ & $ 0 $ 
 & $ 0 $ 
 & $ -\sqrt{\frac{ 10 }{ 9 }} 
   $ \\
$ \Sigmac \omega $& $ 0 $ 
 & $ -\sqrt{\frac{ 2 }{ 9 }} 
   $ & $ -\sqrt{\frac{ 4 }{ 3 
     }} $ & $ 0 $ 
 & $ -\sqrt{\frac{ 8 }{ 3 }} 
   $ & $ -\sqrt{\frac{ 1 }{ 3 
     }} $ & $ \sqrt{\frac{ 40 
    }{ 9 }} $ 
 & $ \sqrt{\frac{ 32 }{ 9 }} 
   $ & $ \frac{ 4 }{ 3  } 
   $ & $ -\sqrt{\frac{ 40 }{ 9 
     }} $ 
 & $ -\sqrt{\frac{ 20 }{ 9 }} 
   $ & $ 0 $ & $ 0 $ 
 & $ \sqrt{\frac{ 1 }{ 3 }} 
   $ & $ -\frac{ 1 }{ 3  } 
   $ & $ 0 $ & $ 0 $ 
 & $ 0 $ & $ 0 $ 
 & $ -\sqrt{\frac{ 5 }{ 9 }} 
   $ \\
$ \SigmacS \rho $
 & $ -\sqrt{\frac{ 20 }{ 3 }} 
   $ & $ \sqrt{\frac{ 20 }{ 9 
    }} $ & $ 0 $ & $ 0 $ 
 & $ \sqrt{\frac{ 5 }{ 3 }} 
   $ & $ -\sqrt{\frac{ 10 }{ 3 
     }} $ & $ \frac{ 1 }{ 3 
     } $ 
 & $ -\sqrt{\frac{ 20 }{ 9 }} 
   $ & $ -\sqrt{\frac{ 40 }{ 9 
     }} $ 
 & $ -\frac{ 10 }{ 3  } $ 
 & $ -\sqrt{\frac{ 32 }{ 9 }} 
   $ & $ 0 $ & $ 0 $ 
 & $ \sqrt{\frac{ 10 }{ 3 }} 
   $ & $ -\sqrt{\frac{ 10 }{ 9 
     }} $ & $ 0 $ 
 & $ 0 $ & $ 0 $ 
 & $ 0 $ 
 & $ -\sqrt{\frac{ 50 }{ 9 }} 
   $ \\
$ \SigmacS \omega $& $ 0 $ 
 & $ -\sqrt{\frac{ 10 }{ 9 }} 
   $ & $ -\sqrt{\frac{ 20 }{ 3 
     }} $ & $ 0 $ 
 & $ \sqrt{\frac{ 10 }{ 3 }} 
   $ & $ -\sqrt{\frac{ 5 }{ 3 
     }} $ & $ \sqrt{\frac{ 2 
    }{ 9 }} $ 
 & $ -\sqrt{\frac{ 40 }{ 9 }} 
   $ & $ -\sqrt{\frac{ 20 }{ 9 
     }} $ 
 & $ -\sqrt{\frac{ 32 }{ 9 }} 
   $ & $ -\frac{ 4 }{ 3  } 
   $ & $ 0 $ & $ 0 $ 
 & $ \sqrt{\frac{ 5 }{ 3 }} 
   $ & $ -\sqrt{\frac{ 5 }{ 9 
     }} $ & $ 0 $ 
 & $ 0 $ & $ 0 $ 
 & $ 0 $ 
 & $ -\frac{ 5 }{ 3  } $ \\
$ \Sigma \DsS $& $ 0 $ 
 & $ \frac{ 4 }{ 3  } $ 
 & $ 0 $ 
 & $ \sqrt{\frac{ 8 }{ 9 }} 
   $ & $ \sqrt{\frac{ 16 }{ 3 
    }} $ & $ \sqrt{\frac{ 2 
    }{ 3 }} $ 
 & $ \sqrt{\frac{ 80 }{ 9 }} 
   $ & $ 0 $ & $ 0 $ 
 & $ 0 $ & $ 0 $ 
 & $ -\frac{ 4 }{ 3  } $ 
 & $ \sqrt{\frac{ 4 }{ 3 }} 
   $ & $ \sqrt{ 6 } $ 
 & $ -\sqrt{\frac{ 2 }{ 9 }} 
   $ & $ -\frac{ 2 }{ 3  } 
   $ & $ -\frac{ 2 }{ 3  } 
   $ & $ \sqrt{\frac{ 20 }{ 9 
    }} $ 
 & $ -\sqrt{\frac{ 20 }{ 9 }} 
   $ & $ -\sqrt{\frac{ 10 }{ 9 
     }} $ \\
$ \SigmaS \Ds $& $ 0 $ 
 & $ -\sqrt{\frac{ 4 }{ 3 }} 
   $ & $ 0 $ 
 & $ -\sqrt{\frac{ 2 }{ 3 }} 
   $ & $ -2 $ 
 & $ \sqrt{ 2 } $ 
 & $ -\sqrt{\frac{ 20 }{ 3 }} 
   $ & $ 0 $ & $ 0 $ 
 & $ 0 $ & $ 0 $ 
 & $ \sqrt{\frac{ 4 }{ 3 }} 
   $ & $ -1 $ & $ 0 $ 
 & $ -\sqrt{\frac{ 8 }{ 3 }} 
   $ & $ \sqrt{\frac{ 4 }{ 3 
    }} $ & $ \sqrt{\frac{ 1 
    }{ 3 }} $ 
 & $ -\sqrt{\frac{ 5 }{ 3 }} 
   $ & $ -\sqrt{\frac{ 5 }{ 3 
     }} $ & $ \sqrt{\frac{ 10 
    }{ 3 }} $ \\
$ \Cascadac \kaonS $
 & $ \sqrt{ 2 } $ 
 & $ 0 $ & $ 1 $ 
 & $ \sqrt{ 3 } $ 
 & $ 0 $ & $ 2 $ 
 & $ 0 $ 
 & $ \sqrt{\frac{ 2 }{ 3 }} 
   $ & $ \sqrt{\frac{ 1 }{ 3 
    }} $ & $ \sqrt{\frac{ 10 
    }{ 3 }} $ 
 & $ \sqrt{\frac{ 5 }{ 3 }} 
   $ & $ \sqrt{ 6 } $ 
 & $ 0 $ & $ 0 $ 
 & $ 0 $ 
 & $ -\sqrt{\frac{ 2 }{ 3 }} 
   $ & $ 0 $ & $ 0 $ 
 & $ -\sqrt{\frac{ 10 }{ 3 }} 
   $ & $ 0 $ \\
$ \Cascadacp \kaonS $
 & $ -\sqrt{\frac{ 2 }{ 3 }} 
   $ & $ 0 $ 
 & $ -\sqrt{\frac{ 1 }{ 3 }} 
   $ & $ -1 $ & $ 0 $ 
 & $ 0 $ & $ 0 $ 
 & $ -\sqrt{\frac{ 2 }{ 9 }} 
   $ & $ -\frac{ 1 }{ 3  } 
   $ & $ -\sqrt{\frac{ 10 }{ 9 
     }} $ 
 & $ -\sqrt{\frac{ 5 }{ 9 }} 
   $ & $ -\sqrt{\frac{ 2 }{ 9 
     }} $ 
 & $ -\sqrt{\frac{ 8 }{ 3 }} 
   $ & $ 0 $ 
 & $ \frac{ 4 }{ 3  } $ 
 & $ -\sqrt{\frac{ 50 }{ 9 }} 
   $ & $ 0 $ 
 & $ \sqrt{\frac{ 40 }{ 9 }} 
   $ & $ \sqrt{\frac{ 10 }{ 9 
    }} $ 
 & $ -\sqrt{\frac{ 20 }{ 9 }} 
   $ \\
$ \Sigmac \phi $& $ 0 $ 
 & $ 0 $ & $ 0 $ 
 & $ 0 $ & $ 0 $ 
 & $ -\sqrt{\frac{ 2 }{ 3 }} 
   $ & $ 0 $ & $ 0 $ 
 & $ 0 $ & $ 0 $ 
 & $ 0 $ 
 & $ -\frac{ 2 }{ 3  } $ 
 & $ \sqrt{\frac{ 4 }{ 3 }} 
   $ & $ -\sqrt{\frac{ 2 }{ 3 
     }} $ 
 & $ -\sqrt{\frac{ 50 }{ 9 }} 
   $ & $ 0 $ & $ 0 $ 
 & $ -\sqrt{\frac{ 20 }{ 9 }} 
   $ & $ 0 $ 
 & $ \sqrt{\frac{ 10 }{ 9 }} 
   $ \\
$ \SigmacS \etap $& $ 0 $ 
 & $ \frac{ 2 }{ 3  } $ 
 & $ 0 $ & $ 0 $ 
 & $ \sqrt{\frac{ 4 }{ 3 }} 
   $ & $ 0 $ 
 & $ \sqrt{\frac{ 20 }{ 9 }} 
   $ & $ 0 $ & $ 0 $ 
 & $ 0 $ & $ 0 $ 
 & $ -\frac{ 2 }{ 3  } $ 
 & $ \sqrt{\frac{ 1 }{ 3 }} 
   $ & $ 0 $ & $ 0 $ 
 & $ 0 $ & $ 0 $ 
 & $ \sqrt{\frac{ 5 }{ 9 }} 
   $ & $ 0 $ & $ 0 $ \\
$ \SigmaS \DsS $& $ 0 $ 
 & $ -\sqrt{\frac{ 20 }{ 9 }} 
   $ & $ 0 $ 
 & $ -\sqrt{\frac{ 10 }{ 9 }} 
   $ & $ -\sqrt{\frac{ 20 }{ 3 
     }} $ & $ \sqrt{\frac{ 10 
    }{ 3 }} $ 
 & $ -\frac{ 10 }{ 3  } $ 
 & $ 0 $ & $ 0 $ 
 & $ 0 $ & $ 0 $ 
 & $ \sqrt{\frac{ 20 }{ 9 }} 
   $ & $ -\sqrt{\frac{ 5 }{ 3 
     }} $ & $ 0 $ 
 & $ \sqrt{\frac{ 40 }{ 9 }} 
   $ & $ -\sqrt{\frac{ 20 }{ 9 
     }} $ & $ \sqrt{\frac{ 5 
    }{ 9 }} $ 
 & $ -\frac{ 5 }{ 3  } $ 
 & $ -\frac{ 1 }{ 3  } $ 
 & $ \sqrt{\frac{ 2 }{ 9 }} 
   $ \\
$ \SigmacS \phi $& $ 0 $ 
 & $ 0 $ & $ 0 $ 
 & $ 0 $ & $ 0 $ 
 & $ -\sqrt{\frac{ 10 }{ 3 }} 
   $ & $ 0 $ & $ 0 $ 
 & $ 0 $ & $ 0 $ 
 & $ 0 $ 
 & $ -\sqrt{\frac{ 20 }{ 9 }} 
   $ & $ -\sqrt{\frac{ 5 }{ 3 
     }} $ 
 & $ -\sqrt{\frac{ 10 }{ 3 }} 
   $ & $ \sqrt{\frac{ 10 }{ 9 
    }} $ & $ 0 $ & $ 0 $ 
 & $ -\frac{ 1 }{ 3  } $ 
 & $ 0 $ 
 & $ -\sqrt{\frac{ 2 }{ 9 }} 
   $ \\
$ \CascadacS \kaonS $
 & $ -\sqrt{\frac{ 10 }{ 3 }} 
   $ & $ 0 $ 
 & $ -\sqrt{\frac{ 5 }{ 3 }} 
   $ & $ -\sqrt{ 5 } $ 
 & $ 0 $ & $ 0 $ 
 & $ 0 $ 
 & $ -\sqrt{\frac{ 10 }{ 9 }} 
   $ & $ -\sqrt{\frac{ 5 }{ 9 
     }} $ 
 & $ -\sqrt{\frac{ 50 }{ 9 }} 
   $ & $ -\frac{ 5 }{ 3  } 
   $ & $ -\sqrt{\frac{ 10 }{ 9 
     }} $ & $ \sqrt{\frac{ 10 
    }{ 3 }} $ & $ 0 $ 
 & $ -\sqrt{\frac{ 20 }{ 9 }} 
   $ & $ \sqrt{\frac{ 10 }{ 9 
    }} $ & $ 0 $ 
 & $ \sqrt{\frac{ 2 }{ 9 }} 
   $ & $ -\sqrt{\frac{ 2 }{ 9 
     }} $ 
 & $ -\frac{ 4 }{ 3  } $ \\
\end{tabular}
\end{ruledtabular}
\label{tab:i1j32s0c1}
\end{sidewaystable}
\begin{sidewaystable}
\centering
\caption{
  $ I=2$, $ J=3/2$, $ S=0$, $ C= 1$.
Baryon-Meson states with more
than one $c$ quark have not been included.
}
\begin{ruledtabular}
\begin{tabular}{rrrrrrrrrrrrrrrrrrrrrrrrrrrrr}
 &  $ \SigmacS \pion $ 
 &  $ \Delta \D $ &  $ \Delta \DS $ 
 &  $ \Sigmac \rho $ &  $ \SigmacS \rho 
  $\\
$ \SigmacS \pion $& $ 2 $ 
 & $ \sqrt{ 3 } $ 
 & $ \sqrt{ 5 } $ 
 & $ \sqrt{\frac{ 4 }{ 3 }} 
   $ & $ \sqrt{\frac{ 20 }{ 3 
    }} $ \\
$ \Delta \D $
 & $ \sqrt{ 3 } $ 
 & $ 0 $ & $ 0 $ 
 & $ -2 $ 
 & $ \sqrt{ 5 } $ \\
$ \Delta \DS $
 & $ \sqrt{ 5 } $ 
 & $ 0 $ & $ 0 $ 
 & $ \sqrt{\frac{ 20 }{ 3 }} 
   $ & $ \sqrt{\frac{ 1 }{ 3 
    }} $ \\
$ \Sigmac \rho $
 & $ \sqrt{\frac{ 4 }{ 3 }} 
   $ & $ -2 $ 
 & $ \sqrt{\frac{ 20 }{ 3 }} 
   $ & $ \frac{ 10 }{ 3  } 
   $ & $ -\sqrt{\frac{ 20 }{ 9 
     }} $ \\
$ \SigmacS \rho $
 & $ \sqrt{\frac{ 20 }{ 3 }} 
   $ & $ \sqrt{ 5 } $ 
 & $ \sqrt{\frac{ 1 }{ 3 }} 
   $ & $ -\sqrt{\frac{ 20 }{ 9 
     }} $ & $ \frac{ 2 }{ 3 
     } $ \\
\end{tabular}
\end{ruledtabular}
\label{tab:i2j32s0c1}
\end{sidewaystable}

\begin{sidewaystable}
\centering
\caption{
  $ I=0$, $ J=5/2$, $ S=0$, $ C= 1$.
Baryon-Meson states with more
than one $c$ quark have not been included.
}
\begin{ruledtabular}
\begin{tabular}{rrrrrrrrrrrrrrrrrrrrrrrrrrrrr}
 &  $ \SigmacS \rho $ &  $ \CascadacS 
  \kaonS $\\
$ \SigmacS \rho $& $ -2 $ 
 & $ 0 $ \\
$ \CascadacS \kaonS $& $ 0 $ 
 & $ -2 $ \\
\end{tabular}
\end{ruledtabular}
\label{tab:i0j52s0c1}
\end{sidewaystable}

\begin{sidewaystable}
\centering
\caption{
  $ I=1$, $ J=5/2$, $ S=0$, $ C= 1$.
Baryon-Meson states with more
than one $c$ quark have not been included.
}
\begin{ruledtabular}
\begin{tabular}{rrrrrrrrrrrrrrrrrrrrrrrrrrrrr}
 &  $ \Delta \DS $ &  $ \SigmacS 
  \rho $ &  $ \SigmacS \omega $ 
 &  $ \SigmaS \DsS $ &  $ \SigmacS \phi 
  $ &  $ \CascadacS \kaonS $\\
$ \Delta \DS $& $ 0 $ 
 & $ 2 $ & $ \sqrt{ 8 } 
   $ & $ 0 $ & $ 0 $ 
 & $ 0 $ \\
$ \SigmacS \rho $& $ 2 $ 
 & $ 0 $ & $ \sqrt{ 8 } 
   $ & $ 0 $ & $ 0 $ 
 & $ 0 $ \\
$ \SigmacS \omega $
 & $ \sqrt{ 8 } $ 
 & $ \sqrt{ 8 } $ 
 & $ 2 $ & $ 0 $ 
 & $ 0 $ & $ 0 $ \\
$ \SigmaS \DsS $& $ 0 $ 
 & $ 0 $ & $ 0 $ 
 & $ 0 $ & $ -2 $ 
 & $ \sqrt{ 8 } $ \\
$ \SigmacS \phi $& $ 0 $ 
 & $ 0 $ & $ 0 $ 
 & $ -2 $ & $ 0 $ 
 & $ -\sqrt{ 8 } $ \\
$ \CascadacS \kaonS $& $ 0 $ 
 & $ 0 $ & $ 0 $ 
 & $ \sqrt{ 8 } $ 
 & $ -\sqrt{ 8 } $ 
 & $ 2 $ \\
\end{tabular}
\end{ruledtabular}
\label{tab:i1j52s0c1}
\end{sidewaystable}

\begin{sidewaystable}
\centering
\caption{
  $ I=2$, $ J=5/2$, $ S=0$, $ C= 1$.
Baryon-Meson  states with more
than one $c$ quark have not been included.
}
\begin{ruledtabular}
\begin{tabular}{rrrrrrrrrrrrrrrrrrrrrrrrrrrrr}
 &  $ \Delta \DS $ &  $ \SigmacS 
  \rho $\\
$ \Delta \DS $& $ 0 $ 
 & $ \sqrt{ 12 } $ \\
$ \SigmacS \rho $
 & $ \sqrt{ 12 } $ 
 & $ 4 $ \\
\end{tabular}
\end{ruledtabular}
\label{tab:i2j52s0c1}
\end{sidewaystable}

\end{document}